\documentclass[a4paper,onecolumn]{quantumarticle}
\pdfoutput=1

\usepackage[utf8]{inputenc}
\usepackage[english]{babel}
\usepackage[T1]{fontenc}
\usepackage{amsmath}
\usepackage{amssymb}
\usepackage{bm}  % Bold symbols, e.g. greek letters
\usepackage{graphicx}
\usepackage{subcaption}
\usepackage{xcolor}
\usepackage{pgfplots}
\usepackage[section]{placeins} %forcing figures to subsection it belongs
\DeclareUnicodeCharacter{2212}{−}
\DeclareUnicodeCharacter{0002}{-}
\usepgfplotslibrary{groupplots,dateplot}
\usetikzlibrary{patterns,shapes.arrows}
\pgfplotsset{compat=newest}
\usepackage{algorithm}
\usepackage{algpseudocode}
\usepackage{tikz}
\usetikzlibrary{quantikz2}
\usepackage{hyperref}
\usepackage[square,numbers,sort]{natbib}

\definecolor{fraunhofergreen}{RGB}{23,156,125} 
\definecolor{fraunhoferblue}{RGB}{0,91,127} 
\definecolor{fraunhofergrey}{RGB}{166,187,200} 
\definecolor{fraunhoferorange}{RGB}{245,130,32} 
\definecolor{fraunhofergraphit}{RGB}{28,63,82} 
\definecolor{fraunhofersand}{RGB}{211,199,174} 
\definecolor{fraunhoferpetrol}{RGB}{0,133,152}
\definecolor{fraunhoferaqua}{RGB}{57,193,205}  
\definecolor{fraunhoferlime}{RGB}{178,210,53}
\definecolor{fraunhoferyellow}{RGB}{253,185,19}  
\definecolor{fraunhoferred}{RGB}{187,0,86}
\definecolor{fraunhoferwinered}{RGB}{124,21,77} 
\definecolor{plotlyblue}{HTML}{636EFA}  % \definecolor{mediumslateblue99110250}{RGB}{99,110,250}
\definecolor{plotlyred}{HTML}{EF553B}
\definecolor{plotlygreen}{HTML}{00CC96}  % \definecolor{lightseagreen0204150}{RGB}{0,204,150}
\definecolor{plotlypurple}{HTML}{AB63FA}
\definecolor{plotlyorange}{HTML}{FFA15A}  % \definecolor{sandybrown25516190}{RGB}{255,161,90}
\definecolor{plotlycyan}{HTML}{19D3F3}
\definecolor{plotlyrose}{HTML}{FF6692}
\definecolor{plotlygrassgreen}{HTML}{B6E880}
\definecolor{plotlypink}{HTML}{FF97FF}
\definecolor{plotlyyellow}{HTML}{FECB52}
\definecolor{pastelblue}{RGB}{102,197,204} 
\usetikzlibrary{quantikz2}
\newcommand{\qiskit}{\texttt{qiskit}}
\newcommand{\optimizationLevel}{\texttt{optimization\_level}}
\newcommand{\resilienceLevel}{\texttt{resilience\_level}}
\newcommand{\circuitScore}{\texttt{circuit\_score}}
\newcommand{\ibmqehningen}{\texttt{ibmq\_ehningen}}
\newcommand{\ibmtorino}{\texttt{ibm\_torino}}
\newcommand{\ibmsherbrooke}{\texttt{ibm\_sherbrooke}}
\newcommand{\ibmcairo}{\texttt{ibm\_cairo}}
\newcommand{\dwaveAdvantage}{\texttt{Advantage 4.1}}
\renewcommand{\i}{\mathrm{i}}

\newcommand{\C}{\mathbb{C}}

\newcommand{\R}{\mathbb{R}}
\newcommand{\Z}{\mathbb{Z}}
\DeclareMathOperator*{\argmin}{arg\,min}
\newcommand{\pmat}[1]{\begin{pmatrix} #1 \end{pmatrix}}
\newcommand{\dimQcio}{N}
\newcommand{\varQcioEntry}{n}
\newcommand{\varQcio}{\mathbf{\varQcioEntry}}
\newcommand{\rhsQcio}{\mathbf{r}}
\newcommand{\costQcio}{f_1}
\newcommand{\matQcio}{\mathbf{A}}
\newcommand{\linQcio}{\mathbf{L}}
\newcommand{\constQcio}{c}
\newcommand{\constraintQcio}{\mathbf{C}}
\newcommand{\myPenalty}{\varrho}
\newcommand{\costQuio}{f_2}
\newcommand{\matQuio}{\widetilde{\matQcio}}
\newcommand{\linQuio}{\widetilde{\linQcio}}
\newcommand{\constQuio}{\widetilde{\constQcio}}
\newcommand{\matBinToInt}{\mathbf{B}}
\newcommand{\dimQubo}{M}
\newcommand{\varQuboEntry}{b}
\newcommand{\varQubo}{\mathbf{\varQuboEntry}}
\newcommand{\varQuboOpt}{\varQubo^\star}
\newcommand{\costQubo}{f_3}
\newcommand{\costQuboGeneric}{f}
\newcommand{\costQuboGenericOpt}{f^\star}
\newcommand{\matQubo}{\mathbf{Q}}
\newcommand{\matQuboEntry}{q}
\newcommand{\constQuboGeneric}{d}
\newcommand{\numQubits}{\dimQubo}
\newcommand{\problemHam}{H_P}
\newcommand{\mixingHam}{H_M}
\newcommand{\psiOpt}{\psi^\diamond}
\newcommand{\amplOpt}{a^\diamond}
\renewcommand{\ket}[1]{\left| #1 \right\rangle}
\newcommand{\ketSimple}[1]{| #1 \rangle}
\renewcommand{\bra}[1]{\left\langle #1 \right|}
\newcommand{\expect}[1]{\left\langle #1 \right\rangle}
\newcommand{\ampl}{a}
\newcommand{\amplHat}{\widehat{a}}

\newcommand{\numShots}{S}
\newcommand{\prob}{p}
\newcommand{\probHat}{\widehat{p}}
\newcommand{\probTilde}{\widetilde{p}}
\newcommand{\psiHat}{\widehat{\psi}}
\newcommand{\psiTilde}{\widetilde{\psi}}
\newcommand{\fidelity}{F}
\newcommand{\relCostError}{E}
\newcommand{\psiQaoa}{\psi_{\mathrm{QAOA}}}
\newcommand{\bfbeta}{\bm{\beta}}
\newcommand{\bfgamma}{\bm{\gamma}}
\newcommand{\bfbetaOpt}{\bm{\beta}^\diamond}
\newcommand{\bfgammaOpt}{\bm{\gamma}^\diamond}
\newcommand{\repsQaoa}{L}
\newcommand{\numQaoaParams}{P}
\newcommand{\problemOperator}{U_P}
\newcommand{\mixingOperator}{U_M}
\newcommand{\vqeParam}{\theta}
\newcommand{\vqeParams}{\bm{\vqeParam}}
\newcommand{\vqeParamsOpt}{\bm{\vqeParam}^\diamond}
\newcommand{\numVqeParams}{P}
\newcommand{\repsVqe}{L}
\newcommand{\psiVqe}{\psi_{\mathrm{VQE}}}
\newcommand{\vqeAnsatz}{U_{\mathrm{VQE}}}
\newcommand{\cost}{e}  % In the sense of the expectation value <H_P>
\newcommand{\param}{\vartheta}
\newcommand{\params}{\bm{\param}}
\newcommand{\paramsOpt}{\params^\diamond}

\DeclareMathOperator{\hadamard}{H}
\DeclareMathOperator{\RX}{RX}
\DeclareMathOperator{\RY}{RY}
\DeclareMathOperator{\RZ}{RZ}
\DeclareMathOperator{\RZZ}{RZZ}
\DeclareMathOperator{\pauliX}{X}

\DeclareMathOperator{\pauliZ}{Z}
\DeclareMathOperator{\SX}{SX}
\DeclareMathOperator{\CX}{CX}
\DeclareMathOperator{\CZ}{CZ}
\DeclareMathOperator{\ECR}{ECR}
\DeclareMathOperator{\SWAP}{SWAP}
\newcommand{\numTimeslots}{T}
\newcommand{\numCars}{K}
\newcommand{\lamaSeries}[1]{\textsf{lama\_#1p*}}  % e.g. example_1
\newcommand{\lamaEx}[2]{\textsf{lama\_#1p#2}}   % e.g. example_1p2
\newcommand{\trpasym}[1]{\textsf{a-trp\_#1}}
\newcommand{\trpsym}[1]{\textsf{s-trp\_#1}}

\begin{document}

\title{Unlocking Quantum Optimization: A Use Case Study on NISQ Systems}

\author{Andreas Sturm}
\email{andreas.sturm@iao.fraunhofer.de}
\author{Bharadwaj Mummaneni}
\email{bharadwaj.chowdary.mummaneni@iao.fraunhofer.de}
\author{Leon Rullk\"otter}
\email{leon.rullkoetter@iao.fraunhofer.de}
\affiliation{%
	Fraunhofer IAO, 
	Fraunhofer-Institut für Arbeitswirtschaft und Organisation IAO,
	Nobelstra{\ss}e 12,
	70569 Stuttgart,
	Germany}
\maketitle

\begin{abstract}
The major advances in quantum computing over the last few decades have
sparked great interest in applying it to solve
the most challenging computational problems in a wide variety of areas.
One of the most pronounced domains here are optimization problems and a number
of algorithmic approaches have been proposed for their solution. For the current
noisy intermediate-scale quantum (NISQ) computers
the quantum approximate optimization algorithm (QAOA), the variational
quantum eigensolver (VQE),
and quantum annealing (QA) are the central algorithms for this problem class.
The two former can be executed on digital gate-model quantum computers, whereas
the latter requires a quantum annealer. Across all hardware architectures
and manufactures, the quantum computers available today share the property of being
too error-prone to reliably execute involved quantum circuits 
as they typically arise from quantum optimization algorithms.
In order to characterize the limits of existing quantum computers,
many component and system level benchmarks have been proposed.
However, owing to the complex 
nature of the errors in quantum
systems these benchmark fail to provide predictive power beyond simple
quantum circuits and small examples. Application oriented benchmarks
have been proposed to remedy this problem, but both, results from real quantum systems
as well as use cases beyond constructed academic examples, remain very rare.
This paper addresses precisely this gap by considering two industrial relevant
use cases: one in the realm of optimizing charging schedules for electric vehicles,
the other concerned with the optimization of truck routes.
Our central contribution are systematic series of examples derived from these
uses cases that we execute on different processors of the gate-based
quantum computers of IBM as well as on the quantum annealer of D-Wave. From
different quality measures, circuit metrics, as well as execution types and dates
we provide a comprehensive insight into the current state of the quantum computing
technology.
\end{abstract}

\section{Introduction}
\label{sec:introduction}
With the advent of quantum computers in the last decades, a novel paradigm
has emerged that holds the promise to solve computational tasks that are
challenging or even considered intractable for classical computers. 
One of the most prominent such tasks
are integer optimization problems \cite{Chen.2010,Schrijver.2000}, which are
of the utmost significance in a wide range of applications
\cite{Bayerstadler.2021,Luckow.2021,Yarkoni.2022}. Hence, speeding up their
solution could revolutionize numerous domains in science and industry.

In the current era of noisy intermediate-scale quantum (NISQ) computers, hybrid
quantum-classical algorithms have gained tremendous interest due to their 
conformity to work with (strongly) limited quantum computing resources.
Within this class,
the quantum approximate optimization algorithm (QAOA) \cite{Farhi.2014} and 
the variational quantum eigensolver (VQE) \cite{Peruzzo.2014} are the most
prominent algorithms that can be employed
on gate-based universal quantum computers
to find approximate solutions to discrete optimization problems.
As an analog computation technology,
quantum annealing (QA) allows the solution of such problems based on quantum
fluctuations.%

Numerous efforts have been undertaken to analyze, improve, and extend these
algorithms, see for example \cite{Abbas.2023} for a current review. 
However, an equally important concern is to characterize 
wether and to what quality these algorithms can be
executed on today's available quantum systems. 
A variety of benchmarks are available on the component and system level, as for
example randomized benchmarking \cite{Knill.2008,Magesan.2011}, 
gate set tomography \cite{Nielsen.2021}, or 
quantum volume \cite{Cross.2019} for gate-based computers and 
single-qubit assessment \cite{Nelson.2021} or 
Hamiltonian noise \cite{Zaborniak.2021} for quantum annealing processors.
However, quantum computers are highly susceptible to a wide range of complex errors
so that the predictive power of these low-level benchmarks 
remains quite limited for involved quantum algorithms, let alone for quantum
applications. In order to fill this gap, application oriented benchmarks 
\cite{Finzgar.2022,Lubinski.2023,Lubinski.2023.2,Tomesh.2022}
have been proposed, but the amount of experiments executed on
real (and not simulated) quantum hardware as well as the availability of 
real-world examples remain scarce. This particularly hinders 
users outside the quantum
research community in assessing the current
status of quantum computing in view of its applicability for their respective
domains. If quantum computing is to be used for more than academic examples 
this community and their uses cases have to be addressed.%

In this work, we illustrate the current status of quantum computing for optimization
problems on the basis of two industry relevant use cases. For this, we consider
systematic series of examples that differ in the number of required qubits as
well as in their coupling strengths. We do not consider problems
that are explicitly constructed to match a certain hardware technology 
or employ expensive post-processing techniques to give an unbiased view. 
We execute our example series on different quantum computing systems 
and provide a detailed analysis as well as a statistical study of different
quality metrics of the obtained results. As gate-based quantum computing platform
we used IBM quantum processors of different generations and sizes.
This also allows us to demonstrate the technological
progress of this hardware architecture over the last years.
On the side of quantum annealing we focused on the
Advantage 4.1 of D-Wave.

This paper is organized as follows: In 
Section~\ref{sec:optimization-problems-algorithms} we briefly introduce
quadratic integer optimization problems and show how they can be transformed
such that they can be processed with quantum computers. Then, we present
the theory of the
QAOA and the VQE algorithms, which we use on the gate-based IBM quantum computers,
and of the quantum annealing protocol for the D-Wave system. 
Section~\ref{sec:quantum-systems} further introduces these
quantum systems and explains their most important characteristics in view of
the aforementioned algorithms. The following two sections are devoted to our
industry use cases. 
Our first application is contained
in Section~\ref{sec:lama-use-case} and aims at optimizing charging schedules
for electric vehicles. In this use case we discuss the optimization of
the variational parameters in QAOA and VQE with a classical optimizer,
the transpilation of the QAOA and VQE circuits to the IBM quantum backends, as
well as different quality measures for the results obtained from these backends.
This leads us to examine different metrics for the transpiled circuits and
their ability to predict the quality we can expect from their execution. Moreover,
we investigate how the results of the same experiment vary for different dates
and how different patches of the quantum processor perform for the same circuit.
We close this use case by presenting results from a D-Wave quantum annealer.
In Section \ref{sec: Annealing for Routing} we present our second use case
that is concerned with computing optimal truck routes between a non-trivial
amount of cities. Owing to the large requirements in quantum computing resources
we focus on the D-Wave quantum annealer for this use case. We provide a systematic
analysis of central hardware parameters such as chain strength, anneal schedules, 
or embedding schemes and discuss their impact on the solution quality.
Last, we give a short conclusion in Section~\ref{sec:conclusion}.

\section{Optimization Problems and Algorithms}
\label{sec:optimization-problems-algorithms}
In this section we give a brief overview of the mathematical framework in which
our use cases can be formulated in the form of optimization problems.
Subsequently, we show how these optimization problems can be transformed into
an Ising formulation, which is the starting point for applying quantum algorithms
for their solution. Last, we will discuss the algorithms QAOA and VQE as well
as the method of quantum annealing. These are the specific quantum algorithms
which we will use to solve our use cases in this paper.
\subsection{Quadratic integer optimization problems}
\label{sec:quadratic-optimization-problems}
Let $\varQcio = (\varQcio_1, \dots, \varQcio_\dimQcio)^T \in \Z^\dimQcio$ be 
a vector of $\dimQcio$ integer variables and $\costQcio: \Z^\dimQcio \to \R$
be a quadratic cost function of the form
\begin{equation}
    \costQcio(\varQcio)
    = \varQcio^T \matQcio \varQcio
    + \linQcio \varQcio
    + \constQcio \, .
    \label{eq:cost-qcio}
\end{equation}
Here, $\matQcio \in \R^{\dimQcio \times \dimQcio}$ is a matrix,
$\linQcio \in \R^{1 \times \dimQcio}$ is a row vector, and
$\constQcio \in \R$ is a constant.
Associated to this cost function we consider the following quadratic 
constrained integer optimization problem
\begin{subequations}
    \begin{align}
        &\min_{\varQcio \in \Z^\dimQcio} \costQcio(\varQcio) \, ,
        \label{eq:qcio-1}
        \\[0.2cm]
        &\text{such that } \constraintQcio \varQcio = \rhsQcio \, .
        \label{eq:qcio-2}
    \end{align}
    Here, we restrict ourselves to a linear constraint given by 
    the matrix $\constraintQcio \in \R^{\dimQcio \times \dimQcio}$
    and the vector $\rhsQcio \in \R^\dimQcio$.
\label{eq:qcio}
\end{subequations}

The hard (i.e. algebraically enforced) constraint \eqref{eq:qcio-2} can be
included into the minimization task by penalizing deviations from it.
This can be done with a new cost function of the form
\begin{equation}
    \costQuio(\varQcio ; \myPenalty)
    = \costQcio(\varQcio) 
    + \varrho \|\constraintQcio \varQcio - \rhsQcio \|^2 \, .
    \label{eq:cost-quio}
\end{equation}
Here, $\| \cdot \|$ is the Euclidean norm and $\myPenalty \ge 0$ is a parameter
that controls the penalty assigned to a deviation from the hard constraint.
Clearly, in this construction the constraint is not enforced algebraically
anymore and thus it is usually referred to as a soft constraint. Note that
we can write the cost function $\costQuio$ as
\begin{equation*}
    \costQuio(\varQcio ; \myPenalty)
    = \varQcio^T \matQuio_\myPenalty \varQcio
    + \linQuio_\myPenalty \varQcio
    + \constQuio_\myPenalty \, ,
\end{equation*}
where
\begin{equation*}
    \matQuio_\myPenalty
    = \matQcio + \myPenalty \constraintQcio^T \constraintQcio \, ,
    \quad
    \linQcio_\myPenalty
    = \linQuio - 2 \myPenalty \rhsQcio^T \constraintQcio \, ,
    \quad
    \constQuio_\myPenalty
    = \constQcio + \myPenalty \|\rhsQcio\|^2 \, .
\end{equation*}
Thus, the cost function $\costQuio$ induces the quadratic unconstrained
integer optimization problem
\begin{equation}
    \min_{\varQcio \in \Z^\dimQcio} \costQuio(\varQcio ; \myPenalty) \, .
    \label{eq:quio}
\end{equation}
It is clear that, if the penalty parameter $\myPenalty$ is large enough, the
solution of \eqref{eq:quio} agrees with the solution of our initial problem
\eqref{eq:qcio}.
However, note that for numerical solvers a too large penalty parameter can be
a disadvantage since the cost function is then dominated by the constraint rather
than the original minimization task.

The integer vector $\varQcio \in \Z^\dimQcio$ can be encoded into
a binary vector $\varQubo \in \{0, 1\}^\dimQubo$ with a
transformation matrix $\matBinToInt \in \R^{\dimQcio \times \dimQubo}$ via
$\varQcio = \matBinToInt \varQubo$. Details on different binary encodings
and the transformation can be found in \cite{Karimi.2019}. 
In general, we have that $\dimQubo > \dimQcio$, i.e. the number of binary variables
is larger than the number of integer ones.
Employing such a binary encoding in our unconstrained optimization problem
\eqref{eq:quio} we obtain a quadratic unconstrained binary optimization (QUBO)
problem
\begin{equation}
    \min_{\varQubo \in \{0, 1\}^\dimQubo} 
    \costQubo(\varQubo; \myPenalty) \ .
    \label{eq:qubo-1}
\end{equation}
\begin{subequations}
    Here, the cost function is given by
    \begin{equation}
        \costQubo(\varQubo; \myPenalty)
        = \varQubo^T \matQubo_{\myPenalty} \varQubo 
        + \constQuio_{\myPenalty} \, ,
        \label{eq:qubo-2}
    \end{equation}
    where we have
    \begin{equation}
        \matQubo_{\myPenalty}
        = \matBinToInt^T \matQuio_\myPenalty \matBinToInt
        + \mathrm{diag}(\linQuio_\myPenalty \matBinToInt) \, ,
        \label{eq:qubo-3}
    \end{equation}
    and where $\mathrm{diag}(\mathbf{v})$ is the matrix with
    $\mathbf{v}$ on its diagonal and otherwise zero entries.
\label{eq:qubo}
\end{subequations}
In \eqref{eq:qubo-3} we could merge the quadratic and linear part
since for binary variables $\varQuboEntry \in \{0, 1\}$ it holds 
that $\varQuboEntry^2 = \varQuboEntry$.

Let us close this section with two remarks:
First, if it is clear from the context
we will drop the subscript $\myPenalty$.
Second, without loss of generality, we can assume that the matrices in
our cost functions
are upper triangular. If this is not the case we can always replace
them with an upper triangular one without changing the cost function due to
\begin{equation*}
    x^T M x 
    = x^T \tilde{M} x,
    \qquad
    \tilde{m}_{ij} = 
    \begin{cases}
        m_{ij} + m_{ji}, & i < j \, , \\
        m_{ij}, & i = j\, , \\
        0, & i > j \, .
    \end{cases}
\end{equation*}

\subsection{Quantum algorithms for optimization problems}
\label{sec:quantum-algorithms-optimization-problems}
From now on, let us consider the generic QUBO problem
\begin{equation}
    \min_{\varQubo \in \{0, 1\}^\dimQubo} 
    \costQuboGeneric(\varQubo) \, ,
    \qquad
    \costQuboGeneric(\varQubo)
    = \varQubo^T \matQubo \varQubo
    + \constQuboGeneric \, ,
    \label{eq:qubo-generic}
\end{equation}
and let us for simplicity assume that we have exactly one solution
\begin{equation*}
    \varQuboOpt = 
    \argmin_{\varQubo \in \{0, 1\}^\dimQubo} \costQuboGeneric(\varQubo) \, .
\end{equation*}
Our aim is to transform the QUBO cost function $\costQuboGeneric$ to an Ising
model in an $\numQubits$ qubit system. This will enable us to formulate the
upper optimization problem as a minimization of an expectation value.
For this, we write the cost function $\costQuboGeneric$ as
\begin{equation}
    \costQuboGeneric(\varQubo)
    = \sum_{i=1}^{\dimQubo} \sum_{j = i}^{\dimQubo}
    \matQuboEntry_{ij} \varQuboEntry_i \varQuboEntry_j
    + \constQuboGeneric \, , 
    \label{eq:qubo-generic-cost}
\end{equation}
where $\matQuboEntry_{ij}$ is the $(i,j)$th entry of the QUBO matrix
$\matQubo$ and $\varQuboEntry_i$ denotes he $i$th entry of the binary vector
$\varQubo$.
Next, we define the operator
\begin{equation*}
    V_i = \tfrac12 (I^{\otimes \numQubits} - Z_i) \, ,
\end{equation*}
where $I^{\otimes \numQubits}$ is the identity operator on all qubits, i.e.
\begin{equation*}
    I^{\otimes \numQubits}
    = \bigotimes_{i=1}^{\numQubits} I \, ,
    \qquad
    I = \pmat{1 & 0 \\ 0 & 1} \, ,
\end{equation*}
and $Z_i$ is the Pauli-$Z$ operator applied to the $i$th qubit:
\begin{equation*}
    Z_i
    = I^{\otimes (i - 1)} \otimes Z \otimes I^{\otimes (\numQubits - i)} \, ,
    \qquad
    Z = \pmat{1 & 0 \\ 0 & -1} \, .
\end{equation*}
Clearly, for $b \in \{0, 1\}$ we have $I \ket{\varQuboEntry} = \ket{\varQuboEntry}$ 
and $Z \ket{\varQuboEntry} = (-1)^\varQuboEntry \ket{\varQuboEntry}$.
Thus, for every computational basis state $\ket{\varQubo} 
= \ket{\varQuboEntry_1  \cdots \varQuboEntry_{\numQubits}}$ we have
\begin{equation*}
    V_i \ket{\varQubo}
    = \varQuboEntry_i \ket{\varQubo} \
    \qquad \text{ and } \qquad
    \bra{\varQubo} V_i \ket{\varQubo}
    = \varQuboEntry_i \ .
\end{equation*}
This motivates to replace $\varQuboEntry_i$ by $V_i$ in
\eqref{eq:qubo-generic-cost}, whence we obtain
the so-called cost (or problem) Hamiltonian $\problemHam$ as
\begin{equation*}
    \problemHam
    = \sum_{i=1}^{\dimQubo} \sum_{j = i}^{\dimQubo}
    \matQuboEntry_{ij} V_i V_j
    + \constQuboGeneric I^{\otimes \dimQubo} \ .
\end{equation*}
It satisfies
\begin{equation}
    \bra{\varQubo} \problemHam \ket{\varQubo}
    = \costQuboGeneric(\varQubo)
    \qquad
    \text{for all }
    \varQubo \in \{0, 1\}^\dimQubo \, .
    \label{eq:connection-qubo-cost-expectation-basis-state}
\end{equation}
Now, consider a general quantum state $\ket{\psi}$ with amplitudes
$\{\ampl_{\varQubo}\}$, i.e.
\begin{equation*}
    \ket{\psi}
    = \sum_{\varQubo \in \{0, 1\}^\numQubits}
    \ampl_{\varQubo} \ket{\varQubo} \ ,
    \qquad
    \ampl_{\varQubo} \in \C \, ,
    \sum_{\varQubo \in \{0, 1\}^\numQubits}
    |\ampl_{\varQubo}|^2 = 1 \, .
\end{equation*}
From \eqref{eq:connection-qubo-cost-expectation-basis-state}
it immediately follows that for the expectation value
$\expect{\problemHam}_{\ket{\psi}} 
= \bra{\psi} \problemHam \ket{\psi}$ we have
\begin{equation}
    \expect{\problemHam}_{\ket{\psi}}
    = \sum_{\varQubo \in \{0, 1\}^\numQubits}
    |\ampl_\varQubo|^2 \costQuboGeneric(\varQubo) \ .
    \label{eq:connection-qubo-cost-expectation}
\end{equation}
From this equation we see that minimizing the expectation value
$\expect{\problemHam}_{\ket{\psi}}$ over all
states $\ket{\psi}$ has to result in the state $\ket{\varQuboOpt}$, 
where $\varQuboOpt$ is the solution of the QUBO \eqref{eq:qubo-generic}:
\begin{equation*}
    \argmin_{\ket{\psi}} \expect{\problemHam}_{\ket{\psi}}
    = \ket{\varQuboOpt} \, .
\end{equation*}
Usually, it is not possible to minimize over the whole state space and one
is satisfied with quantum algorithms that compute approximations to 
$\ket{\varQuboOpt}$ in the sense that they generate a state 
$\ket{\psiOpt}$ which amplitudes $\amplOpt_{\varQubo}$ satisfy
\begin{subequations}
    \begin{equation}
        |\amplOpt_{\varQuboOpt}|^2
        \gg |\amplOpt_{\varQubo}|^2
        \qquad
        \text{for all }
        \varQubo \in \{0, 1\}^\dimQubo \setminus \{\varQuboOpt\} \, .
        \label{eq:approximate-solution-quantum-algorithm-1}
    \end{equation}
    In the best case we would have 
    \begin{equation}
        |\amplOpt_{\varQuboOpt}|^2 \approx 1
        \qquad
        \text{and}
        \qquad
        |\amplOpt_{\varQubo}|^2 \approx 0 
        \ \
        \text{for all }
        \varQubo \in \{0, 1\}^\numQubits \setminus \{\varQuboOpt\} \, .
        \label{eq:approximate-solution-quantum-algorithm-2}
    \end{equation}
    \label{eq:approximate-solution-quantum-algorithm}
\end{subequations}
Let us recall from the postulates of quantum mechanics \cite{Nielsen.2010}
that for a state
$\ket{\psi} = \sum_{\varQubo \in \{0, 1\}^\numQubits} \ampl_{\varQubo} \ket{\varQubo}$
the square of the absolute value of an amplitude, 
$\prob_\varQubo = |\ampl_{\varQubo}|^2$, encodes the probability
that we obtain the bitstring $\varQubo$ as a result from a measurement in the 
computational basis. Thus, \eqref{eq:approximate-solution-quantum-algorithm}
means that from $\numShots$ preparations and measurements of the state 
$\ket{\psiOpt}$
the majority should have the exact solution
bitstring $\varQuboOpt$ as result. One pass of state preparation 
(i.e. execution of a quantum circuit) and measurement is called a shot.
So, only a few shots are needed to reveal $\varQuboOpt$ with high probability.

In the case that we have several optimal solutions
$\varQuboOpt_1, \dots, \varQuboOpt_I$ the upper conditions become
\begin{equation*}
    |\amplOpt_{\varQuboOpt_i}|^2
    \gg |\amplOpt_{\varQubo}|^2
    \qquad
    \text{for all }
    \varQubo \in \{0, 1\}^\dimQubo 
    \setminus \{\varQuboOpt_1, \dots, \varQuboOpt_I\} \, ,
    \qquad
    i \in \{1, \dots, I\} \, ,
\end{equation*}
and
\begin{equation*}
    \sum_{i=1}^I |\amplOpt_{\varQuboOpt_i}|^2 \approx 1 \, ,
    \qquad
    |\amplOpt_{\varQubo}|^2 \approx 0 
    \ \
    \text{for all }
    \varQubo \in \{0, 1\}^\numQubits 
    \setminus \{\varQuboOpt_1, \dots, \varQuboOpt_I\} \, .
\end{equation*}

Let us close this section with three remarks:
First, the cost Hamiltonian $\problemHam$ can be written as
\begin{equation}
    \problemHam
    = \sum_{i=1}^{\numQubits} \sum_{j = i}^{\numQubits} h_{ij} Z_i Z_j
    + \sum_{i=1}^{\numQubits} h^\prime_{i} Z_i
    + h^{\prime \prime} I^{\otimes \numQubits} \, ,
    \label{eq:problem-hamiltonian}
\end{equation}
where the coefficients $h_{ij}, h^\prime_{i},$ and $h^{\prime \prime}$ can be 
computed from $\matQuboEntry_{ij}$ and $\constQuboGeneric$. Second,
if the QUBO cost function depends on the penalty parameter $\myPenalty$
so does $\problemHam$. And last, note that $\problemHam$ 
is a diagonal matrix of dimension $2^\numQubits \times 2^\numQubits$. 

\subsubsection{QAOA}
\label{sec:qaoa}
\begin{figure}
    \centering
    \input{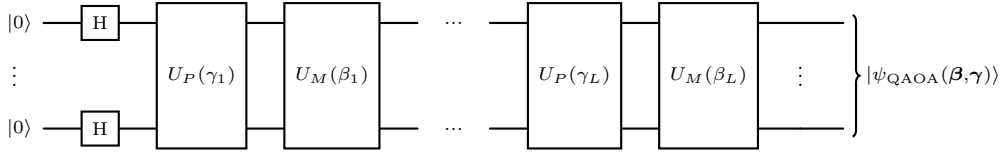}
    \caption{
        QAOA circuit with $\repsQaoa$ layers and parameters 
        $\bfbeta = (\beta_1, \dots, \beta_\repsQaoa)$
        and $\bfgamma = (\gamma_1, \dots, \gamma_\repsQaoa)$.}
    \label{fig:qaoa-circuit}
\end{figure}

The most prominent algorithm for computing states that approximately minimize
the expectation value $\expect{\problemHam}_{\ket{\psi}}$ (and as pointed out 
above thus approximately solve the QUBO problem \eqref{eq:qubo-generic}) is 
the quantum approximate optimization algorithm (QAOA) \cite{Farhi.2014}. This
algorithm generates a parametrized quantum state in the following steps:
\begin{subequations}
    It starts at the uniform superposition
    \begin{equation}
        \ket{+}^{\otimes \numQubits}
        = \bigotimes_{i=1}^{\numQubits} \ket{+}
        % = \ket{+} \otimes \cdots \otimes \ket{+}
        = \tfrac{1}{\sqrt{2^\numQubits}} 
        \sum_{\varQubo \in \{0, 1\}^\numQubits} \ket{\varQubo} \, ,
        \label{eq:qaoa-0}
    \end{equation}
    and then applies $\repsQaoa$ layers of alternating applications of
    the phase operator 
    \begin{equation}
        \problemOperator(\gamma)
        = \exp \bigl(-\i \gamma \problemHam \bigr) \ ,
        \label{eq:qaoa-1}
    \end{equation}
    and the mixing operator
    \begin{equation}
        \mixingOperator(\beta)
        = \exp \bigl(-\i \beta \mixingHam \bigr) \ .
        \label{eq:qaoa-2}
    \end{equation}
    Here, the mixer $\mixingHam$ is given by
    \begin{equation}
        \mixingHam = \sum_{i=1}^\numQubits X_i \ ,
        \qquad
        X_i
        = I^{\otimes (i - 1)} \otimes X \otimes I^{\otimes (\numQubits - i)} \ ,
        \qquad
        X = \pmat{0 & 1 \\ 1 & 0} \ ,
        \label{eq:qaoa-3}
    \end{equation}
    and $\beta$ and $\gamma$ are real parameters. Altogether, the parametrized 
    QAOA state is given by
    \begin{equation}
        \ket{\psiQaoa(\bfbeta, \bfgamma)}
        = \sideset{}{^\prime}\prod_{\ell=1}^{\repsQaoa}
        % \bigl( 
            \mixingOperator(\beta_\ell)
            \problemOperator(\gamma_\ell)
        % \bigr)
        \ket{+}^{\otimes \numQubits} \ ,
        \label{eq:qaoa}
    \end{equation}
    where we collect the parameters in $\bfbeta = (\beta_1, \dots, \beta_\repsQaoa)$
    and $\bfgamma = (\gamma_1, \dots, \gamma_\repsQaoa)$.
    The product symbol with a prime indicates a reversed multiplication order
    \begin{equation*}
        \sideset{}{^\prime}\prod_{j=1}^{J} A_j
        = A_J A_{J-1} \cdots A_{1} \, .
    \end{equation*}
    A visualization of the QAOA circuit can be found in Figure~\ref{fig:qaoa-circuit}.
    The values of the parameters $\bfbeta$ and $\bfgamma$
    are determined with a classical optimizer such that
    \begin{equation}
        (\bfbetaOpt, \bfgammaOpt)
        = \argmin_{(\bfbeta, \bfgamma)}
        \expect{\problemHam}_{\ket{\psiQaoa(\bfbeta, \bfgamma)}} \ .
        \label{eq:qaoa-optimal-parameters}
    \end{equation}
    \label{eq:qaoa-all}
\end{subequations}
Let us write the QAOA state with optimal parameters $\bfbetaOpt$ and $\bfgammaOpt$
as
\begin{equation*}
    \ket{\psiQaoa(\bfbetaOpt, \bfgammaOpt)}
    = \sum_{\varQubo \in \{0, 1\}^\numQubits}
    \amplOpt_{\varQubo, \mathrm{QAOA}}
    \ket{\varQubo} \ ,
    \qquad
    \amplOpt_{\varQubo, \mathrm{QAOA}}
    = \ampl_{\varQubo, \mathrm{QAOA}}(\bfbetaOpt, \bfgammaOpt) \, .
\end{equation*}
Then, due to \eqref{eq:connection-qubo-cost-expectation}
and for large enough $\repsQaoa$, the 
amplitudes $\amplOpt_{\varQubo, \mathrm{QAOA}}$ should satisfy 
\eqref{eq:approximate-solution-quantum-algorithm}.
As pointed out above this means that in the
measurement statistics of the state $\ket{\psiQaoa(\bfbetaOpt, \bfgammaOpt)}$ 
the exact solution bitstring $\varQuboOpt$ appears with highest probability.

Last, we give a gate representation of the phase and mixing operator. Details
can be found in \cite{Sturm.2023}. For the mixing operator we have
\begin{equation}
    \mixingOperator(\beta)
    = \prod_{i=1}^{\numQubits} \RX_i (2 \beta) \ ,
    \label{eq:mixing-operator-gates}
\end{equation}
where $\RX_i(\theta) = \exp \bigl(- \i \tfrac\theta2 X_i \bigr)$
is a rotation of the $i$th qubit about the $x$-axis with
angle $\theta$. Note that in \eqref{eq:mixing-operator-gates} 
all qubits are rotated with the same angle $2 \beta$.
For the phase operator we have
\begin{equation}
    \problemOperator(\gamma)
    = \exp(-\i \gamma h^{\prime \prime})
    \prod_{i=1}^{\numQubits} \RZ_i(2 \gamma h_i^\prime)
    \prod_{i=1}^{\numQubits} \prod_{j=i}^{\numQubits}
    \RZZ_{i, j}(2 \gamma h_{i,j}) \ ,
    \label{eq:phase-operator-decomposition}
\end{equation}
where $\RZ_i(\theta) = \exp \bigl(- \i \tfrac\theta2 Z_i \bigr)$
is a rotation of the $i$th qubit about the $z$-axis with
angle $\theta$, and 
$\RZZ_{i, j}(\theta) = \exp \bigl(- \i \tfrac\theta2 Z_i Z_j \bigr)$
is a parametric two qubit $Z \otimes Z$
interaction with parameter $\theta$. 
Note that the parameter $\gamma$ enters in all gates but is weighted with 
the coefficients $h_{i,j}, h_i^\prime,$ and $h^{\prime \prime}$ of the problem
Hamiltonian $\problemHam$.
Finally, note that the initial state \eqref{eq:qaoa-0} can be generated
with Hadamard gates
applied to the ground state $\ket{0}^{\otimes \numQubits}$
\begin{equation}
    \ket{+}^{\otimes \numQubits}
    = \hadamard^{\otimes \numQubits} \ket{0}^{\otimes \numQubits} \ ,
    \qquad
    \hadamard
    = \tfrac1{\sqrt2} \pmat{1 & 1 \\ 1 & -1} \ .
    \label{eq:qaoa-state-preparation}
\end{equation}
\subsubsection{VQE}
Initially, the variational quantum eigensolver (VQE) \cite{Peruzzo.2014} was proposed
to compute the ground state energy $e_\mathrm{min}$ of a Hamiltonian $H$
\begin{equation*}
    e_\mathrm{min} = \min_{\ket{\psi}} \expect{H}_{\ket{\psi}} \, .
\end{equation*}
For this, the expectation value of $H$ over 
a trial state $\ket{\psiVqe(\vqeParams)}$ is minimized
\begin{equation*}
    \min_{\vqeParams} \expect{H}_{| \psiVqe(\vqeParams) \rangle} \, ,
    \qquad
    \ket{\psiVqe(\vqeParams)}
    = \vqeAnsatz(\vqeParams) \ket{\psi_0} \, .
\end{equation*}
Here, $\vqeAnsatz(\vqeParams)$ is a variational ansatz depending on 
$\vqeParams = (\vqeParam_1, \dots, \vqeParam_\numVqeParams)$ real parameters
and $\ket{\psi_0}$ is an initial state. Contrary to QAOA, the variational ansatz
and the initial state are not predetermined in VQE but there a variety of options
as for example problem-inspired or hardware efficient ansatzes \cite{Cerezo.2021}.

Clearly, we can use VQE for our optimization problem by applying it to the 
cost Hamiltonian $\problemHam$. Then, the state $\ket{\psiVqe(\vqeParamsOpt)}$
with optimized parameters
\begin{equation}
    \vqeParamsOpt
    = \argmin_{\vqeParams}
    \expect{\problemHam}_{| \psiVqe(\vqeParams) \rangle} \, ,
    \label{eq:vqe-optimal-parameters}
\end{equation}
should be an approximate solution to our optimization problem in the sense that
the amplitudes
$\amplOpt_{\varQubo, \mathrm{VQE}} 
= \ampl_{\varQubo, \mathrm{VQE}}(\vqeParamsOpt)$ satisfy
\eqref{eq:approximate-solution-quantum-algorithm}. 

For our use cases we use a two-local ansatz VQE in the following form:
\begin{equation}
    \ket{\psiVqe(\vqeParams)}
    =
    \prod_{i=1}^\numQubits \RY_i(\vqeParam_{(\repsVqe-1)\numQubits})
    \sideset{}{^\prime} \prod_{\ell=1}^{\repsVqe} 
    \Bigl(
        \sideset{}{^\prime} \prod_{i=1}^{\numQubits-1} \CX_{i, i+1}
        \prod_{i=1}^\numQubits \RY_i(\vqeParam_{i+(\ell-1)\numQubits})
    \Bigr)
    \ket{0}^{\otimes \numQubits} \ .
    \label{eq:vqe}
\end{equation}
Here, $\RY_i(\theta) = \exp \bigl(- \i \tfrac\theta2 Y_i \bigr)$
is a rotation of the $i$th qubit about the $y$-axis with
angle $\theta$,
$\CX_{i, j}$ is the controlled-$\pauliX$ (or CNOT) gate between qubits
$i$ and $j$ given by
\begin{equation}
    \CX_{i, j}
    = \tfrac12 \bigl( I^{\otimes \numQubits} + Z_i \bigr)
    + \tfrac12 \bigl( I^{\otimes \numQubits} - Z_i \bigr) X_j \, ,
    \qquad
    i \neq j \, ,
    \label{eq:cx-gate}
\end{equation}
and where we denote with $\repsVqe$ the number of layers.
In Figure~\ref{fig:vqe-circuit} we give an example of \eqref{eq:vqe} for
$\numQubits = 4$ qubits and $\repsVqe = 2$ layers. We see that our two-local 
ansatz consists of alternating single qubit rotation blocks and
next-neighbor two qubit entanglement blocks.
Note that the VQE circuits
we use in this paper have $\numVqeParams = \numQubits (\repsVqe+1)$ parameters.
So, contrary to QAOA the number of parameters depends on the number of qubits.
\begin{figure}
    \centering
    \input{figures/circuits/vqe_two_local_ry_cx_lin.tex}
    \caption{
        Example of a two-local VQE circuit.
        In this case we have $\numQubits = 4$ qubits, $\repsVqe = 2$ layers,
        and $\numVqeParams = 12$ parameters 
        $\vqeParams = (\vqeParam_1, \dots, \vqeParam_{12})$.
    }
    \label{fig:vqe-circuit}
\end{figure}

\subsubsection{Quantum annealing}
\label{quantum annealing theory}

Quantum annealing (QA) \cite{Kadowaki1998} is a metaheuristic for finding the
global minimum of a given objective function by a process using quantum fluctuations. It is a quantum analog to simulated annealing
(SA) \cite{Kirkpatrick1983}, which is a probabilistic technique for approximating 
the global optimum of a given function. Recent studies have compared the 
performance of QA and SA in terms of computational time for obtaining high-accuracy 
solutions. While most studies have shown QA to be superior to SA \cite{Santoro2001, QA_TSP, QIP},
some have suggested the opposite \cite{Battaglia2005}. 
The development of commercial quantum annealers, such as 
those based on superconducting flux qubits by D-Wave, has led
to experimental studies of QA and demonstrations of the applicability of quantum 
annealers to practical problems \cite{Boyda2017, Neukart2017, Daniel_O}.
%It ends here
%This is from https://arxiv.org/pdf/2303.11705.pdf

In adiabatic quantum computing (AQC) \cite{McGeoch2014, Albash2018} the forces
acting in a quantum system are described by a time-varying Hamiltonian 
$\mathcal{H}(t)$ of the form
\begin{equation*}
    \mathcal{H}(t)
    = -\frac{a(t)}{2} \mixingHam
    + \frac{b(t)}{2} \problemHam \, ,
    \qquad
    a(0) = 1 \, , \ b(0) = 0 \, ,
    \; \; 
    a(T) = 0 \, , \ b(T) = 1 \, ,
\end{equation*}
where $\mixingHam$ and $\problemHam$ where defined in \eqref{eq:qaoa-3} 
and \eqref{eq:problem-hamiltonian}, respectively,
$a(t)$ and $b(t)$ are annealing functions, and $T$ is the anneal time. 
The first summand on the 
right-hand side (in this context called the tunneling Hamiltonian) corresponds to
the system's initial state, where $\ket{+}^{\otimes \numQubits}$ is the ground 
state, i.e. the eigenstate corresponding to the smallest eigenvalue. The second
summand corresponds to the system's final state which has, as we discussed in 
Section~\ref{sec:quantum-algorithms-optimization-problems}, the optimal solution
$\ket{\varQuboOpt}$ as its ground state. According to the adiabatic theorem \cite{Born1928}
the system will remain in its ground state if the annealing process is sufficiently slow.
Thus, by adjusting the annealing functions $a(t)$ and $b(t)$ to ensure 
the problem Hamiltonian is introduced progressively, the system evolves from
$\ket{+}^{\otimes \numQubits}$ to $\ket{\varQuboOpt}$. Or put in another way:
quantum annealing solves our optimization problem.

Let us remark that on D-Wave quantum annealers, the evolution 
schedule is adjustable by modifying $t$. With an appropriate annealing 
schedule \cite{Kadowaki1998}, QA has demonstrated superiority over SA for 
certain problems, including Ising spin glasses \cite{Heim2015}, 
the traveling salesman problem \cite{PhysRevE.70.057701}, 
and specific non-convex problems \cite{Baldassi2018}.

\section{Quantum Systems}
\label{sec:quantum-systems}
In this section we briefly introduce the quantum computing systems that we
use for this article.

\begin{table}
    \centering
    \begin{tabular}{llcc}
        name & processor type & \# qubits & two qubit gate
        \\
        \hline \hline
        \ibmqehningen & Falcon r5.11 (January 2021) & 27 & $\CX$
        \\
        \ibmcairo & Falcon r5.11 (January 2021) & 27 & $\CX$
        \\
        \ibmsherbrooke & Eagle r3 (December 2022) & 127 & $\ECR$
        \\
        \ibmtorino & Heron r1 (December 2023) & 133 & $\CZ$
    \end{tabular}
    \caption{
        Specifications of the IBM quantum systems used in this paper
        \cite{IbmProcessorTypes.2024, IbmComputeResources.2024}.
    }
    \label{tab:ibm-quantum-systems}
\end{table}
\begin{figure}
    \centering
    \includegraphics[scale=0.5]{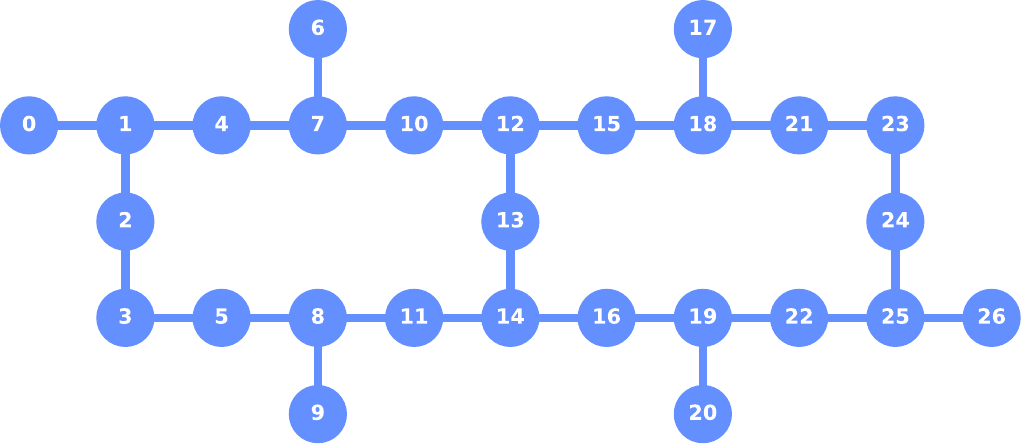}
    \caption{
        Topology of \ibmqehningen. The two qubit gate $\CX$ can only be executed
        between connected qubits.}
    \label{fig:topology-ibmq-ehningen}
\end{figure}

\subsection{IBM quantum backends}
\label{sec:quantum-systems-ibmq}
As gate-based, universal quantum computers we utilized the superconducting 
quantum backends of IBM.
We mainly employed the \ibmqehningen\ backend and for comparison ran 
our experiments also on
the backends \ibmtorino , \ibmsherbrooke , and \ibmcairo. Details on the
quantum processors of these backends are provided in
Table~\ref{tab:ibm-quantum-systems}.
Two important characteristics of the IBM quantum backends is first that they have
a limited set of gates that can be executed on the hardware. We will call this
set the hardware gates. For all backends that we consider in this article 
this set consists of the single qubit gates
$\RZ$, $\SX= \sqrt{\pauliX}$, and $\pauliX$. Depending on the backend, the two qubit 
gate is either the $\CX$, $\CZ$, or $\ECR$ gate, see
Table~\ref{tab:ibm-quantum-systems}. Here, $\CX$ is the
controlled-$\pauliX$ gate, which we already introduced in \eqref{eq:cx-gate},
$\CZ$ is the analogously defined controlled-$\pauliZ$ gate, and $\ECR$ is 
the echoed cross-resonance gate \cite{IbmECR.2024}.
The second key characteristic is that the quantum processors have a topology
with limited connectivity in the sense that the hardware two qubit gate can only be
executed on certain qubit pairs. We give the topology of \ibmqehningen\ in
Figure~\ref{fig:topology-ibmq-ehningen} and refer to 
\cite{IbmComputeResources.2024} for the other backends.
More details about the backends such as gate errors or coherence times can be
found in Figures~\ref{fig:ibmq-coherence-times} and \ref{fig:ibmq-gate-errors}
in the appendix.

\subsection{D-Wave backends}
\label{sec:quantum-systems-dwave}
\begin{figure}[t]
	\centering
	\begin{subfigure}{0.3\textwidth}
		\centering
		\includegraphics[scale=1]{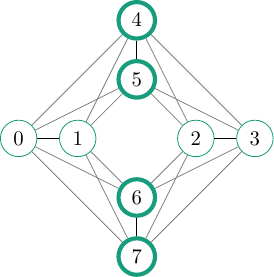}
		\caption{Pegasus single cell.}
	\end{subfigure}
	\; \;
	\begin{subfigure}{0.6\textwidth}
		\centering
		\includegraphics[scale=0.5]{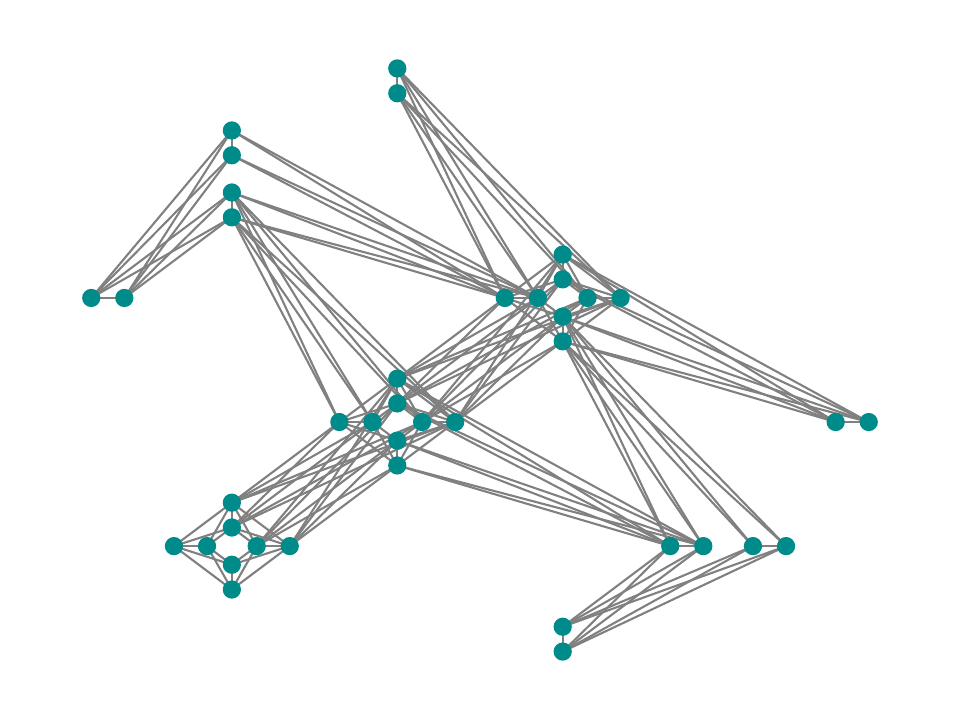}
		\caption{Repetition pattern of Pegasus single cell which forms one unit cell.}
	\end{subfigure}
	\\[0.3cm]
	\begin{subfigure}{0.7\textwidth}
		\centering
		\includegraphics[scale=1.2]{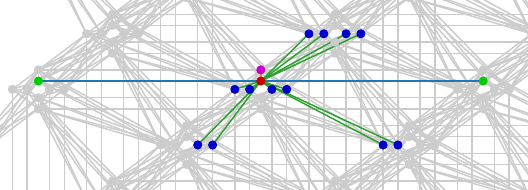}
		\caption{Single qubit connectivity (degree = 15).}
	\end{subfigure}
	\caption{Architecture of the D-Wave \dwaveAdvantage. }
	\label{Dwave Pegasus architecture}
\end{figure}

For our quantum annealing experiments we used the D-Wave \dwaveAdvantage. The
specifications of this system are given in Table~\ref{tab:specifications-dwave}.
The Pegasus features qubits of degree 15, i.e. each qubit is coupled to 15 
different qubits, and native $K_4$ and $K_{6,6}$ subgraphs,
see Figure~\ref{Dwave Pegasus architecture}. The Pegasus qubits have three 
different types of couplers which are internal, external and odd: 
The Pegasus has nominal length of 12 meaning each qubit is connected to
12 orthogonal qubits through internal couplers indicated with green lines
in Figure~\ref{Dwave Pegasus architecture} (c). Orthogonal qubits are those 
that are not connected to each other. The external couplers connect the 
qubits with the adjacent Pegasus single cells shown with light blue lines 
in Figure~\ref{Dwave Pegasus architecture} (c). The Pegasus single cell is 
shown in Figure~\ref{Dwave Pegasus architecture} (a). An odd coupler can 
only couple to one qubit, which is in the same single Pegasus cell. For example,
qubit 4 and 5 are coupled via an odd coupler in
Figure~\ref{Dwave Pegasus architecture} (a), also colored pink in
Figure~\ref{Dwave Pegasus architecture} (c).

\begin{table}
	\centering
	\begin{tabular}{cccccc}
		name 
		& topology 
		& \# qubits 
		& $\begin{array}{c} \text{annealing} \\ \text{range [$\mu$s] } \end{array}$
		& $\begin{array}{c} \text{connectivity} \\ \text{degree} \end{array}$
		& $\begin{array}{c} \text{nominal} \\ \text{length} \end{array}$
		\\
		\hline \hline
		\dwaveAdvantage & Pegasus& 5627 & 0.5 -- 2000 & 15 & 12\\
	\end{tabular}
	\caption{Specifications of the D-Wave quantum annealer that was used for the
		experiments in this article.}
	% \label{Table of Dwave advantage system details}
	\label{tab:specifications-dwave}
\end{table}

Now, let us introduce the embedding which is is a fundamental aspect
of utilizing D-Wave quantum annealers. As we have seen, the \dwaveAdvantage\ has
a limited connectivity between qubits so that not all problem Hamiltonians
can be directly represented on the quantum
processing unit (QPU).
Minor embedding \cite{Choi2008} is the process of mapping a problem's logical qubits onto the annealer's physical qubits, forming chains, where necessary. This technique ensures that the logical problem graph is represented within the constraints of the QPU's architecture, such as Pegasus topologies \cite{Dwave_topology}. Usually, there are many possible embeddings that differ in their quality.
Poor embeddings can lead to longer chains of physical qubits, which are more
susceptible to errors such as broken chains. These errors occur when the
physical qubits in a chain do not all agree on a value, potentially 
deteriorating the solution quality \cite{Outeiral2022}. 
Conversely, a good embedding ensures that the problem's structure is 
well-represented on the QPU, allowing the quantum annealer to effectively 
explore the problem's solution space \cite{embedding_subproblems}. However, 
finding an optimal embedding is challenging and known to be NP-hard, adding 
complexity to the problem-solving process on D-Wave quantum annealers.

Intimately connected to the embedding is the chain strength. A chain in this 
context refers to a set of qubits that collectively represent a single variable
(i.e. a logical qubit) from the problem being solved. As we have discussed above,
chains are required due to the limited connectivity in the QPU.
The primary purpose of setting the chain strength is to ensure that all
qubits within a chain yield the same value when a computation is performed. 
This is achieved by applying a strong ferromagnetic coupling between the qubits
in a chain, making it energetically favorable for them to align in the same 
state \cite{Dwave_minor}. The chain strength must be carefully balanced.
If it is too small the qubits within a chain may not align, leading to broken
chains, where the value of the logical qubit is ambiguous. Conversely, if
the chain strength is too high, it can dominate over the problem's
intrinsic couplings which can potentially skew the 
solution landscape and lead to suboptimal solutions \cite{Dwave_chain}.

Last, the annealing time schedule defines the temporal profile of the quantum 
annealing process, where the system transitions from an initial quantum 
state to a final state that ideally represents the solution of the optimization 
problem \cite{Dwave_annealer_1}. The schedule is typically a 
function that describes how the system's Hamiltonian changes over time, 
affecting the dynamics of the annealing process and the likelihood of the 
system settling into the global minimum energy state. A well-chosen annealing 
schedule can enhance the probability of finding the optimal solution by 
carefully managing the rate at which quantum fluctuations are reduced during 
the annealing process \cite{Oscar}. If the annealing is too rapid, the system 
may not maintain its ground state, leading to suboptimal solutions. 
Conversely, if the annealing is too slow, it may become computationally 
inefficient \cite{Carugno2022}. Experimentation with different annealing 
schedules, such as introducing pauses or changes in the annealing rate, 
can improve performance for certain types of problems \cite{Pelofske2020}.
The optimal annealing schedule may vary depending on the specific problem instance, and finding 
the right balance is often a matter of empirical investigation \cite{Oscar}.

% LamA Part
\section{Use Case 1: Optimization of Charging Schedules}
\label{sec:lama-use-case}
In our first use case we study the optimization of charging schedules for electric
vehicles. More precisely, this use case stems from the project
LamA (Laden am Arbeitsplatz, in English: charging at the workplace) where a charging 
infrastructure is built up at 37 institutes of the Fraunhofer society,
on which employees can charge their electric vehicles \cite{LamaWeb.2023}. In this paper
we consider a proof of concept (PoC) model that is suitable for the current NISQ era
quantum computers. We will only give a short overview here and refer the reader to
\cite{Sturm.2023} where  the details are elaborated.
Owing to the background of this use case we will refer to it as
the ``LamA use case'' throughout this paper.
\begin{figure}[t]
    \centering
    \small
    \begin{subfigure}{0.45\linewidth}
        \includegraphics{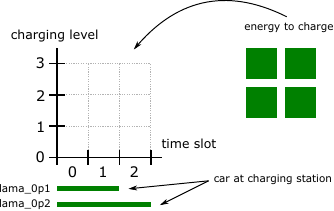}
        \caption{Example series $\lamaSeries{0}$ with one car and three time
        slots.}
    \end{subfigure}
    \; \;
    \begin{subfigure}{0.45\linewidth}
        \includegraphics{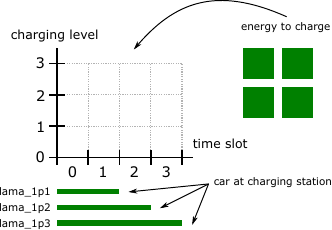}
        \caption{Example series $\lamaSeries{1}$ with one car and four time
        slots.}
    \end{subfigure}
    \\[0.3cm]
    \begin{subfigure}{0.75\linewidth}
        \centering
        \includegraphics{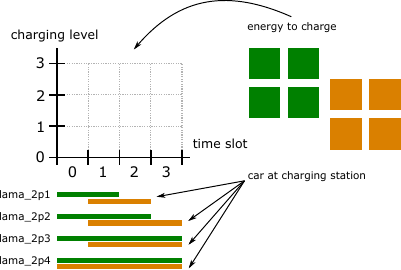}
        \caption{Example series $\lamaSeries{2}$ with two cars and four time
        slots.}
    \end{subfigure}
    \caption{
        Visualization of the three example series for the LamA use case.
        All series are based on one charging station with four charging levels
        but differ in the number of cars that are at the charging station
        and the number of time slots when the car(s) can charge at this station.
        Within an example series we consider different examples that vary in the
        time slots when the car(s) are at the charging station.}
    \label{fig:lama-example-series}
\end{figure}

We consider the PoC model with the following setting: We have a single
charging station with four charging levels 0, 1, 2, and 3, and $\numTimeslots$ time
slots 0, 1, \dots, $\numTimeslots-1$ when cars can charge. Moreover, we assume that
$\numCars$ cars are charging energy and that the required amount of energy
as well as the time slots when each car is at the charging station are known.
The aim of
our optimization is to calculate a charging schedule that minimizes the peak load.
As presented in \cite{Sturm.2023} this can be modeled by a quadratic 
constrained integer optimization
problem of the form \eqref{eq:qcio}. Here, the minimization part \eqref{eq:qcio-1}
is responsible to minimize the peak load wheres the constraint part \eqref{eq:qcio-2}
ensures that each car charges the right amount of energy in valid time slots.
Since we have four charging levels the integer vector $\varQcio$ in \eqref{eq:qcio}
is taken from the set $\{0, 1, 2, 3\}^{\dimQcio}$
where $\dimQcio = \numCars \numTimeslots$. As we have discussed in
Section~\ref{sec:quadratic-optimization-problems} such an optimization problem
can be transformed to a QUBO \eqref{eq:qubo}. Since
we can encode the four charing levels into two qubits we have 
$\dimQubo = 2 \numCars \numTimeslots$ binary variables
$\varQubo \in \{0, 1\}^{\dimQubo}$.
\begin{table}
    \centering
    \begin{subtable}[t]{0.65\linewidth}
        \begin{tabular}{ccccc}
                $\begin{array}{c} \text{example} \\ \text{series} \end{array}$
                & $\begin{array}{c} \text{\# time slots} \\ \numTimeslots \end{array}$
                & $\begin{array}{c} \text{\# cars} \\ \numCars \end{array}$
                & $\begin{array}{c} \text{\# qubits} \\ \numQubits \end{array}$
            \\
            \hline \hline
            \lamaSeries{0} & 3 & 1 & 6
            \\
            \phantom{\lamaSeries{0}}
            \\
            \hline
            \lamaSeries{1} & 4 & 1 & 8
            \\
            \phantom{\lamaSeries{0}}
            \\
            \phantom{\lamaSeries{0}}
            \\
            \hline
            \lamaSeries{2} & 4 & 2 & 16
            \\
            \phantom{\lamaSeries{0}}
            \\
            \phantom{\lamaSeries{0}}
            \\
            \phantom{\lamaSeries{0}}
        \end{tabular}
        \caption{
            Number of time slots $\numTimeslots$ when cars can charge at the
            charging station, number of cars $\numCars$ that are at the 
            station, and number of qubits $\numQubits$ that are needed to model
            the example series.}
        \label{tab:lama-example-series-problem-parameters}
    \end{subtable}
    \quad
    \begin{subtable}[t]{0.3\linewidth}
        \begin{tabular}{cc}
            example
            & $\begin{array}{c} \text{\# optimal } \\ \text{solutions}\end{array}$
            \\
            \hline \hline
            \lamaEx{0}{1} & 1
            \\
            \lamaEx{0}{2} & 3
            \\
            \hline
            \lamaEx{1}{1} & 1
            \\
            \lamaEx{1}{2} & 3
            \\
            \lamaEx{1}{3} & 1
            \\
            \hline
            \lamaEx{2}{1} & 3
            \\
            \lamaEx{2}{2} & 3
            \\
            \lamaEx{2}{3} & 6
            \\
            \lamaEx{2}{4} & 19
        \end{tabular}
        \caption{Number of optimal solutions.}
    \end{subtable}
    \caption{
        Description of the example series for the LamA use case (a)
        and number of optimal charging schedules for each example (b).
    }
    \label{tab:lama-example-series}
\end{table}

\begin{figure}
    \small
    \begin{subfigure}{0.2\textwidth}
        \centering
        \includegraphics{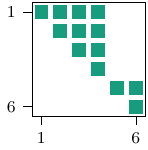}
        \caption{\lamaEx{0}{1}.}
        \label{fig:lama-sparsity-qubo-ex0p1}
    \end{subfigure}
    \begin{subfigure}{0.2\textwidth}
        \centering
        \includegraphics{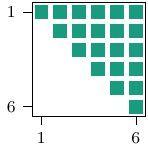}
        \caption{\lamaEx{0}{2}.}
        \label{fig:lama-sparsity-qubo-ex0p2}
    \end{subfigure}
    \\[0.2cm]
    \begin{subfigure}{0.2\textwidth}
        \centering
        \includegraphics{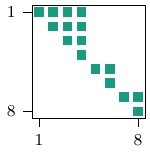}
        \caption{\lamaEx{1}{1}.}
        \label{fig:lama-sparsity-qubo-ex1p1}
    \end{subfigure}
    \begin{subfigure}{0.2\textwidth}
        \centering
        \includegraphics{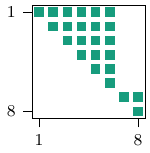}
        \caption{\lamaEx{1}{2}.}
        \label{fig:lama-sparsity-qubo-ex1p2}
    \end{subfigure}
    \begin{subfigure}{0.2\textwidth}
        \centering
        \includegraphics{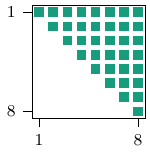}
        \caption{\lamaEx{1}{3}.}
        \label{fig:lama-sparsity-qubo-ex1p3}
    \end{subfigure}
    \\[0.2cm]
    \begin{subfigure}{0.2\textwidth}
        \centering
        \includegraphics{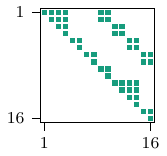}
        \caption{\lamaEx{2}{1}.}
        \label{fig:lama-sparsity-qubo-ex2p1}
    \end{subfigure}
    \begin{subfigure}{0.2\textwidth}
        \centering
        \includegraphics{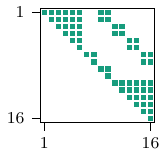}
        \caption{\lamaEx{2}{2}.}
        \label{fig:lama-sparsity-qubo-ex2p2}
    \end{subfigure}
    \begin{subfigure}{0.2\textwidth}
        \centering
        \includegraphics{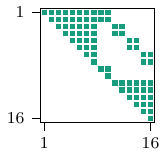}
        \caption{\lamaEx{2}{3}.}
        \label{fig:lama-sparsity-qubo-ex2p3}
    \end{subfigure}
    \begin{subfigure}{0.2\textwidth}
        \centering
        \includegraphics{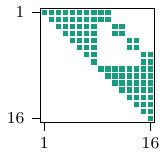}
        \caption{\lamaEx{2}{4}.}
        \label{fig:lama-sparsity-qubo-ex2p4}
    \end{subfigure}
    \caption{
        Spy plot of the QUBO matrix $\matQubo = (q_{ij})_{i,j=1}^{\numQubits}$
        for the
        examples in the LamA use case. A marker indicates a non-zero
        matrix entry $q_{ij} \neq 0$, whereas a missing marker means
        $q_{ij} = 0$.
    }
    \label{fig:lama-sparsity-qubo}
\end{figure}

In this setting we construct three example series \lamaSeries{0}, 
\lamaSeries{1}, and \lamaSeries{2} of increasing problem sizes (i.e. number of qubits).
Within each series we
consider different examples of increasing coupling strength which we obtain from
iteratively increasing the overlap of time slots when cars charge in parallel. In 
Figure~\ref{fig:lama-example-series} we visualize the different examples and in
Table~\ref{tab:lama-example-series} we give the problem parameters and
required number of qubits. The different coupling strengths
of the examples in every example series is directly reflected in
the sparsity pattern of the associated QUBO matrix $\matQubo$,
see Figure~\ref{fig:lama-sparsity-qubo}.
We already note here that the sparsity pattern will have a strong impact
on the depth of the QAOA circuits and through this on the quality of results from
real quantum backends.

\subsection{Classical optimization}
\label{sec:lama-classical-optimization}
\begin{figure}[t]
    \small
    \centering
    \begin{subfigure}{0.33\textwidth}
        \includegraphics{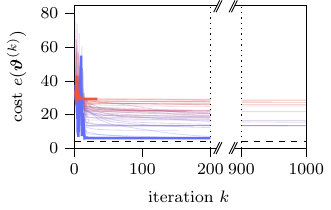}
        \caption{QAOA, $\repsQaoa = 1$, $\numQaoaParams = 2$.}
    \end{subfigure}
    \; \;
    \begin{subfigure}{0.3\textwidth}
        \includegraphics{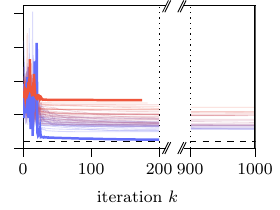}
        \caption{QAOA, $\repsQaoa = 2$, $\numQaoaParams = 4$.}
    \end{subfigure}
    \begin{subfigure}{0.3\textwidth}
        \includegraphics{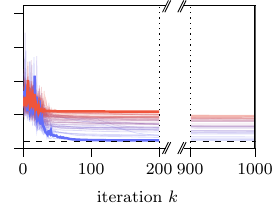}
        \caption{QAOA, $\repsQaoa = 3$, $\numQaoaParams = 6$.}
    \end{subfigure}
    \\[0.2cm]
    \begin{subfigure}{0.33\textwidth}
        \includegraphics{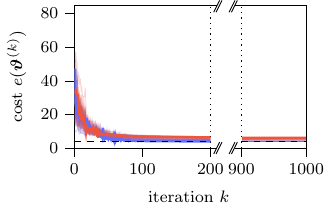}
        \caption{VQE, $\repsVqe = 1$, $\numVqeParams = 16$.}
    \end{subfigure}
    \; \;
    \begin{subfigure}{0.3\textwidth}
        \includegraphics{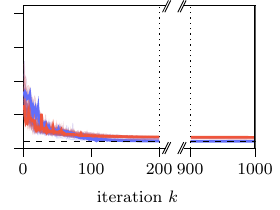}
        \caption{VQE, $\repsVqe = 2$, $\numVqeParams = 24$.}
    \end{subfigure}
    \begin{subfigure}{0.3\textwidth}
        \includegraphics{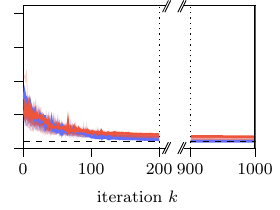}
        \caption{VQE, $\repsVqe = 3$, $\numVqeParams = 32$.}
    \end{subfigure}
    \caption{
        Convergence of the classical optimizer COBYLA used to minimize the cost
        $\cost(\params)$ for QAOA (upper plots) and VQE (lower plots) applied to
        \lamaEx{1}{3}. For both algorithms different number of layers $\repsVqe$
        have been used (left to right), which result in different numbers 
        $\numVqeParams$ of parameters that are optimized.
        Each line stems from one of 50 optimization runs, where each run
        was provided with a different, random initial guess $\params^{(0)}$. 
        The color of
        the lines indicate the final cost $\cost(\params^{\mathrm{final}})$.
        The best and the worst run are indicated with a bold line. The optimization
        was stopped when the default
        termination criterion of COBYLA was satisfied or after
        the default maximum number of 1000 iterations. The quantum circuits were
        executed with an exact simulation. The dashed line indicates the cost
        of the exact solution $\varQuboOpt$.
    }
    \label{fig:convergence-cobyla-exp1p3-exact-simulation}
\end{figure}
As described in Section~\ref{sec:quantum-algorithms-optimization-problems},
we can derive a problem Hamiltonian $\problemHam$ from the QUBO formulation of
our LamA use case examples. Then, we have all ingredients to built the QAOA and 
VQE circuits given in \eqref{eq:qaoa} and \eqref{eq:vqe}, respectively.
In this section we are concerned with the finding of optimal variational parameters,
i.e. $(\bfbetaOpt, \bfgammaOpt)$ for QAOA and $\vqeParamsOpt$ 
for VQE. Recall that these parameters are the solutions of the minimization problems
\eqref{eq:qaoa-optimal-parameters} and \eqref{eq:vqe-optimal-parameters}, respectively.
In the subsequence we just write $\params$ and 
\begin{equation}
    \paramsOpt = \argmin_{\params} \cost(\params) \, ,
    \qquad
    \cost(\params) = \expect{\problemHam}_{\ket{\psi(\params)}} \, ,
    \label{eq:minimization-problem-generic}
\end{equation}
as a generic minimization problem fitting both for QAOA and VQE. For this minimization
problem $\cost$ plays the role of the cost function. Its connection to the
QUBO cost function $\costQuboGeneric$ is given by 
\eqref{eq:connection-qubo-cost-expectation}.

Among the wide variety of classical optimizers that can be used to (approximately)
solve the minimization task \eqref{eq:minimization-problem-generic}
we selected the local optimizer COBYLA (as it is implemented in \qiskit\ with
its default parameters) for the LamA use case. This optimizer computes iteratively 
a sequence of parameters $\{\params^{(j)}\}$ that should approximate the minimum
of $\cost$, i.e. $\cost(\params^{(j)}) \approx \cost(\paramsOpt)$ for $j$ large
enough. To start this sequence, the user has to provide an initial guess $\params^{(0)}$.
As we will see below, this choice can have a strong influence on the sequence
of parameters $\{\params^{(j)}\}$.
In order to compute a new iterate $\params^{(j+1)}$ the cost function $\cost$
has to be evaluated. It is important to note that evaluating the cost function
for a certain choice of parameters $\widetilde{\params}$ 
requires that the QAOA or VQE circuit have to be executed
either on a simulator or on a real quantum backend. However, only on an exact 
simulator $|\psi(\widetilde{\params})\rangle$ and thus 
$\expect{\problemHam}_{|\psi(\widetilde{\params})\rangle}$
can be obtained. For a noisy simulator or a real quantum backend 
only approximations to this expectation can be obtained as we will discuss
further below.
\begin{figure}
    \small
    \centering
    \begin{subfigure}{0.9\textwidth}
        \centering
        \includegraphics{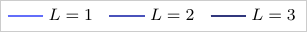}
        \caption*{}
    \end{subfigure}
    \\
    \vspace{-0.4cm}
    \begin{subfigure}{0.325\textwidth}
        \includegraphics{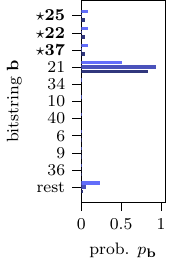}
        \hspace{-0.4cm}
        \includegraphics{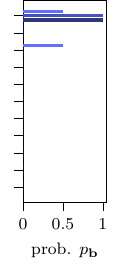}
        \caption{\lamaEx{0}{2}.}
    \end{subfigure}
    \begin{subfigure}{0.295\textwidth}
        \includegraphics{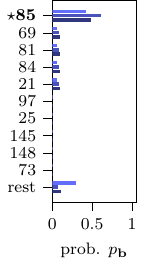}
        \hspace{-0.4cm}
        \includegraphics{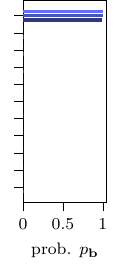}
        \caption{\lamaEx{1}{3}.}
    \end{subfigure}
    \begin{subfigure}{0.32\textwidth}
        \includegraphics{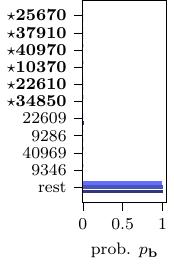}
        \hspace{-0.4cm}
        \includegraphics{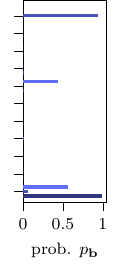}
        \caption{\lamaEx{2}{3}.}
    \end{subfigure}
    \caption{
        Probabilities $\prob_\varQubo(\paramsOpt)$ for the best variational
        parameters $\paramsOpt$ found with the optimizer
        COBYLA.
        In each subplot the left figure stems
        from the QAOA circuit and the right from the VQE circuit,
        where the number of layers $\repsQaoa$
        is indicated by the color of the bar.
        The bitstrings $\varQubo$ are sorted by cost (lower cost on top),
        denoted in integer notation, and only the ten with the lowest cost 
        are displayed. The optimal QUBO solution(s) are marked
        in bold and with a $\star$.
        Note that the convergence plots for the variational parameters
        can be found in Figure~\ref{fig:convergence-cobyla-exp1p3-exact-simulation}
        for \lamaEx{1}{3} and 
        in Figure~\ref{fig:convergence-cobyla-ex0p2-ex2p3-exact-simulation}
        for \lamaEx{0}{2} and \lamaEx{2}{3}.
        }
    \label{fig:figure-probability-distribution-cobyla-exact-simulation}
\end{figure}

The first experiment we want to report here is concerned with optimizing the
QAOA and VQE parameters for the
fully coupled, eight qubit example \lamaEx{1}{3} for circuits 
with $\repsQaoa = 1$, $2$, and $3$ layers. We used an exact simulation to execute
the circuits and ran the COBYLA optimization 50 times, where
in each run we used a different, randomly generated initial guess $\params^{(0)}$.
In Figure~\ref{fig:convergence-cobyla-exp1p3-exact-simulation} we show the
cost $\cost(\params^{(k)})$ versus the iteration number $k$, where every line stems
from one of the 50 initial guesses. Thereby, the color scale indicates the 
final cost $\cost(\params^{\mathrm{final}})$, which is either when the (default) 
termination criterion is satisfied or after the
default maximum number of 1000 iterations.
We use a color scale from blue (lowest cost)
to red (highest cost). Additionally, we marked the two iterations that lead
to the lowest and to the highest final cost with a bold line.
For QAOA we can see a strong dependency of the final cost on the initial guess.
Moreover, we see that for QAOA the optimizer
converged fast and the optimization only continued because the 
default stopping criterion does not work well.
For VQE the convergence is on the one hand slower but on the other hand
the dependency on the initial guess is much less pronounced compared to
QAOA. The reason for both observations could be that in QAOA only 
$\numQaoaParams = 2 \repsQaoa$ parameters have to be optimized,
whereas in VQE we have
$\numVqeParams = \numQubits (\repsVqe + 1) = 8 (\repsVqe + 1)$ parameters. For 
VQE this gives the optimizer the freedom to drive any given initial guess to a 
good solution, however at the cost of having to search in a much larger space of 
parameters.
The convergence plots for \lamaEx{0}{2} and \lamaEx{2}{3} are given in 
Figure \ref{fig:convergence-cobyla-ex0p2-ex2p3-exact-simulation}
in the Appendix. 

Recall from \eqref{eq:approximate-solution-quantum-algorithm} that the aim
of optimizing the variational parameters is to obtain a circuit that generates
a state $\ket{\psi(\paramsOpt)} 
= \sum_{\varQubo \in \{0,1\}^\numQubits} \ampl_{\varQubo}(\paramsOpt) \ket{\varQubo}$
which has an amplitude 
$\ampl_{\varQuboOpt}(\paramsOpt)$ with large absolute value for the basis state 
$\ket{\varQuboOpt}$ of the exact solution of our QUBO.
In order examine if our optimization was able to achieve this we order
the bitstrings $\varQubo_1, \varQubo_2, \dots, \varQubo_{2^\numQubits}$
in $\{0, 1\}^\numQubits$
with respect to their cost, i.e. $\cost(\varQubo_1) \le \cost(\varQubo_2) \le \dots
\le \cost(\varQubo_{2^\numQubits})$. 
For a shorter notation we write the bitstrings in integer notation, i.e. for 
$\varQubo = (\varQuboEntry_1, \dots, \varQuboEntry_\numQubits)$ we write
$n = \sum_{i=1}^\numQubits \varQuboEntry_i 2^{i-1}$.
As optimal parameters $\paramsOpt$ we take
$\params^\mathrm{final}$ from the optimization pass that yielded the lowest 
cost, i.e. we take the best parameters that were found in the 50 optimizations.
In Figure~\ref{fig:figure-probability-distribution-cobyla-exact-simulation} we
show $\prob_{\varQubo_i}(\paramsOpt) = |\ampl_{\varQubo_i}(\paramsOpt)|^2$ 
for the ten best bitstrings 
$\varQubo_i$, $i=1, \dots, 10$, for \lamaEx{0}{2}, \lamaEx{1}{3}, and \lamaEx{2}{3},
and for both QAOA and VQE with $\repsQaoa = 1$, $2$, and $3$ layers. 
We see that for the two smaller examples
VQE is able to generate states which solely
contain optimal solution bitstrings (marked in bold and with 
a $\star$ in the plots). Even for the largest example
and $\repsVqe=2$ this is almost the case. On the other hand, QAOA generates
for \lamaEx{0}{2} and all layer numbers states that have their largest amplitude
for a bitstring that has the second smallest cost. For \lamaEx{1}{3} the largest 
amplitude is for the correct bitstring but with a smaller absolute value
square than for VQE. Finally, for \lamaEx{2}{3} QAOA is not able to generate a state with 
amplitudes that have an absolute value notably above zero for the ten bitstrings
with the lowest cost.

\begin{figure}
    \small
    \centering
    \begin{subfigure}{0.33\textwidth}
        \includegraphics{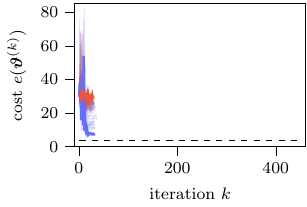}
        \caption{QAOA, $\repsQaoa = 1$, $\numQaoaParams = 2$.}
    \end{subfigure}
    \; \;
    \begin{subfigure}{0.3\textwidth}
        \includegraphics{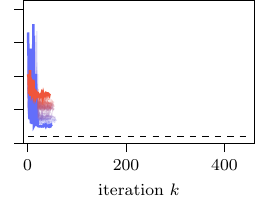}
        \caption{QAOA, $\repsQaoa = 2$, $\numQaoaParams = 4$.}
    \end{subfigure}
    \begin{subfigure}{0.3\textwidth}
        \includegraphics{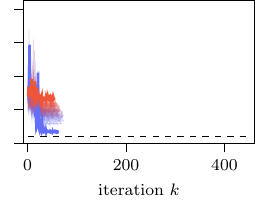}
        \caption{QAOA, $\repsQaoa = 3$, $\numQaoaParams = 6$.}
    \end{subfigure}
    \\[0.2cm]
    \begin{subfigure}{0.33\textwidth}
        \includegraphics{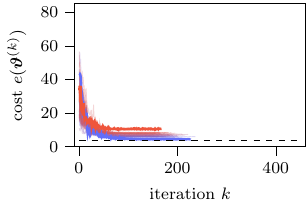}
        \caption{VQE, $\repsVqe = 1$, $\numVqeParams = 12$.}
    \end{subfigure}
    \; \;
    \begin{subfigure}{0.3\textwidth}
        \includegraphics{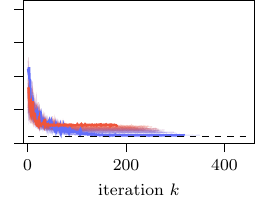}
        \caption{VQE, $\repsVqe = 2$, $\numVqeParams = 18$.}
    \end{subfigure}
    \begin{subfigure}{0.3\textwidth}
        \includegraphics{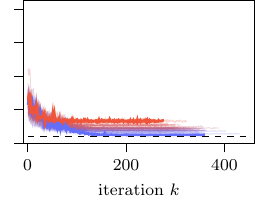}
        \caption{VQE, $\repsVqe = 3$, $\numVqeParams = 24$.}
    \end{subfigure}
    \caption{
        Convergence of the classical optimizer COBYLA for \lamaEx{1}{3} in the
        same situation as in
        Figure~\ref{fig:convergence-cobyla-exp1p3-exact-simulation}
        except that an exact simulation with shot noise is used to execute the
        quantum circuits. The number of shots is set to $\numShots = 1000$.
    }
    \label{fig:convergence-cobyla-exp1p3-shot-noise-simulation}
\end{figure}

Last in this section, let us recall that on real quantum backends we do not have
access to the amplitudes of a quantum state but only to an approximation
$\probTilde_\varQubo$ of the probabilities $\prob_\varQubo = |\ampl_\varQubo|^2$.
The probability distribution $\{\probTilde_\varQubo\}$ is obtained from
executing and measuring the quantum circuit for $\numShots$ times, where each
cycle is called a shot. Thus, the error between $\probTilde_\varQubo$ 
and $\prob_\varQubo$ is called shot noise. This is not connected to hardware
errors but would be present even for an error free quantum computer. Such an error
free quantum computer can be emulated with an exact simulator with shot noise. 
For such an simulator we show in
Figure~\ref{fig:convergence-cobyla-exp1p3-shot-noise-simulation}
the results for $\numShots = 1000$ shots of the exact same experiments
as above with the exact, shot noise free simulator. We see that for QAOA the optimizer terminates much earlier
but with a higher cost function value. The strong dependency on the initial guess
is still present and even more pronounced towards optimization passes that end
with a high cost function value. For VQE we observe that the initial value now has
a slightly stronger influence and that the final cost is also slightly higher compared
to the shot noise free simulation.

We close this section with remarking that we ran all the upper classical 
optimizations for 10 different values of the penalty parameter $\myPenalty$.
More precisely, we calculated the minimum penalty $\myPenalty_\mathrm{min}$
that ensures that the solution
of the QUBO \eqref{eq:qubo} is also the solution of the original problem 
\eqref{eq:qcio} and considered the values
$\myPenalty = \myPenalty_\mathrm{min} + 0.1k$, $k=0, \dots, 9$. The results from
above stem from the penalty that gave the lowest cost in the classical 
optimization. This penalty is also used
for all results on quantum computing backends that follow in the subsequence.
\subsection{Experiments on \ibmqehningen}
\label{sec:lama-ibmq-ehningen}
\begin{figure}
    \centering
    \input{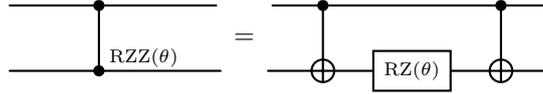}
    \caption{
        Decomposition of the $\RZZ$ gate into one $\RZ$ and two $\CX$ gates.
    }
    \label{fig:rzz-transpiled}
\end{figure}

In this section we present experiments on the 27 qubit backend \ibmqehningen.
Experiments on the backends \ibmtorino , \ibmsherbrooke , and \ibmcairo
can be found in Section~\ref{sec:lama-other-ibmq}. 
\paragraph{Transpilation}
As a first step the logical QAOA and VQE circuits, recall 
Figures~\ref{fig:qaoa-circuit} and \ref{fig:vqe-circuit}, respectively,
have to be transpiled
to circuits that can be executed on \ibmqehningen. Recalling 
Section~\ref{sec:quantum-systems-ibmq} on the IBM quantum systems, among others,
the transpilation has to route multi-qubit gates such that they match the topology
of \ibmqehningen, see Figure~\ref{fig:topology-ibmq-ehningen}, and it has
to synthesize the logical gates into the hardware gates
$\RZ$, $\pauliX$, $\SX$, and $\CX$.
Of all hardware gates the $\CX$ gate has by far the largest error, 
see Figure~\ref{fig:ibmq-gate-errors}, and thus will contribute the
dominant gate error part of a quantum circuit. Consequently, 
we are mostly interested in the $\CX$ gates in the transpiled quantum
circuit and will neglect the transpilation of single qubit gates.
\begin{figure}
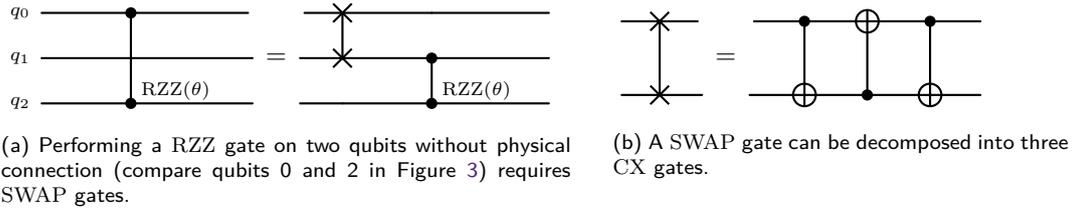

    \centering
    \begin{subfigure}[t]{0.5\textwidth}
        \input{figures/circuits/rzz_unconnected_qubits.tex}
        \caption{
            Performing a $\RZZ$ gate on two qubits without physical
            connection (compare qubits 0 and 2 in
            Figure~\ref{fig:topology-ibmq-ehningen})
            requires $\SWAP$ gates.
        }
        \label{fig:rzz-unconnected-qubits-1}
    \end{subfigure}
    \; \;
    \begin{subfigure}[t]{0.4\textwidth}
        \input{figures/circuits/swap_cnots.tex}
        \caption{A $\SWAP$ gate can be decomposed into three $\CX$ gates.}
        \label{fig:rzz-unconnected-qubits-2}
    \end{subfigure}
    \caption{
        A $\RZZ$ gate for unconnected qubits requires the insertion of 
        $\SWAP$ gates which can be decomposed into $\CX$ gates.
    }
    \label{fig:rzz-unconnected-qubits}
\end{figure}
Now, let us look deeper into the transpilation of our QAOA and VQE circuits:
Recall that for the
QAOA algorithm \eqref{eq:qaoa-all} we derived the gate representation
of the involved operators in
\eqref{eq:mixing-operator-gates}, \eqref{eq:phase-operator-decomposition}, and 
\eqref{eq:qaoa-state-preparation}. There, we see that the only multi-qubit gate
in the logical quantum circuit is the $\RZZ$ gate. For the VQE algorithm with
the two-local ansatz \eqref{eq:vqe} the only multi-qubit gate is the $\CX$ gate.
To match the topology of \ibmqehningen\ the transpilation has to add $\SWAP$
gates for all two qubit gates that act on two qubits that are not connected. 
For example an $\RZZ$ gate on qubits 0 and 2 requires a $\SWAP$ gate between 
qubits 0 and 1, compare 
Figure~\ref{fig:rzz-unconnected-qubits-1}.
Each $\SWAP$ gate in turn can be decomposed into three $\CX$ gates, see 
Figure~\ref{fig:rzz-unconnected-qubits-2}.
Moreover, an $\RZZ$ gate can be decomposed
into two $\CX$ gates and one $\RZ$ gate,
see Figure~\ref{fig:rzz-transpiled}. 

\begin{figure}
    \small
    \centering
    \begin{subfigure}[t]{0.4\textwidth}
        \includegraphics{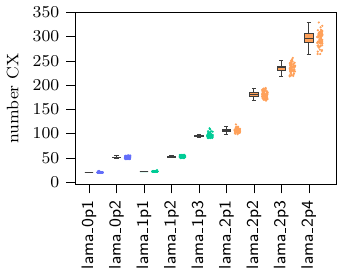}
        \caption{QAOA.}
        \label{fig:qaoa-transpilation-CNOTs}
    \end{subfigure}
    \qquad
    \begin{subfigure}[t]{0.4\textwidth}
        \includegraphics{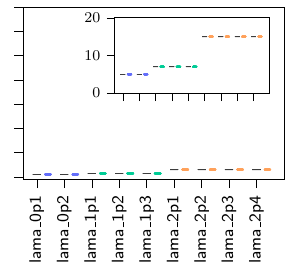}
        \caption{VQE.}
        \label{fig:vqe-transpilation-CNOTs}
    \end{subfigure}
    \caption{
        Number of $\CX$ gates in the transpilations of the 
        QAOA (left) and VQE (right) circuits to the backend \ibmqehningen\
        for the LamA use case examples.
        The transpilation was carried out with the \qiskit\ transpiler on
        \optimizationLevel\ 3. Every circuit was transpiled 75 times, where
        for each transpilation a different transpilation seed was used.
        For every example each dot in the figures represents the number
        of $\CX$ gates from one such transpilation.
    }
    \label{fig:transpilation-CNOTs}
\end{figure}
With these considerations in mind, let us now look at the transpilations of the
logical QAOA and VQE circuits for all our LamA examples which we obtained
by using the \qiskit\ transpiler. Clearly, there are many possibilities how our
logical circuits can be transpiled to \ibmqehningen\ and finding the best transpilation
is a non-trivial task. In order to control how much optimization is performed
on this task the \qiskit\ transpiler offers four \optimizationLevel s  0, 1, 2, and 3.
Moreover, it is important to know
that the transpiler has stochastic components which can be controlled with a
so-called random seed. In Figure~\ref{fig:transpilation-CNOTs} we show the number of
$\CX$ gates for the transpilations
of the QAOA and VQE circuits to \ibmqehningen\ for 75 different seeds
and for the highest \optimizationLevel\ 3.
Let us first look at the VQE circuits in Figure~\ref{fig:vqe-transpilation-CNOTs}.
There, we see that the number of $\CX$ gates is the same for every example
of an example series. This is due to the fact that only the number of qubits
enters in our VQE algorithm \eqref{eq:vqe} and this number is the same
for all examples of an example series, see 
Table~\ref{tab:lama-example-series-problem-parameters}. 
Moreover, we only observe a slight increase in the number of $\CX$ gates
from $\lamaSeries{0}$ to $\lamaSeries{1}$ to $\lamaSeries{2}$. Besides the increasing
qubit number the reason lies in the insertion of $\SWAP$ gates which are necessary
to perform $\CX$ gates between unconnected qubits. However, due to the linear 
entanglement structure in our VQE ansatz only few $\SWAP$ gates are needed. The
simplicity of this ansatz is also reflected in the fact that every transpilation
seed results in a circuit with the same number of $\CX$ gates.
For the QAOA circuits we observe a completely different behavior, see 
Figure~\ref{fig:qaoa-transpilation-CNOTs}. First of all, the (median) number 
of $\CX$ gates is not constant within the example series but different for
each individual example. In order to understand this, first note that the
problem Hamiltonian $\problemHam$ enters in the phase operator $\problemOperator$,
see \eqref{eq:qaoa-1}, and thus the  QAOA circuit depends on it. In particular, 
by recalling \eqref{eq:phase-operator-decomposition}
we see that the off-diagonal non-zero entries $h_{ij} \neq 0$, $i \neq j$, 
of the problem 
Hamiltonian $\problemHam$ induce the $\RZZ_{i,j}$ gates in the QAOA circuits.
It is easy to show
that $h_{ij} \neq 0$ if and only if $\matQuboEntry_{ij} \neq 0$, 
where $\matQuboEntry_{ij}$ is the $(i, j)$th entry
of the QUBO matrix $\matQubo$, see \eqref{eq:qubo-generic-cost}. This means that
the sparsity pattern (the non-zero entries) of the QUBO matrix is directly connected
to the number of $\RZZ$ gates in the logical QAOA circuits. 
Recall from above that it is precisely the $\RZZ$ gates that generate the
$\CX$ gates in the transpilation. Thus, one reason for the different
number of $\CX$ gates for every example is that the sparsity pattern of the QUBO
matrices of our examples, see Figure~\ref{fig:lama-sparsity-qubo}, are
all different. Moreover, the insertion of $\SWAP$ gates that are needed to realize
$\RZZ$ gates between unconnected qubits increases the number of $\CX$ gates in the
transpiled circuits. Note that due to the low connectivity of \ibmqehningen, recall
Figure~\ref{fig:topology-ibmq-ehningen}, the denser the QUBO matrices are (the 
more non-zero entries they have) the stronger this effect is. We can see this e.g.
by comparing \lamaEx{0}{2} and \lamaEx{1}{2} or \lamaEx{1}{3} and \lamaEx{2}{1},
where the example with less qubits but denser QUBO matrix
has a similar number of $\CX$ gates than the example with the higher qubit number
but sparser QUBO matrix.
Last, let us remark that the deviation of the $\CX$ numbers for a fixed example 
is due to the optimization potential and the stochastic part in the transpilation.
This optimization potential is the higher the deeper the circuits get as can
be seen by comparing the variances from the smaller example series with the larger
ones.

\begin{figure}
    \small
    \centering
    \begin{subfigure}[t]{0.3\textwidth}
        \centering
        \includegraphics{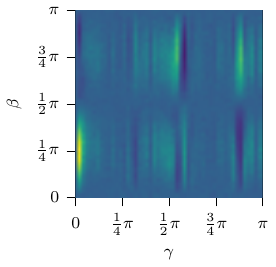}
        \caption{Exact simulation with shot noise.}
        \label{fig:energy-landscape-exact-simulation}
    \end{subfigure}
    \;
    \begin{subfigure}[t]{0.28\textwidth}
        \centering
        \includegraphics{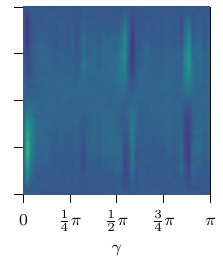}
        \caption{Simulation with noise model of \ibmqehningen.}
        \label{fig:energy-landscape-noisy-simulation}
    \end{subfigure}
    \begin{subfigure}[t]{0.3\textwidth}
        \centering
        \includegraphics{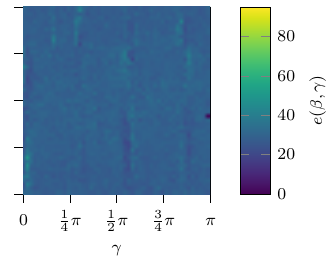}
        \caption{\ibmqehningen.}
        \label{fig:energy-landscape-ibmq-ehningen}
    \end{subfigure}
    \\[0.2cm]
    \begin{subfigure}[t]{0.28\textwidth}
        \centering
        \phantom{\includegraphics{figures/lama_energy_landscape/energy_heatmap_exact_simulation.pdf}}
    \end{subfigure}
    \hspace{-0.3cm}
    \begin{subfigure}[t]{0.3\textwidth}
        \centering
        \includegraphics{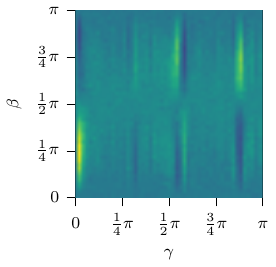}
        \caption{Simulation with noise model \\ of \ibmqehningen.}
        \label{fig:energy-landscape-noisy-simulation-zoom}
    \end{subfigure}
    \; \;
    \begin{subfigure}[t]{0.3\textwidth}
        \centering
        \includegraphics{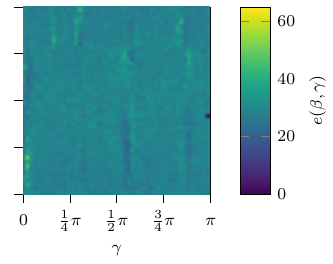}
        \caption{\ibmqehningen.}
        \label{fig:energy-landscape-ibmq-ehningen-zoom}
    \end{subfigure}
    \caption{Heat map of the cost $\cost(\beta, \gamma)$, 
        for $\beta, \gamma \in [0, \pi]$,
        QAOA with $\repsQaoa = 1$ layer,
        and \lamaEx{1}{3}. The interval $[0, \pi]$
        was discretized in 50 grid points for both $\beta$ and $\gamma$. The QAOA
        circuit was executed with three different backends: exact simulator with
        shot noise (a), simulator with noise model of \ibmqehningen\ (b) and (d),
        and \ibmqehningen\ (c) and (e). For all backends the number of shots
        was set to $\numShots = 1000$. The results for \ibmqehningen\ stem from
        three sessions that ran on 2023/11/08, 2023/11/09, and 2023/11/10. The noise
        model is from 2023/11/10. For the (a) -- (c) the color scale ranges from
        $0$ to $95$ and for (d) and (e) from $0$ to $65$.}
    \label{fig:energy-landscape}
\end{figure}

\begin{figure}
    \small
    \centering
    \begin{subfigure}{0.32\textwidth}
        \centering
        \includegraphics{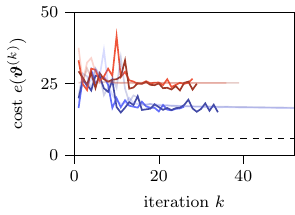}
        \caption{\lamaEx{0}{2}, QAOA.}
        \label{fig:lama-class-optimization-ehningen-qaoa-ex0p2}
    \end{subfigure}
    \;
    \begin{subfigure}{0.3\textwidth}
        \centering
        \includegraphics{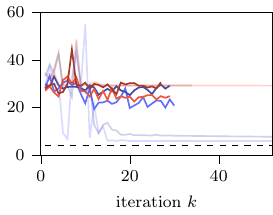}
        \caption{\lamaEx{1}{3}, QAOA.}
        \label{fig:lama-class-optimization-ehningen-qaoa-ex1p3}
    \end{subfigure}
    \;
    \begin{subfigure}{0.3\textwidth}
        \centering
        \includegraphics{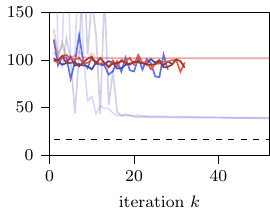}
        \caption{\lamaEx{2}{4}, QAOA.}
        \label{fig:lama-class-optimization-ehningen-qaoa-ex2p4}
    \end{subfigure}
    \\[0.2cm]
    \begin{subfigure}{0.32\textwidth}
        \centering
        \includegraphics{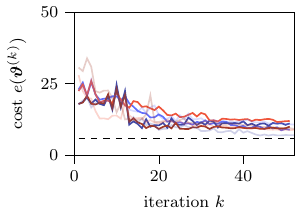}
        \caption{\lamaEx{0}{2}, VQE.}
        \label{fig:lama-class-optimization-ehningen-vqe-ex0p2}
    \end{subfigure}
    \;
    \begin{subfigure}{0.3\textwidth}
        \centering
        \includegraphics{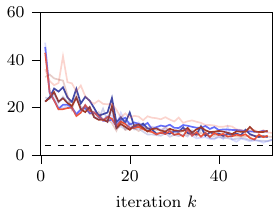}
        \caption{\lamaEx{1}{3}, VQE.}
        \label{fig:lama-class-optimization-ehningen-vqe-ex1p3}
    \end{subfigure}
    \;
    \begin{subfigure}{0.3\textwidth}
        \centering
        \includegraphics{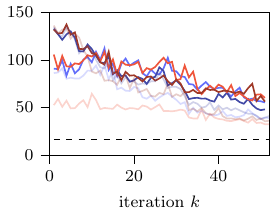}
        \caption{\lamaEx{2}{4}, VQE.}
        \label{fig:lama-class-optimization-ehningen-vqe-ex2p4}
    \end{subfigure}
    \caption{
        Convergence of the classical optimizer COBYLA used to minimize the cost
        $\cost(\params)$ for QAOA (upper plots) and VQE (lower plots) applied to
        \lamaEx{0}{2}, \lamaEx{1}{3}, and \lamaEx{2}{4} (left to right). 
        For both algorithms $\repsVqe = 1$ layer was used. As initial guess
        $\params^{(0)}$ the two best (blue lines) and the two worst (red lines)
        choices, that have been
        identified with an exact simulation (compare 
        Figure~\ref{fig:convergence-cobyla-exp1p3-exact-simulation}), were used.
        The quantum circuits were executed on 
        \ibmqehningen\ using \qiskit's \texttt{Estimator} primitive
        with \resilienceLevel\ 1 and \optimizationLevel\ 3.
        In order to save quantum computing resources
        the maximum number of iterations was limited
        to 50 and the number of shots was chosen as 1000. 
        The optimization was stopped when the default
        termination criterion of COBYLA was satisfied or after
        the maximum number iterations was reached.
        The dashed line indicates the cost
        of the exact solution $\varQuboOpt$ and the transparent lines show the
        convergence of the exact simulation.
        The circuits for the plots (a) -- (f) were executed on 2023/10/24 11:13,
        2023/10/23 22:48, 2023/10/26 20:26, 2023/10/25 11:24,
        2023/10/24 12:39, and 2023/10/27 21:37, respectively.
        }
    \label{fig:lama-class-optimization-ehningen}
\end{figure}

\paragraph{Cost landscape}
Recall from \eqref{eq:qaoa-optimal-parameters} and 
\eqref{eq:minimization-problem-generic} that we are interested
in finding the QAOA parameters $\bfbeta$ and $\bfgamma$ that minimize 
$\cost(\bfbeta, \bfgamma) 
= \expect{\problemHam}_{\ket{\psiQaoa(\bfbeta, \bfgamma)}}$. For QAOA with 
$\repsQaoa = 1$ layer we only have two parameters $\bfbeta = \beta$ 
and $\bfgamma = \gamma$,
and thus can visualize the cost
\begin{equation*}
    \cost(\beta, \gamma) 
    = \expect{\problemHam}_{|\psiQaoa(\beta, \gamma)\rangle} \, ,
    \qquad
    \beta, \gamma \in [0, \pi] \, ,
\end{equation*}
as a heat map. For this we discretized the interval $[0, \pi]$ into
$50$ grid points for both $\beta$ and $\gamma$, and then computed 
$\cost(\beta, \gamma)$ for each of the $50 \cdot 50 = 2500$ grid points. Note that
each evaluation of $\cost(\beta, \gamma)$ requires the execution of one QAOA circuit.
In Figure~\ref{fig:energy-landscape} we show the resulting cost landscape
for \lamaEx{1}{3}, where the QAOA circuits were executed either with 
an exact simulation with shot noise 
(Figure~\ref{fig:energy-landscape-exact-simulation}),
a simulation with a noise model of \ibmqehningen\ 
(Figure~\ref{fig:energy-landscape-noisy-simulation}),
or on the \ibmqehningen\ backend. In all three cases the number of shots was set to 
$\numShots = 1000$. For \ibmqehningen\ a transpilation with \optimizationLevel\ 3
and measurement error mitigation was used. Comparing the result from \ibmqehningen\
with the exact simulation we clearly see that most of the signal is lost and 
only few details of the cost landscape can be resolved. The simulation with the
noise model predicts a loss of signal and details but by far not as strong as
obtained from the real backend. Moreover, we see that the cost landscapes have different
scales. In fact, the minimum and maximum for the exact simulation are $8.04$ and 
$95.56$, for the simulation with the noise model $14.51$ and $63.53$, and for \ibmqehningen\
$0.0$ and $52.52$. In Figures~\ref{fig:energy-landscape-noisy-simulation-zoom} and
\ref{fig:energy-landscape-ibmq-ehningen-zoom} we give the cost landscapes for
the simulation with the noise model and \ibmqehningen\ on a zoomed color scale. Also
on this scale the signal of \ibmqehningen\ is weak and much noisier than
the simulation with the noise model predicts. This already hints that optimizing the variational parameters
$\bfbeta$ and $\bfgamma$ on the real quantum computer is a difficult task.
We will take a closer look at this in the next paragraph.
\paragraph{Classical optimization on \ibmqehningen}
In Section~\ref{sec:lama-classical-optimization} we studied 
the optimization of
the parameters $\bfbeta$ and $\bfgamma$ for QAOA and $\vqeParams$ for VQE 
with the COBYLA optimizer in combination with an 
exact simulator with and without shot noise to execute the quantum circuits.
Now, we use the same optimizer with the same settings (except we limit the maximum 
number of function evaluations to 50) but execute the circuits on the \ibmqehningen\ 
backend. We do this only for the two best and two worst initial guesses, that we
found when using the exact simulator. In 
Figure~\ref{fig:lama-class-optimization-ehningen}
we show the results for \lamaEx{0}{2}, \lamaEx{1}{3}, and \lamaEx{2}{4} for
QAOA and VQE. In low opacity we added the result of the exact simulator. We 
can clearly see that for the smallest example the QAOA parameters
can be determined with \ibmqehningen\ with a  cost function value comparable to
the one of the exact simulator. However, for the two larger examples
we see (nearly) no convergence and no agreement with
the exact simulation. A meaningful optimization does not take place. In contrary,
we see a good agreement for VQE for the smaller examples, and even for the largest
example we can recognize a convergence behavior.
\subsubsection{Quality of results}
In the next paragraphs we are concerned with the question of how well our QAOA and
VQE circuits can be executed on today's error-prone quantum hardware.
For this purpose we will only execute fully trained circuits, where we take the best
values for our variational parameters that we have found in the above optimization
runs. Before we present the results let us introduce some
notations: Let us consider a fully trained circuit and let us denote with 
\begin{equation*}
    \ketSimple{\psiHat}
    = \sum_{\varQubo \in \{0,1\}^\numQubits}
    \amplHat_\varQubo \ket{\varQubo}
\end{equation*}
the state that such a circuit generates. Using an exact simulation
we can access $\ketSimple{\psiHat}$ and, in particular, we can calculate the
probability distribution
$\{\probHat_\varQubo\}$ by $\probHat_\varQubo = |\amplHat_\varQubo|^2$.
Now, let us denote with $\ketSimple{\psiTilde}$ the state that a real quantum
computer generates when it executes the same circuit. Then, we do not have 
access to 
the quantum state $\ketSimple{\widetilde{\psi}}$ but only to the
probability distribution $\{\probTilde_\varQubo\}$ that we
obtain from $\numShots$ shots of executions and measurements of the circuit.
In this paper we will consider two metrics to measure the quality of 
$\{\probTilde_\varQubo\}$.

The first metric is the Hellinger fidelity defined by
\begin{equation}
    \fidelity
    = \fidelity \bigl( 
        \{\probHat_\varQubo\}, \{\probTilde_\varQubo\} 
    \bigr)
    = \left(
        \sum_{\varQubo \in \{0, 1\}^\numQubits}
        \sqrt{\probHat_\varQubo \; \probTilde_\varQubo}
    \right)^2 \, .
    \label{eq:fidelity}
\end{equation}
For this metric we have $\fidelity \in [0, 1]$, where
$\fidelity \approx 1$ indicates
a high agreement of the result from the real computer with the simulation whereas
$\fidelity \approx 0$ indicates a poor agreement. Note that the fidelity is problem
agnostic, so that a high fidelity does not necessarily mean that 
$\ketSimple{\widetilde{\psi}}$ yields a good solution for our
underlying QUBO problem. 
This is different for the second metric we consider, namely 
the relative error
\begin{subequations}
    \begin{equation}
        \relCostError\bigl( \{\prob_\varQubo\} \bigr)
        = \frac{1}{\costQuboGenericOpt}
        \bigl|
            \cost(\{\prob_\varQubo\}) 
            - \costQuboGenericOpt
        \bigr| \, ,
        \qquad
        \costQuboGenericOpt = \costQuboGeneric(\varQuboOpt) \, ,
    \end{equation}
    where $\varQuboOpt$ is the solution of our QUBO \eqref{eq:qubo-generic} and
    where in analogy to \eqref{eq:connection-qubo-cost-expectation} we define
    \begin{equation}
        \cost(\{\prob_\varQubo\})
        = \sum_{\varQubo \in \{0, 1\}^\numQubits}
        \prob_\varQubo \costQuboGeneric(\varQubo) \, .
    \end{equation}
    \label{eq:relative-cost-error}
\end{subequations}
If $\relCostError\bigl( \{\prob_\varQubo\} \bigr) \approx 0$ then we have a 
good approximation to the solution of our QUBO in the sense of 
\eqref{eq:approximate-solution-quantum-algorithm}.

In the now following we will report and discuss results from \ibmqehningen.
Results from other quantum backends then follow in Section~\ref{sec:lama-other-ibmq}.

\paragraph{Setup}
As we have discussed further above, before circuits can be executed on a 
quantum computer they have to be transpiled. We have seen that this step is not
deterministic and, depending on the transpilation seed, we can obtain circuits with 
different numbers of gates. Moreover, all our examples use less than the 27 qubits 
that \ibmqehningen\ provides. Thus, transpilations with different seeds could
use different qubits, which in turn
can have different error rates. In order to account for this we transpile
every logical circuit 75 times, where we use a different transpilation seed each time.
We run all of the resulting 75 transpiled circuits and will always report
on all their outcomes in the subsequence.
We always use measurement error mitigation as provided by \resilienceLevel\ 1
of \qiskit's \texttt{Sampler} primitive and $\numShots = 10000$ shots.
Moreover, we use a $\pauliX$--$\pauliX$ dynamical decoupling sequence
as a simple but powerful technique to
strengthen the resilience of the qubits \cite{Ezzell.2023,Kim.2023}. 
In order to show the differences we always report results once without
and once with dynamical decoupling.
%

% Fidelity boxplots, reps=1, 2023-10-23 12:03
\begin{figure}
    \small
    \centering
    \begin{subfigure}{0.45\textwidth}
        \centering
        \includegraphics{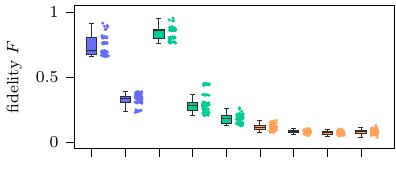}
        \caption{QAOA, w/o dynamical decoupling.}
    \end{subfigure}
    \; \;
    \begin{subfigure}{0.45\textwidth}
        \centering
        \includegraphics{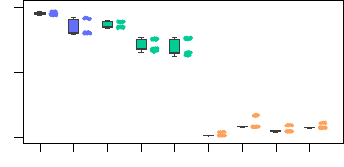}
        \caption{VQE, w/o dynamical decoupling.}
    \end{subfigure}
    \\[0.3cm]
    \begin{subfigure}{0.45\textwidth}
        \centering
        \includegraphics{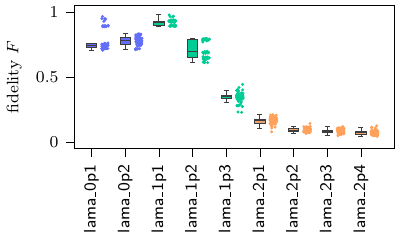}
        \caption{QAOA, w/ dynamical decoupling.}
        \label{fig:lama-fidelity-boxplots-qaoa-dd}
    \end{subfigure}
    \; \;
    \begin{subfigure}{0.45\textwidth}
        \centering
        \includegraphics{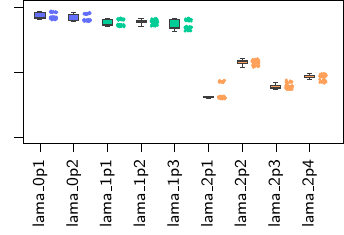}
        \caption{VQE, w/ dynamical decoupling.}
        \label{fig:lama-fidelity-boxplots-vqe-dd}
    \end{subfigure}
    \caption{
        Fidelity $\fidelity$, as defined in \eqref{eq:fidelity},
        obtained from executing the QAOA (left) and VQE 
        (right) circuits with $\repsQaoa = 1$ layer
        for all LamA use case examples on \ibmqehningen.
        We used
        the best values for the variational parameters that we found
        with the optimizer COBYLA and an exact simulation,
        see Section~\ref{sec:lama-classical-optimization}.
        For each example, the logical circuit was transpiled 75 times,
        using a different transpilation seed each time. Each dot stems from 
        such a transpilation. We ran the resulting circuits with (bottom) and
        without (top) dynamical decoupling. The number of shots was set to
        $\numShots = 10000$. All circuits were ran within a single \qiskit\ 
        \texttt{session}, where the first circuit ran on 
        2023/10/23 12:03.
    }
    \label{fig:lama-fidelity-boxplots}
\end{figure}

\paragraph{Fidelity}
In Figure~\ref{fig:lama-fidelity-boxplots} we show the fidelity $\fidelity$
obtained from running the QAOA and VQE
circuits on \ibmqehningen\ for all LamA examples. For QAOA we observe that without
dynamical decoupling only \lamaEx{0}{1} and \lamaEx{1}{1}, i.e. the small and weakly
coupled examples, can be carried out with high fidelity. The larger examples, in particular
all examples in example series \lamaSeries{2} and the stronger coupled examples
of series \lamaSeries{0} and \lamaSeries{1} only achieve a poor fidelity. 
This can be improved by using dynamical decoupling
so that also \lamaEx{0}{2} and \lamaEx{1}{2} can be carried out
with higher fidelity. However, example series \lamaSeries{2} stays infeasible
for \ibmqehningen.
On the other hand, the VQE circuits can be carried out on \ibmqehningen\
with a rather high
fidelity for both example series \lamaSeries{0} and \lamaSeries{1}. In
particular, with dynamical decoupling the fidelity is always close to $1$. However,
for all examples from \lamaSeries{2} the fidelity drops significantly.
Only with dynamical decoupling we obtain a fidelity around 0.5. Recalling from
Figure~\ref{fig:transpilation-CNOTs} that all VQE circuits have nearly 
the same amount of $\CX$ gates
it is surprising that the \lamaSeries{2} series yields such bad fidelities.
%

% Bar plots, probability
\begin{figure}
    \small
    \centering
    \begin{subfigure}[t]{0.33\textwidth}
        % \centering
        \includegraphics{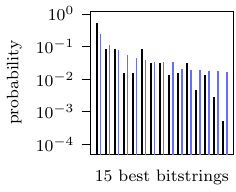}
        \includegraphics{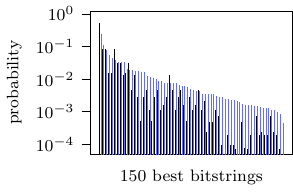}
        \caption{QAOA, \lamaEx{0}{2}.}
    \end{subfigure}
    \begin{subfigure}[t]{0.3\textwidth}
        % \centering
        \includegraphics{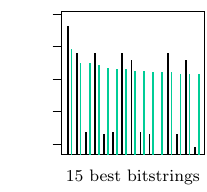}
        \includegraphics{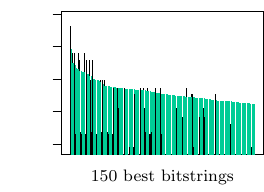}
        \caption{QAOA, \lamaEx{1}{3}.}
    \end{subfigure}
    \begin{subfigure}[t]{0.3\textwidth}
        % \centering
        \includegraphics{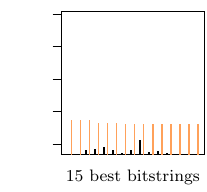}
        \includegraphics{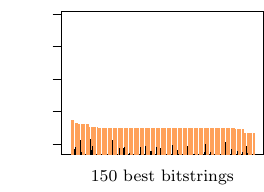}
        \caption{QAOA, \lamaEx{2}{3}.}
    \end{subfigure}
    \\[0.3cm]
    \begin{subfigure}[t]{0.33\textwidth}
        % \centering
        \includegraphics{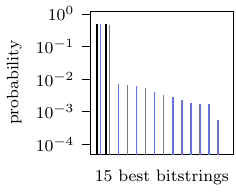}
        \includegraphics{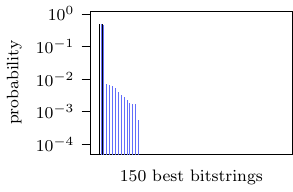}
        \caption{VQE, \lamaEx{0}{2}.}
    \end{subfigure}
    \begin{subfigure}[t]{0.3\textwidth}
        % \centering
        \includegraphics{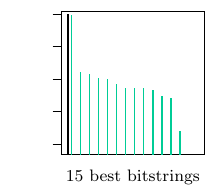}
        \includegraphics{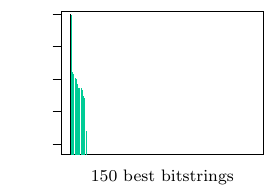}
        \caption{VQE, \lamaEx{1}{3}.}
    \end{subfigure}
    \begin{subfigure}[t]{0.3\textwidth}
        % \centering
        \includegraphics{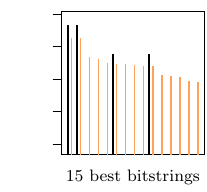}
        \includegraphics{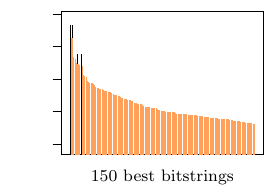}
        \caption{VQE, \lamaEx{2}{3}.}
    \end{subfigure}
    \caption{
        Probabilities $\probTilde_\varQubo$ resulting from executing the QAOA
        (top) and VQE (bottom) circuits with $\repsQaoa = 1$ layer for
        different LamA use case examples on 
        \ibmqehningen\ (colored bars). The results stem from the experiment described in 
        Figure~\ref{fig:lama-fidelity-boxplots} and we show the case of the best
        transpilation seed (with respect to the fidelity) and with dynamical
        decoupling. In each subfigure we show the 15 (upper plot) and 150 
        (lower plot) highest probabilities. For clarity we do not show the 
        associated bitstrings but only refer to them as the 
        15 or 150 best bitstrings. The black bars are the probabilities
        $\probHat_{\varQubo}$ for those bitstrings but stemming from using an
        exact simulator instead of \ibmqehningen.
    }
    \label{fig:lama-probability-distributions}
\end{figure}

This becomes clearer when analyzing the probability distributions 
$\{\probTilde_\varQubo\}$
that we obtain from 
\\
\ibmqehningen\ for the different examples. We show them 
in Figure~\ref{fig:lama-probability-distributions} 
for \lamaEx{0}{2}, \lamaEx{1}{3}, and \lamaEx{2}{3}, for the best transpilation seed
and with dynamical decoupling. More precisely, we show 
the 150 of the
$2^\numQubits$ pairs $(\varQubo, \probTilde_{\varQubo})$ with the highest
probability $\probTilde_{\varQubo}$. On the $x$-axis we have the bitstring $\varQubo$ whose
value we do not show to keep a clear figure. On the $y$-axis we plot the respective
probability $\probTilde_{\varQubo}$. Note that on this axis we are using a 
logarithmic scale. As comparison we plot in black the probability 
$\probHat_{\varQubo}$ of the exact simulation. 
In good accordance with the QAOA fidelity plot in
Figure~\ref{fig:lama-fidelity-boxplots-qaoa-dd} we see that for \lamaEx{0}{2} 
we obtain a probability distribution 
$\{\probTilde_{\varQubo}\}$ from \ibmqehningen\ that has large values on similar
bitstrings as the exact simulation. However, it is not able to resolve
the details correctly. For \lamaEx{1}{3} we see that the peak probabilities of the 
exact simulation can only be reached very poorly. Finally, for \lamaEx{2}{3} 
we see that the probability distribution of \ibmqehningen\ does not match the
simulation at all. Let us also emphasize that the probability distributions
generated by the QAOA circuits are complicated and that 
for example a probability of $10^{-3}$ corresponds to only $10$ of our $10000$ shots. 
In the case of VQE we see that the probability distributions of \ibmqehningen\
match very well with those of the simulation for \lamaEx{0}{2} and \lamaEx{1}{3}.
Only a few $\probTilde_\varQubo$
are incorrectly non-zero, however their size is at least two magnitudes lower
than the $\probTilde_\varQubo$ that have correctly non-zero values. This matches 
perfectly with the fidelity plots in Figure~\ref{fig:lama-fidelity-boxplots-vqe-dd}.
For \lamaEx{2}{3} we see that the two best bitstrings agree with the simulation but
their probability is much lower. Furthermore, we see many incorrect probabilities
also in the regime of $10^{-1}$ to $10^{-2}$. This explains the bad fidelity we
observed in Figure~\ref{fig:lama-fidelity-boxplots-vqe-dd}. It is rather the number
of qubits and the associated chances for errors than the number of $\CX$ gates
(which are only 15)
that cause the bad fidelity. One can make this clear by computing that for 
\lamaEx{0}{2} and \lamaEx{1}{3} we only have $2^6 = 64$ and $2^8 = 256$, 
respectively,
possible bitstrings, whereas for \lamaEx{2}{3} there are $2^{16} = 65536$
possibilities.
%

% Relative error in cost, boxplots, reps=1, 2023-10-23 12:03
\begin{figure}
    \small
    \centering
    \begin{subfigure}{0.45\textwidth}
        \centering
        \includegraphics{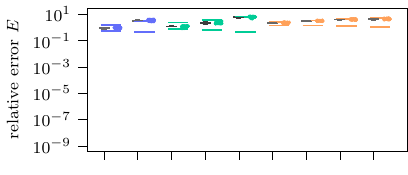}
        \caption{QAOA, w/o dynamical decoupling.}
    \end{subfigure}
    \; \;
    \begin{subfigure}{0.45\textwidth}
        \centering
        \includegraphics{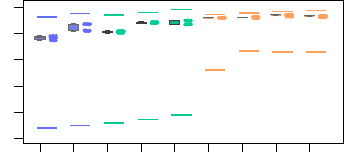}
        \caption{VQE, w/o dynamical decoupling.}
    \end{subfigure}
    \\[0.3cm]
    \centering
    \begin{subfigure}{0.45\textwidth}
        \centering
        \includegraphics{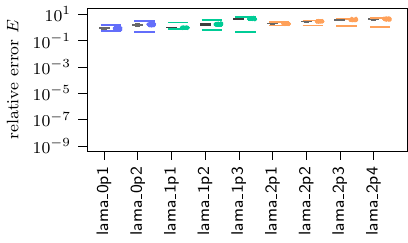}
        \caption{QAOA, w/ dynamical decoupling.}
    \end{subfigure}
    \; \;
    \begin{subfigure}{0.45\textwidth}
        \centering
        \includegraphics{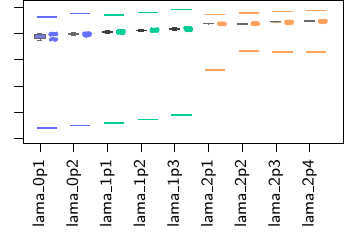}
        \caption{VQE, w/ dynamical decoupling.}
    \end{subfigure}
    \caption{
        Relative error $\relCostError$, as defined in \eqref{eq:relative-cost-error},
        obtained from executing the QAOA (left) and VQE 
        (right) circuits with $\repsQaoa = 1$ layer
        for all LamA use case examples on \ibmqehningen.
        The results stem from
        the same experiment as described in
        Figure~\ref{fig:lama-fidelity-boxplots}. The lower horizontal lines indicate
        the relative error when using an exact simulator instead of \ibmqehningen.
        The upper lines stem from the mean of the relative error
        of the probability distributions of 50 random statevectors.
    }
    \label{fig:lama-relative-cost-error}
\end{figure}
%

% Relative error in cost, zoom boxplots, reps=1, 2023-10-23 12:03
\begin{figure}
    \small
    \centering
    \begin{subfigure}{0.45\textwidth}
        \centering
        \includegraphics{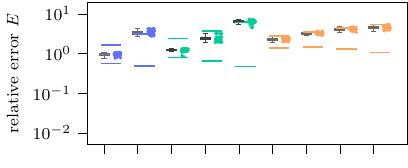}
        \caption{QAOA, w/o dynamical decoupling.}
    \end{subfigure}
    \; \;
    \begin{subfigure}{0.45\textwidth}
        \centering
        \includegraphics{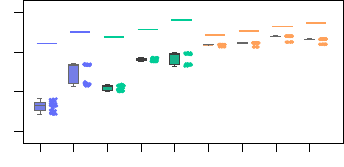}
        \caption{VQE, w/o dynamical decoupling.}
    \end{subfigure}
    \\[0.3cm]
    \centering
    \begin{subfigure}{0.45\textwidth}
        \centering
        \includegraphics{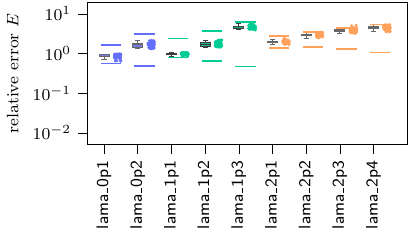}
        \caption{QAOA, w/ dynamical decoupling.}
    \end{subfigure}
    \; \;
    \begin{subfigure}{0.45\textwidth}
        \centering
        \includegraphics{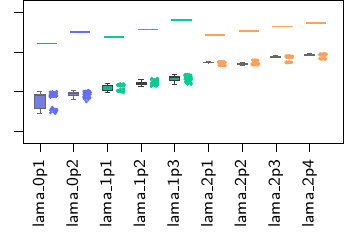}
        \caption{VQE, w/ dynamical decoupling.}
    \end{subfigure}
    \caption{Zoom-in for Figure~\ref{fig:lama-relative-cost-error}.}
    \label{fig:lama-relative-cost-error-zoom}
\end{figure}
%

% Legend for scatter plots
\begin{figure}
    \small
    \centering
    \includegraphics{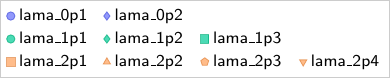}
    \caption{Legend for scatter plots.}
    \label{fig:legend-lama-examples}
\end{figure}

\paragraph{Relative error}
Now, let us look at the relative error $\relCostError$ 
that we defined in \eqref{eq:relative-cost-error}.
In Figure~\ref{fig:lama-relative-cost-error} we show 
$\relCostError(\{\probTilde_\varQubo\})$ for all LamA examples
for QAOA and for VQE, and in Figure~\ref{fig:lama-relative-cost-error-zoom} we
show the same data but with a zoomed-in $y$-axis. For comparison we also added
the error for the probability distribution obtained with an exact simulation
$\relCostError(\{\probHat_\varQubo\})$ (the lower lines in the plots). Moreover,
we added the error of a random probability distribution. For this we used the 
probabilities of 50 random statevector and calculated the mean of the relative error
of each. In the plots these are the upper lines. In 
Figure~\ref{fig:lama-relative-cost-error} we can see clearly that already the 
error for the exact simulation of QAOA are magnitudes larger than for VQE. However,
this small error for VQE cannot be obtained from \ibmqehningen. Nevertheless
it is much smaller than the error obtained from QAOA on \ibmqehningen.
Looking at the details
in Figure~\ref{fig:lama-relative-cost-error-zoom} we see that except for the small 
and weakly coupled examples \lamaEx{0}{1} and \lamaEx{1}{1} the error of QAOA
is rather in the realm of the random guess. VQE stays away from the random guess,
but we can see how the error increases with the size of the examples.

\begin{figure}
    \small
    \centering
    \begin{subfigure}[t]{0.48\textwidth}
        \centering
        \includegraphics{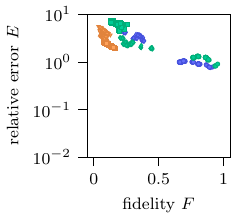}
        \includegraphics{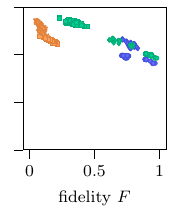}
        \caption{QAOA, w/o (left) and w/  (right) dynamical decoupling.}
    \end{subfigure}
    \;
    \begin{subfigure}[t]{0.46\textwidth}
        \centering
        \includegraphics{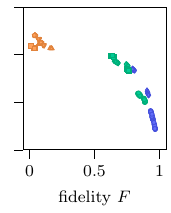}
        \includegraphics{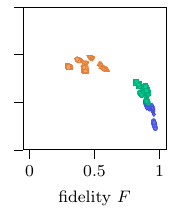}
        \caption{VQE, w/o (left) and w/  (right) dynamical decoupling.}
    \end{subfigure}
    \caption{
        Fidelity $\fidelity$ versus relative error $\relCostError$
        for the QAOA and VQE 
        circuits with $\repsQaoa = 1$ layer
        for all LamA use case examples executed on \ibmqehningen. 
        The results stem from
        the same experiment as described in
        Figure~\ref{fig:lama-fidelity-boxplots}.
        The legend for the marker symbols can be found 
        in Figure~\ref{fig:legend-lama-examples}.
    }
    \label{fig:lama-fidelity-vs-rel-cost-error}
\end{figure}

\paragraph{Relation between fidelity and relative error}
Now, let us look at the relation between the relative error
$\relCostError$ and the fidelity $\fidelity$. In 
Figure~\ref{fig:lama-fidelity-vs-rel-cost-error} we show this for all LamA examples.
For VQE we can clearly see that in the regime $\fidelity > 0.5$ a higher 
fidelity $\fidelity$ yields a better error $\relCostError$. For fidelities below
$0.5$ we see that the error stays (approximately) on a plateau. For QAOA the situation
is more complicated. In the case of dynamical decoupling we can again see that a
higher fidelity gives a smaller error. In the case without dynamical decoupling
we observe a plateau for the error in the regime $\fidelity > 0.5$ and a cluster
for $\fidelity < 0.5$.

%
% Scatter plots, fidelity vs. cost error
\paragraph{Metrics to predict quality}
In the upper paragraph we have seen that the quality of the quantum computer results
strongly differ for the different examples but also for different transpilations
of the same example, see for example Figure~\ref{fig:lama-fidelity-boxplots} and 
in particular the variances for each example. Thus, an interesting question
is wether we can predict the expected quality (for example in terms of the fidelity)
from the transpiled quantum circuits.
%and the calibration data of the backend it should be run on.
%

% Scatter plot: number CNOTs vs fidelity
\begin{figure}
    \small
    \centering
    %
    % \begin{subfigure}[t]{0.9\textwidth}
    %     \includegraphics{figures/lama_trained_circuits_ibmq_ehningen/scatterplots_cnots/legend.pdf}
    %     \caption*{}
    % \end{subfigure}
    % %
    % \\
    %
    \begin{subfigure}[t]{0.55\textwidth}
        % \centering
        \includegraphics{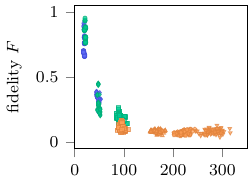}
        \includegraphics{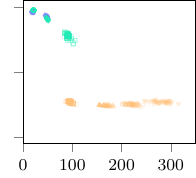}
        \caption{
            QAOA, w/o dynamical decoupling,
            \ibmqehningen\ (left), simulator with noise model (right).
        }
        \label{fig:lama-fidelity-vs-cnots-ehningen-qaoa}
    \end{subfigure}
    \; \; \;
    \begin{subfigure}[t]{0.4\textwidth}
        % \centering
        \includegraphics{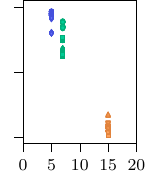}
        \includegraphics{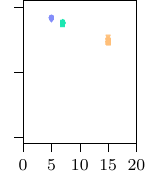}
        \caption{
            VQE, w/o dynamical decoupling,
            \ibmqehningen\ (left), simulator with noise model (right).
        }
        \label{fig:lama-fidelity-vs-cnots-ehningen-vqe}
    \end{subfigure}
    \\[0.3cm]
    \begin{subfigure}[t]{0.55\textwidth}
        % \centering
        \includegraphics{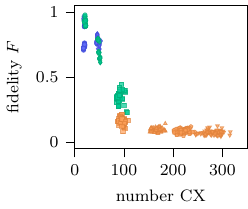}
        \includegraphics{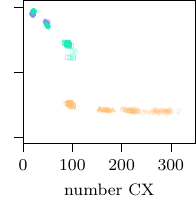}
        \caption{
            QAOA, w/ dynamical decoupling,
            \ibmqehningen\ (left), simulator with noise model (right).
        }
        \label{fig:lama-fidelity-vs-cnots-ehningen-qaoa-dd}
    \end{subfigure}
    \; \; \;
    \begin{subfigure}[t]{0.4\textwidth}
        % \centering
        \includegraphics{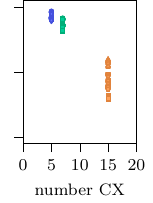}
        \includegraphics{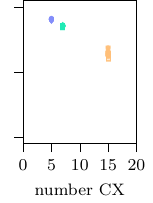}
        \caption{
            VQE, w/ dynamical decoupling,
            \ibmqehningen\ (left), simulator with noise model (right).
        }
        \label{fig:lama-fidelity-vs-cnots-ehningen-vqe-dd}
    \end{subfigure}
    \caption{
        Number of $\CX$ gates versus fidelity $\fidelity$
        for the QAOA and VQE 
        circuits with $\repsQaoa = 1$ layer
        for all LamA use case examples executed on \ibmqehningen\ and 
        on a simulator with the noise model of \ibmqehningen.
        The results stem from
        the same experiment as described in
        Figure~\ref{fig:lama-fidelity-boxplots}. The noise model is built with
        the calibration data from the time when the circuits ran on
        \ibmqehningen.
        The legend for the marker symbols can be found 
        in Figure~\ref{fig:legend-lama-examples}.
    }
    \label{fig:lama-fidelity-vs-cnots-ehningen}
\end{figure}

As we have seen above, the $\CX$ gate is by far the most erroneous hardware gate.
Consequently, a simple metric to assess a quantum circuit is to count the number
of $\CX$ gates of the transpiled version of the circuit. In 
Figure~\ref{fig:lama-fidelity-vs-cnots-ehningen} we show the number of $\CX$ gates
versus the fidelity $\fidelity$ for all our LamA examples for both QAOA and VQE
on \ibmqehningen. Additionally, we provide the fidelities of a noisy simulator 
that is built with a noise model using the calibration data of \ibmqehningen\
from the time when the circuits were executed on the real machine.
We can make the following observations:
First of all, we 
see for QAOA without dynamical decoupling 
(Figure~\ref{fig:lama-fidelity-vs-cnots-ehningen-qaoa})
a drastic drop of the fidelity with increasing
number of $\CX$ gates. In fact, only the examples with around 20 $\CX$ gates 
(\lamaEx{0}{1} and \lamaEx{1}{1}), show an acceptable fidelity, 
the examples with around
50 $\CX$ gates (\lamaEx{0}{2} and \lamaEx{1}{2}) are already much worse, closely
followed by the examples with around 100 $\CX$ gates (\lamaEx{1}{3} and \lamaEx{2}{1}),
and finally all examples with more than 150 $\CX$ (\lamaEx{2}{2} -- \lamaEx{2}{4})
give a fidelity around $0$. Also note that the simulation with the noise model
of the backend does hardly agree with the real results. In particular, note the
strong differences for examples series \lamaSeries{0} and \lamaSeries{1}. 
For QAOA
with dynamical decoupling (Figure~\ref{fig:lama-fidelity-vs-cnots-ehningen-qaoa-dd})
we see an improvement in particular for \lamaEx{0}{2} and \lamaEx{1}{2} 
(both around 50 $\CX$ gates). Also note that the fidelities of \lamaEx{1}{3} 
and \lamaEx{2}{1}
(both around 100 $\CX$ gates) now differ stronger and come closer to the predicted
gap of the simulation. Here, we see the effect of the number of qubits of the different
example series (recall \lamaSeries{1} uses 8 and \lamaSeries{2} uses 16 qubits).
The effect of the qubit number is more drastically seen in the VQE experiments
(Figures~\ref{fig:lama-fidelity-vs-cnots-ehningen-vqe} and 
\ref{fig:lama-fidelity-vs-cnots-ehningen-vqe-dd}). Although the number of $\CX$ gates
is nearly the same the fidelities for example series \lamaSeries{0} is better than
for example series \lamaSeries{1}, and both are much better than 
for example series \lamaSeries{2}. As for QAOA we also observe improvements when
using dynamical decoupling and that simulations and real experiments do not match well.

% Scatter plot: circuit score vs fidelity
\begin{figure}
    \small
    \centering
    %
    % \begin{subfigure}[t]{0.9\textwidth}
    %     \includegraphics{figures/lama_trained_circuits_ibmq_ehningen/scatterplots_cnots/legend.pdf}
    %     \caption*{}
    % \end{subfigure}
    % %
    % \\
    %
    \begin{subfigure}[t]{0.55\textwidth}
        \includegraphics{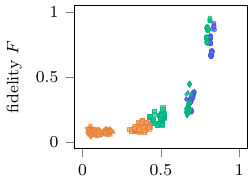}
        \includegraphics{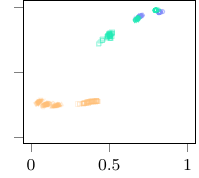}
        \caption{
            QAOA, w/o dynamical decoupling,
            \ibmqehningen\ (left), simulator with noise model (right).
        }
        \label{fig:lama-fidelity-vs-circuit-score-ehningen-qaoa}
    \end{subfigure}
    \; \; \;
    \begin{subfigure}[t]{0.4\textwidth}
        \includegraphics{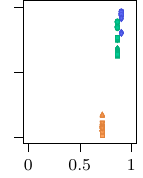}
        \includegraphics{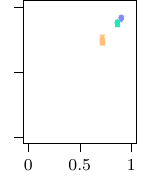}
        \caption{
            VQE, w/o dynamical decoupling,
            \ibmqehningen\ (left), simulator with noise model (right).
        }
        \label{fig:lama-fidelity-vs-circuit-score-ehningen-vqe}
    \end{subfigure}
    \\[0.3cm]
    \begin{subfigure}[t]{0.55\textwidth}
        \includegraphics{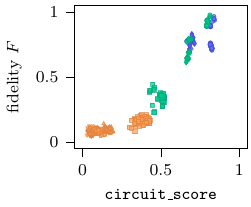}
        \includegraphics{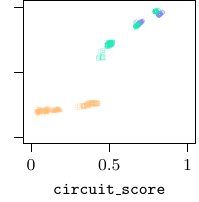}
        \caption{
            QAOA, w/ dynamical decoupling,
            \ibmqehningen\ (left), simulator with noise model (right).
        }
        \label{fig:lama-fidelity-vs-circuit-score-ehningen-qaoa-dd}
    \end{subfigure}
    \; \; \;
    \begin{subfigure}[t]{0.4\textwidth}
        \includegraphics{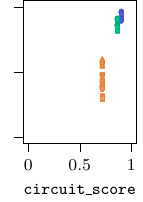}
        \includegraphics{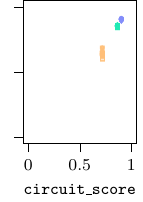}
        \caption{
            VQE, w/ dynamical decoupling,
            \ibmqehningen\ (left), simulator with noise model (right).
        }
        \label{fig:lama-fidelity-vs-circuit-score-ehningen-vqe-dd}
    \end{subfigure}
    \caption{
        \circuitScore, as defined in Algorithm~\ref{alg:circuit-score},
        versus fidelity $\fidelity$
        for the QAOA and VQE 
        circuits with $\repsQaoa = 1$ layer
        for all LamA use case examples executed on \ibmqehningen\ and 
        on a simulator with the noise model of \ibmqehningen.
        The results stem from
        the same experiment as described in
        Figure~\ref{fig:lama-fidelity-boxplots}. The noise model is built with
        the calibration data from the time when the circuits ran on
        \ibmqehningen.
        The legend for the marker symbols can be found 
        in Figure~\ref{fig:legend-lama-examples}.
    }
    \label{fig:lama-fidelity-vs-circuit-score-ehningen}
\end{figure}

\begin{figure}
    \small
    \centering
    \begin{subfigure}{0.45\textwidth}
        \centering
        \includegraphics{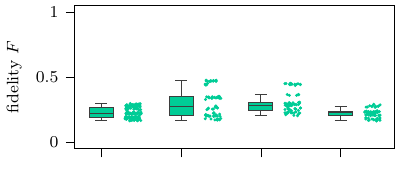}
        \caption{QAOA, w/o dynamical decoupling.}
        \label{fig:lama-fidelity-boxplots-ex1p2-different-dates-qaoa}
    \end{subfigure}
    \; \;
    \begin{subfigure}{0.45\textwidth}
        \centering
        \includegraphics{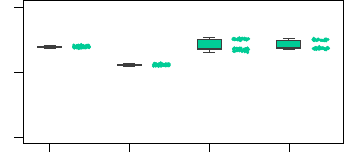}
        \caption{VQE, w/o dynamical decoupling.}
        \label{fig:lama-fidelity-boxplots-ex1p2-different-dates-vqe}
    \end{subfigure}
    \\[0.3cm]
    \begin{subfigure}{0.45\textwidth}
        \centering
        \includegraphics{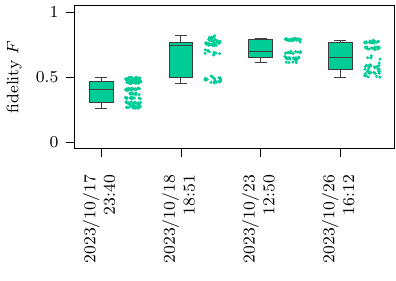}
        \caption{QAOA, w/ dynamical decoupling.}
        \label{fig:lama-fidelity-boxplots-ex1p2-different-dates-qaoa-dd}
    \end{subfigure}
    \; \;
    \begin{subfigure}{0.45\textwidth}
        \centering
        \includegraphics{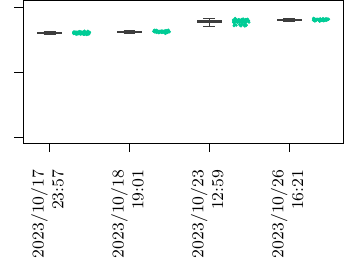}
        \caption{VQE, w/ dynamical decoupling.}
        \label{fig:lama-fidelity-boxplots-ex1p2-different-dates-vqe-dd}
    \end{subfigure}
    \caption{
        Fidelity $\fidelity$ for the QAOA and VQE 
        circuits with $\repsQaoa = 1$ layer
        for \lamaEx{1}{2} executed on \ibmqehningen\
        on different dates.
        The experiment setup as is the same as described in
        Figure~\ref{fig:lama-fidelity-boxplots}.
    }
    \label{fig:lama-fidelity-boxplots-ex1p2-different-dates}
\end{figure}

As we have seen, the number of $\CX$ gates is not able to accurately predict the fidelity.
In particular, we observed that circuits with the same number of $\CX$ gates can result
in strongly different fidelities. In these cases the assumption is obvious that
the circuits use different qubits of \ibmqehningen\ as all qubits and the gates, that
are executed on them, all have with different error rates. A metric that incorporates
the errors of the quantum gates is for example presented in 
\cite[Algorithm 1]{Nation.2023}. We use it in the form of the \circuitScore\
as given in Algorithm~\ref{alg:circuit-score}, where we include besides the quantum
gates also the measurement in the instructions set $\{G\}$. Note that a 
\circuitScore\ $\approx 1$ indicates a circuit with high quality instructions,
whereas \circuitScore\ $\approx 0$ hints at a circuit with low fidelity instructions. 
\begin{algorithm}
    \begin{algorithmic}[1]
        \State Input: transpiled circuit with instructions $\{G\}$, error map $\mathcal{E}$
        \State circuit\_score $\gets$ 1
        \For{$g \in \{G\}$}
            \State circuit\_score $\gets$ circuit\_score $\ast$ $(1 - \mathcal{E}(g))$
        \EndFor
        \State \Return circuit\_score
    \end{algorithmic}
    \caption{Algorithm for \circuitScore}
    \label{alg:circuit-score}
\end{algorithm}
In Figure~\ref{fig:lama-fidelity-vs-circuit-score-ehningen} we report the same
experiments as for Figure~\ref{fig:lama-fidelity-vs-cnots-ehningen} but with
the \circuitScore\ instead of the number of $\CX$ gates. 
In general, we make the same observations we already did for 
the number of $\CX$ gates. In particular, we see that also the \circuitScore\ 
is in many cases not able to accurately predict the fidelity, see for example
the individual transpilations of
\lamaEx{0}{1} and \lamaEx{1}{1} in 
Figure~\ref{fig:lama-fidelity-vs-circuit-score-ehningen-qaoa},
which have approximately the same \circuitScore\ but their fidelity differs strongly.
The same holds for the different transpilations of \lamaEx{0}{2} and \lamaEx{1}{2}.
Only the fidelities of \lamaEx{1}{3} and \lamaEx{2}{1} can be distinguished
a bit better using the \circuitScore\ than compared to the number of $\CX$ gates.
Also note that, by construction, the \circuitScore\ cannot reveal the effect of 
dynamical decoupling as can be seen by comparing
Figure~\ref{fig:lama-fidelity-vs-circuit-score-ehningen-qaoa}
with Figure~\ref{fig:lama-fidelity-vs-circuit-score-ehningen-qaoa-dd} or Figure~\ref{fig:lama-fidelity-vs-circuit-score-ehningen-vqe}
with Figure~\ref{fig:lama-fidelity-vs-circuit-score-ehningen-vqe-dd}. Moreover,
the \circuitScore\ does not incorporate the number of qubits so it cannot by used
to compare examples with different qubit numbers. This is prominently seen in the
VQE results in Figures~\ref{fig:lama-fidelity-vs-circuit-score-ehningen-vqe}
and \ref{fig:lama-fidelity-vs-circuit-score-ehningen-vqe-dd}, where the circuits
all have a similar \circuitScore\ but the fidelities differ strongly.
\paragraph{Reproducibilty over time}
The error rates of today's quantum computers are not constant but change over time.
Thus, it is interesting how the results of our experiments look at different days.
In Figure~\ref{fig:lama-fidelity-boxplots-ex1p2-different-dates} we report on the
fidelity for \lamaEx{1}{2} on four different dates. For QAOA without dynamical
decoupling (Figure~\ref{fig:lama-fidelity-boxplots-ex1p2-different-dates-qaoa})
we see that the median fidelity is nearly constant over the four days but
the variance of the 75 different transpilations differs. For QAOA with dynamical
decoupling (Figure~\ref{fig:lama-fidelity-boxplots-ex1p2-different-dates-qaoa-dd})
the results look different. In particular, the median fidelity
as well as the the distribution of the fidelities of the different transpilations 
differ strongly over the four days. For example, on 2023/10/17 and 2023/10/26 we see a more or less uniform distribution around the median,
whereas on 2023/10/18 we see two clusters -- one around a high fidelity and one around
a low one. For VQE (Figures~\ref{fig:lama-fidelity-boxplots-ex1p2-different-dates-vqe}
and \ref{fig:lama-fidelity-boxplots-ex1p2-different-dates-vqe-dd}) there are far 
less deviations in the four days, in particular in the case with dynamical decoupling.

%
% Fidelity boxplots, different dates, example_1p2
%

% Compare patches of GPU, example_0p2
\begin{figure}
    \small
    \centering
    \begin{subfigure}{0.95\textwidth}
        \centering
        \includegraphics{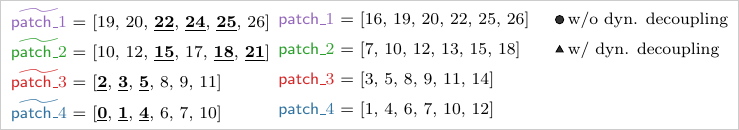}
        \caption*{}
    \end{subfigure}
    \begin{subfigure}{0.45\textwidth}
        \centering
        \includegraphics{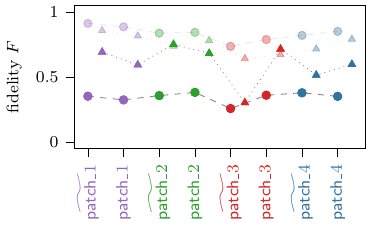}
        \caption{2023/10/27 20:38.}
        \label{fig:lama-fidelity-vs-qpu-patches-ex0p2-27-10-2023}
    \end{subfigure}
    \;
    \begin{subfigure}{0.45\textwidth}
        \centering
        \includegraphics{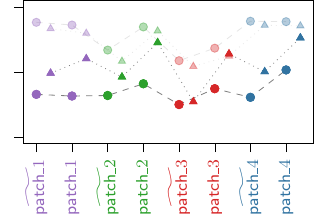}
        \caption{2023/11/22 18:22.}
        \label{fig:lama-fidelity-vs-qpu-patches-ex0p2-22-11-2023}
    \end{subfigure}
    \caption{
        Fidelity $\fidelity$ for the QAOA
        circuit with $\repsQaoa = 1$ layer
        for \lamaEx{0}{2} executed on different qubit patches of \ibmqehningen\
        on two different dates. In all $\widetilde{\textsf{patch}\_\textsf{i}}$
        there is
        a qubit triplet (marked in bold and underlined in the legend) that suffers
        from cross talk.
        In $\textsf{patch}\_\textsf{i}$ one qubit is exchanged such
        that these patches do not contain a cross talk triplet.
        As shown in the legend the marker type indicates wether 
        dynamical decoupling was used in the quantum circuit. The transparent
        markers stem from a simulation of the quantum circuits using a 
        simulator with the noise model of \ibmqehningen. The noise model 
        is built with the calibration data from the time when the circuits ran
        on \ibmqehningen.
    }
    \label{fig:lama-fidelity-vs-qpu-patches-ex0p2}
\end{figure}
%
% Compare patches of GPU, example_1p3
\begin{figure}
    \small
    \centering
    \begin{subfigure}{0.95\textwidth}
        \centering
        \includegraphics{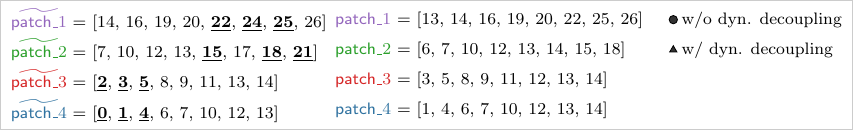}
        \caption*{}
    \end{subfigure}
    \begin{subfigure}{0.45\textwidth}
        \centering
        \includegraphics{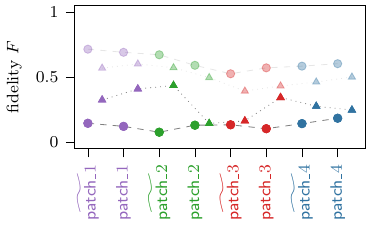}
        \caption{2023/10/27 20:18}
        \label{fig:lama-fidelity-vs-qpu-patches-ex1p3-27-10-2023}
    \end{subfigure}
    \;
    \begin{subfigure}{0.45\textwidth}
        \centering
        \includegraphics{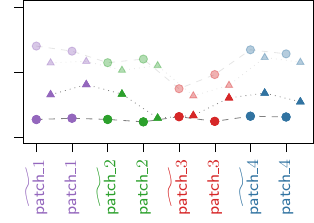}
        \caption{2023/11/22 18:40}
        \label{fig:lama-fidelity-vs-qpu-patches-ex1p3-22-11-2023}
    \end{subfigure}
    \caption{
        Fidelity $\fidelity$ for the QAOA
        circuit with $\repsQaoa = 1$ layer
        for \lamaEx{1}{3} executed on different qubit patches of \ibmqehningen\
        on two different dates. The setup is the same as for 
        Figure~\ref{fig:lama-fidelity-vs-qpu-patches-ex0p2}.
    }
    \label{fig:lama-fidelity-vs-qpu-patches-ex1p3}
\end{figure}

\paragraph{Choice of best qubits and cross talk}
We have seen in the upper experiments that different transpilations of the same
example (i.e. of the same logical QAOA or VQE circuit) can lead to drastically
different fidelities. With the number of $\CX$ gates and the \circuitScore\
we considered two metrics to explain this behavior, but both were not able
to distinguish good from bad transpilations.
Clearly, the question arises if the different fidelities stem from
error sources that are not covered by these two metrics. One candidate for this
error source is cross talk. This phenomenon has been analyzed in \cite{Ketterer.2023}
for \ibmqehningen\ and the qubit triplets that suffer from it have been identified.
Using this information we want to analyze if cross talk is responsible for
the lower fidelities we see in our experiments. For this we chose four cross
talk qubit triplets of \ibmqehningen,
namely [0, 1, 4], [2, 3, 5], [15, 18, 21], and [22, 24, 25]. Then, we selected four
patches of the quantum processor that contain these qubit triplets. We name these patches
$\widetilde{\textsf{patch}\_1}$, \dots, $\widetilde{\textsf{patch}\_4}$. Afterwards, we exchanged
a single qubit in each of these patches so that the cross talk qubit triplet is no
longer part of the patch. We call these patches $\textsf{patch}\_1$, \dots, $\textsf{patch}\_4$.
We did this procedure for patches of sizes 6 and 8 qubits. 
Having these patches of the quantum processor
we transpiled the logical QAOA circuit for \lamaEx{0}{2} and for \lamaEx{1}{3} 10 times
and gave the transpiler our patches as initial layout. We set the \optimizationLevel\
to 3 so that we get optimized circuits. From the resulting transpilations we
removed those where the optimization of the transpiler changed the qubits. We ran
the remaining circuits on \ibmqehningen\ on two different dates. In
Figures~\ref{fig:lama-fidelity-vs-qpu-patches-ex0p2} and 
\ref{fig:lama-fidelity-vs-qpu-patches-ex1p3} we report the mean of the fidelities 
for \lamaEx{0}{2} and for \lamaEx{1}{3}, respectively. 
For QAOA without dynamical decoupling (indicated by the 
circle markers) we see no (Figure~\ref{fig:lama-fidelity-vs-qpu-patches-ex1p3}) or
only a slight (Figure~\ref{fig:lama-fidelity-vs-qpu-patches-ex0p2})
difference between the cross talk and non cross talk patches. In particular, in 
Figure~\ref{fig:lama-fidelity-vs-qpu-patches-ex0p2-22-11-2023} we see a better
fidelity in the non cross talk patches. Considering the circuit with dynamical 
decoupling (triangle markers) we observe large differences for the different patches.
More precisely,
we see for both \lamaEx{0}{2} and \lamaEx{1}{3} a better fidelity for
the non cross talk patch $\textsf{patch}\_3$
compared to its counterpart $\widetilde{\textsf{patch}\_3}$ with the cross talk triplet.
For \lamaEx{0}{2} (Figure~\ref{fig:lama-fidelity-vs-qpu-patches-ex0p2})
the difference is drastic. However, for $\textsf{patch}\_2$ and 
$\widetilde{\textsf{patch}\_2}$ the situation is the other way round (at least for
Figures~\ref{fig:lama-fidelity-vs-qpu-patches-ex0p2-27-10-2023}, 
\ref{fig:lama-fidelity-vs-qpu-patches-ex1p3-27-10-2023}, and 
\ref{fig:lama-fidelity-vs-qpu-patches-ex1p3-22-11-2023}). Here, the cross talk patches
yield a better fidelity. For the other patches there are either no big differences, 
the cross talk patch is a bit better ($\textsf{patch}\_1$ and 
$\widetilde{\textsf{patch}\_1}$ in 
Figure~\ref{fig:lama-fidelity-vs-qpu-patches-ex0p2-27-10-2023}),
or the non cross talk patch is a bit better
($\textsf{patch}\_4$ and $\widetilde{\textsf{patch}\_4}$ in
\ref{fig:lama-fidelity-vs-qpu-patches-ex0p2-22-11-2023}). Thus, we conclude that
cross talk might be one of the error sources leading to the variance in fidelities,
that we see in our experiments, but is not the only missing error source in a metric
to explain the fidelity.

\subsection{Experiments on \ibmcairo, \ibmsherbrooke, and \ibmtorino}
\label{sec:lama-other-ibmq}
In this section we present the same experiments that we ran on \ibmqehningen\
in Section~\ref{sec:lama-ibmq-ehningen} for the backends \ibmcairo, 
\ibmsherbrooke, and \ibmtorino. As it can be seen in
Table~\ref{tab:ibm-quantum-systems} 
the first has the same processor type as \ibmqehningen,
the second one is a 127 qubit processor from 2022, and the last is the newest 
processor available from IBM and has 133 qubits. This choice lets us, on the one
hand, compare two similar processors and, on the other hand, enables us to show 
and assess the technological progress over the last years.
%

% Fidelity boxplots, reps=1, ibm_cairo
\begin{figure}
    \small
    \centering
    \begin{subfigure}{0.45\textwidth}
        \centering
        \includegraphics{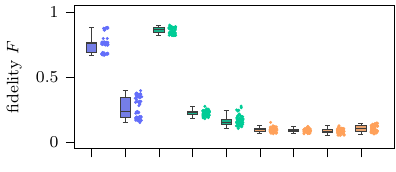}
        \caption{QAOA, w/o dynamical decoupling.}
    \end{subfigure}
    \; \;
    \begin{subfigure}{0.45\textwidth}
        \centering
        \includegraphics{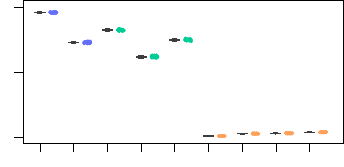}
        \caption{VQE, w/o dynamical decoupling.}
    \end{subfigure}
    \\[0.3cm]
    \centering
    \begin{subfigure}{0.45\textwidth}
        \centering
        \includegraphics{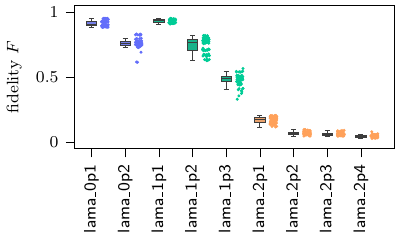}
        \caption{QAOA, w/ dynamical decoupling.}
    \end{subfigure}
    \; \;
    \begin{subfigure}{0.45\textwidth}
        \centering
        \includegraphics{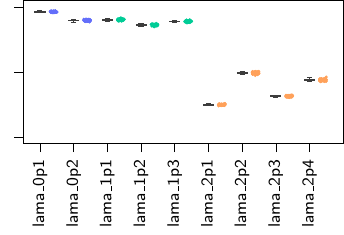}
        \caption{VQE, w/ dynamical decoupling.}
    \end{subfigure}
    \caption{
        Fidelity $\fidelity$ obtained from executing the QAOA (left) and VQE 
        (right) circuits with $\repsQaoa = 1$ layer
        for all LamA use case examples on \ibmcairo.
        The experiment setup is the same as described in
        Figure~\ref{fig:lama-fidelity-boxplots}. The
        fist job of the session ran on 2023/12/13 11:44.
    }
    \label{fig:lama-fidelity-boxplots-ibm-cairo}
\end{figure}
%

% Fidelity boxplots, reps=1, ibm_sherbrooke
\begin{figure}
    \small
    \centering
    \begin{subfigure}{0.45\textwidth}
        \centering
        \includegraphics{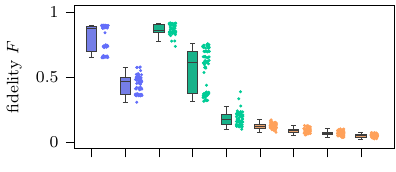}
        \caption{QAOA, w/o dynamical decoupling.}
    \end{subfigure}
    \; \;
    \begin{subfigure}{0.45\textwidth}
        \centering
        \includegraphics{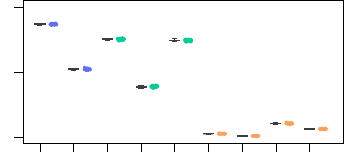}
        \caption{VQE, w/o dynamical decoupling.}
    \end{subfigure}
    \\[0.3cm]
    \centering
    \begin{subfigure}{0.45\textwidth}
        \centering
        \includegraphics{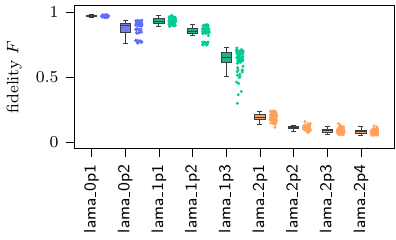}
        \caption{QAOA, w/ dynamical decoupling.}
    \end{subfigure}
    \; \;
    \begin{subfigure}{0.45\textwidth}
        \centering
        \includegraphics{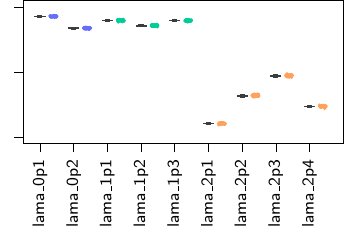}
        \caption{VQE, w/ dynamical decoupling.}
    \end{subfigure}
    \caption{
        Fidelity $\fidelity$ obtained from executing the QAOA (left) and VQE 
        (right) circuits with $\repsQaoa = 1$ layer
        for all LamA use case examples on \ibmsherbrooke.
        The experiment setup is the same as described in
        Figure~\ref{fig:lama-fidelity-boxplots}. The
        fist job of the session ran on 2023/12/15 10:05.
    }
    \label{fig:lama-fidelity-boxplots-ibm-sherbrooke}
\end{figure}
%

% Fidelity boxplots, reps=1, ibm_torino
\begin{figure}
    \small
    \centering
    \begin{subfigure}{0.45\textwidth}
        \centering
        \includegraphics{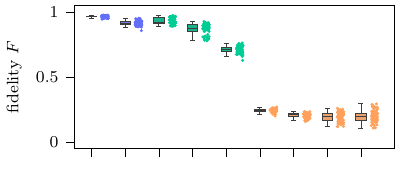}
        \caption{QAOA, w/o dynamical decoupling.}
    \end{subfigure}
    \; \;
    \begin{subfigure}{0.45\textwidth}
        \centering
        \includegraphics{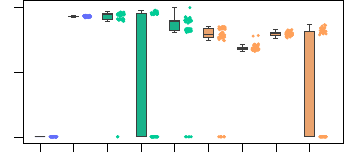}
        \caption{VQE, w/o dynamical decoupling.}
    \end{subfigure}
    \\[0.3cm]
    \centering
    \begin{subfigure}{0.45\textwidth}
        \centering
        \includegraphics{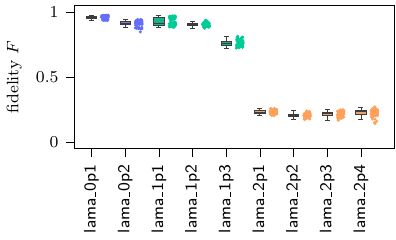}
        \caption{QAOA, w/ dynamical decoupling.}
    \end{subfigure}
    \; \;
    \begin{subfigure}{0.45\textwidth}
        \centering
        \includegraphics{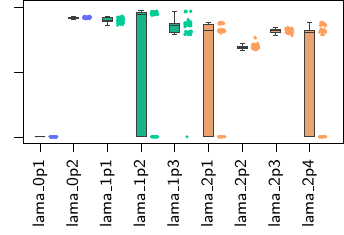}
        \caption{VQE, w/ dynamical decoupling.}
    \end{subfigure}
    \caption{
        Fidelity $\fidelity$ obtained from executing the QAOA (left) and VQE 
        (right) circuits with $\repsQaoa = 1$ layer
        for all LamA use case examples on \ibmtorino.
        The experiment setup is the same as described in
        Figure~\ref{fig:lama-fidelity-boxplots}. The
        fist job of the session ran on 2023/12/11 13:03.
    }
    \label{fig:lama-fidelity-boxplots-ibm-torino}
\end{figure}

\paragraph{Metrics to predict quality}
Let us start with showing the fidelities for all LamA examples. In 
Figure~\ref{fig:lama-fidelity-boxplots-ibm-cairo} we show them for \ibmcairo,
in Figure~\ref{fig:lama-fidelity-boxplots-ibm-sherbrooke} for \ibmsherbrooke,
and in Figure~\ref{fig:lama-fidelity-boxplots-ibm-torino} for \ibmtorino. As expected,
we do not see major differences when comparing the results of \ibmcairo\ with those
of \ibmqehningen, recall Figure~\ref{fig:lama-fidelity-boxplots}. The most notable
difference is that the VQE fidelities have a much smaller variance for \ibmcairo.
This can also be observed by comparing the VQE fidelities of \ibmsherbrooke\ with
\ibmqehningen. Surprisingly, we can also observe that without dynamical decoupling
the fidelities for \lamaEx{0}{2} and \lamaEx{1}{2} are quite a bit lower for
\ibmsherbrooke\ than for \ibmqehningen. Looking at QAOA we see larger variances
on \ibmsherbrooke\ in the case without dynamical decoupling, in particular for
\lamaEx{1}{2}. When using dynamical decoupling we see an improvement in comparison
to \ibmqehningen, particularly for the smaller example series \lamaSeries{0} and
\lamaSeries{1}. Last, for QAOA and \ibmtorino\ we see a significant improvement
in the fidelities. Indeed, all fidelities in example series \lamaSeries{0} and
\lamaSeries{1} are drastically improved and in many examples near to $1$. But also
the fidelities of the largest examples in example series \lamaSeries{2} are higher,
which could not be achieved by the other backends. For the VQE circuits we see many
outliers from unclear error sources. However, for the other circuits we again
see drastically higher fidelities in comparison to \ibmqehningen\ and also
compared to the other backends considered in this paper.
%

% Bar plots, probability, ibm torino
\begin{figure}
    \small
    \centering
    \begin{subfigure}[t]{0.33\textwidth}
        \includegraphics{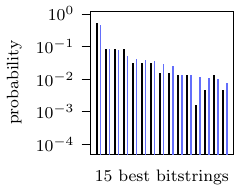}
        \includegraphics{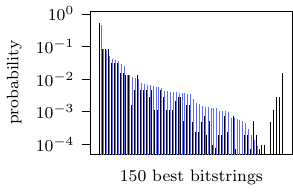}
        \caption{QAOA, \lamaEx{0}{2}.}
    \end{subfigure}
    \begin{subfigure}[t]{0.3\textwidth}
        \includegraphics{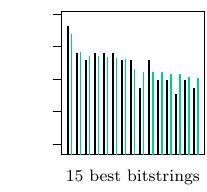}
        \includegraphics{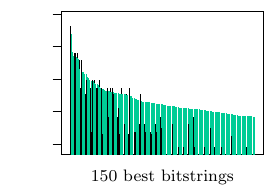}
        \caption{QAOA, \lamaEx{1}{3}.}
    \end{subfigure}
    \begin{subfigure}[t]{0.3\textwidth}
        \includegraphics{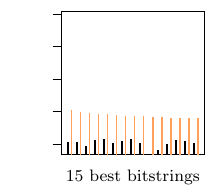}
        \includegraphics{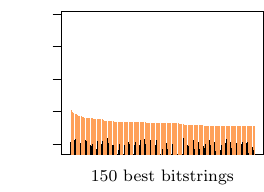}
        \caption{QAOA, \lamaEx{2}{3}}
    \end{subfigure}
    \\[0.3cm]
    \begin{subfigure}[t]{0.33\textwidth}
        \includegraphics{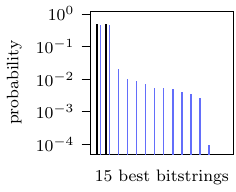}
        \includegraphics{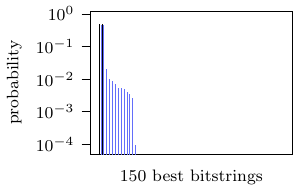}
        \caption{VQE, \lamaEx{0}{2}.}
    \end{subfigure}
    \begin{subfigure}[t]{0.3\textwidth}
        \includegraphics{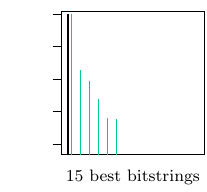}
        \includegraphics{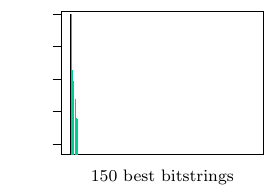}
        \caption{VQE, \lamaEx{1}{3}.}
    \end{subfigure}
    \begin{subfigure}[t]{0.3\textwidth}
        \includegraphics{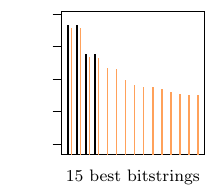}
        \includegraphics{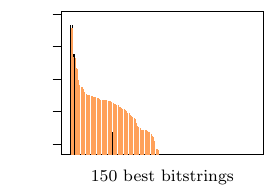}
        \caption{VQE, \lamaEx{2}{3}.}
    \end{subfigure}
    \caption{
        Probabilities resulting from executing the QAOA
        (top) and VQE (bottom) circuits with $\repsQaoa = 1$ layer for
        different LamA use case examples on 
        \ibmtorino\ (colored bars) and on an exact simulator (black bars).
        The experiment setup is the same as described in 
        Figure~\ref{fig:lama-probability-distributions}.
    }
    \label{fig:lama-probability-distributions-ibm-torino}
\end{figure}

As we have seen the most differences in comparison to \ibmqehningen\ are obtained
from \ibmtorino. Thus, we restrict ourselves to show here the probability 
distributions and the relative error for this backend. In 
Figure~\ref{fig:lama-probability-distributions-ibm-torino} we show the probability
distributions $\{\probTilde_\varQubo\}$ from \ibmtorino\ with the same setting as
we had in Figure~\ref{fig:lama-probability-distributions} for \ibmqehningen. For
QAOA and \lamaEx{0}{2} as well as \lamaEx{1}{3} we see a higher probability for
the best bitstring and a high agreement of the best 15 bitstrings for \ibmtorino.
Moreover, we see a faster decay of the probability of the bitstrings that should have
zero probability according to the exact simulation. Only for \lamaEx{2}{3} and QAOA
we do only observe a very small improvement. On the contrary, we observe for all
VQE examples probability distributions from \ibmtorino\ that match with the exact
simulation fairly well. The most significant improvement can be seen for 
\lamaEx{2}{3} which matches well with the improved fidelity we have seen 
in Figure~\ref{fig:lama-fidelity-boxplots-ibm-torino} above.

\begin{figure}
    \small
    \centering
    \begin{subfigure}{0.45\textwidth}
        \centering
        \includegraphics{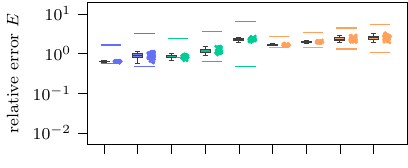}
        \caption{QAOA, w/o dynamical decoupling.}
    \end{subfigure}
    \; \;
    \begin{subfigure}{0.45\textwidth}
        \centering
        \includegraphics{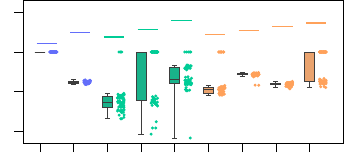}
        \caption{VQE, w/o dynamical decoupling.}
    \end{subfigure}
    \\[0.3cm]
    \centering
    \begin{subfigure}{0.45\textwidth}
        \centering
        \includegraphics{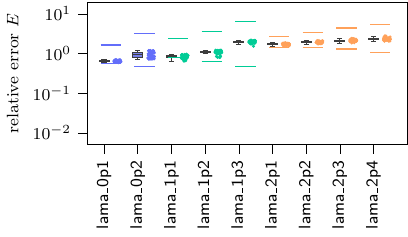}
        \caption{QAOA, w/ dynamical decoupling.}
    \end{subfigure}
    \; \;
    \begin{subfigure}{0.45\textwidth}
        \centering
        \includegraphics{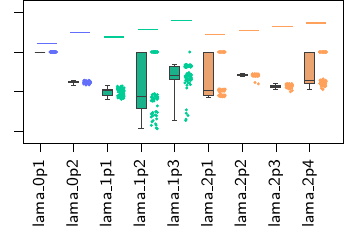}
        \caption{VQE, w/ dynamical decoupling.}
    \end{subfigure}
    \caption{
        Relative error $\relCostError$
        obtained from executing the QAOA (left) and VQE 
        (right) circuits with $\repsQaoa = 1$ layer
        for all LamA use case examples on \ibmtorino.
        The experiment setup is the same as described in 
        Figure~\ref{fig:lama-probability-distributions}.
        The lower horizontal lines indicate
        the relative error when using an exact simulator instead of \ibmtorino.
        The upper lines stem from the mean of the relative error
        of the probability distributions of 50 random statevectors.
    }
    \label{fig:lama-relative-cost-error-ibm-torino}
\end{figure}

Last in this paragraph, we plotted in 
Figure~\ref{fig:lama-relative-cost-error-ibm-torino} the relative error 
$\relCostError$ for \ibmtorino. Comparing with the error of \ibmqehningen,
see Figure~\ref{fig:lama-relative-cost-error-zoom}, we see that the results are
much closer to the simulation for QAOA. This is also true for VQE except of the
outliers that we already saw in Figure~\ref{fig:lama-fidelity-boxplots-ibm-torino}.

% Relative error in cost boxplots, zoom, reps=1, ibm_torino

\paragraph{Metrics to predict quality}
We conclude the LamA use case section by reporting on metrics to predict the 
fidelity. For \ibmqehningen\ we considered the number of $\CX$ gates and the 
\circuitScore. These two metrics can also be used for \ibmcairo. However,
\ibmsherbrooke\ and \ibmtorino\ do not have the $\CX$ gate as two qubit hardware
gate but the $\ECR$ and the $\CZ$ gate, respectively. For both backends
the two qubit hardware gate is the most erroneous so that they play the analog
role (and can be used as the analog metric) as the $\CX$ gate for \ibmqehningen.
By definition the \circuitScore\ is computed from the transpiled circuits and
thus the different hardware gates enter automatically there.

\begin{figure}
    \small
    \centering
    %
    % \begin{subfigure}[t]{0.9\textwidth}
    %     \includegraphics{figures/lama_trained_circuits_ibmq_ehningen/scatterplots_cnots/legend.pdf}
    %     \caption*{}
    % \end{subfigure}
    % %
    % \\
    %
    \begin{subfigure}[t]{0.55\textwidth}
        \includegraphics{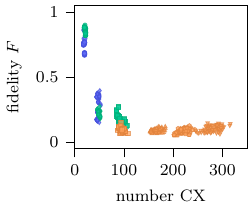}
        \;
        \includegraphics{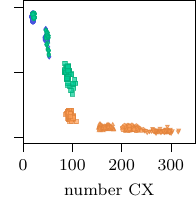}
        \caption{QAOA, w/o (left) and w/ (right) dynamical decoupling.}
    \end{subfigure}
    \;
    \begin{subfigure}[t]{0.4\textwidth}
        \includegraphics{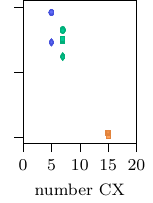}
        \;
        \includegraphics{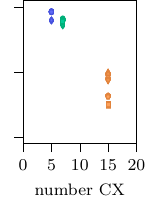}
        \caption{VQE, w/o (left) and w/ (right) dynamical decoupling.}
    \end{subfigure}
    \\[0.2cm]
    \begin{subfigure}[t]{0.55\textwidth}
        \includegraphics{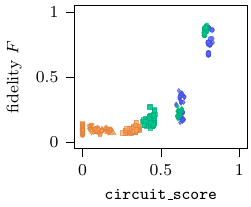}
        \;
        \includegraphics{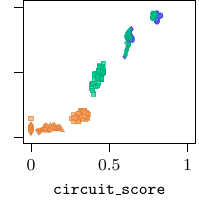}
        \caption{QAOA, w/o (left) and w/ (right) dynamical decoupling.}
    \end{subfigure}
    \;
    \begin{subfigure}[t]{0.4\textwidth}
        \includegraphics{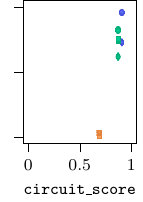}
        \;
        \includegraphics{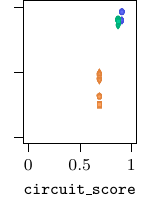}
        \caption{VQE, w/o (left) and w/ (right) dynamical decoupling.}
    \end{subfigure}
    \caption{
        Number of $\CX$ gates (top) and \circuitScore\ (bottom)
        versus fidelity $\fidelity$
        for the QAOA and VQE 
        circuits with $\repsQaoa = 1$ layer
        for all LamA use case examples executed on \ibmcairo.
        The results stem from
        the same experiment as shown in
        Figure~\ref{fig:lama-fidelity-boxplots-ibm-cairo}.
        The legend for the marker symbols can be found 
        in Figure~\ref{fig:legend-lama-examples}.
    }
    \label{fig:lama-fidelity-vs-metrics-ibm-cairo}
\end{figure}

\begin{figure}
    \small
    \centering
    %
    % \begin{subfigure}[t]{0.9\textwidth}
    %     \includegraphics{figures/lama_trained_circuits_ibmq_ehningen/scatterplots_cnots/legend.pdf}
    %     \caption*{}
    % \end{subfigure}
    % %
    % \\
    %
    \begin{subfigure}[t]{0.55\textwidth}
        \includegraphics{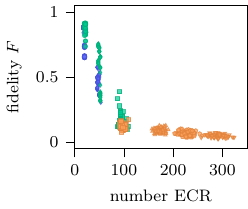}
        \;
        \includegraphics{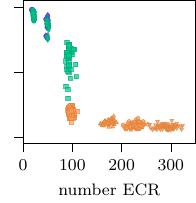}
        \caption{QAOA, w/o (left) and w/ (right) dynamical decoupling.}
    \end{subfigure}
    \; 
    \begin{subfigure}[t]{0.4\textwidth}
        \includegraphics{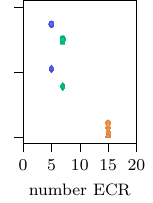}
        \;
        \includegraphics{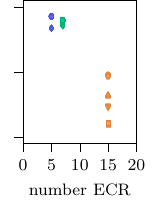}
        \caption{VQE, w/o (left) and w/ (right) dynamical decoupling.}
    \end{subfigure}
    \\[0.2cm]
    \begin{subfigure}[t]{0.55\textwidth}
        \includegraphics{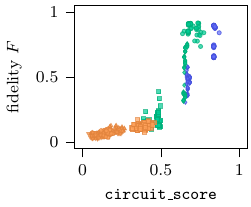}
        \;
        \includegraphics{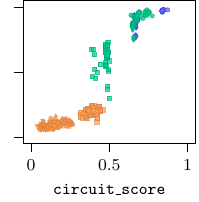}
        \caption{QAOA, w/o (left) and w/ (right) dynamical decoupling.}
    \end{subfigure}
    \;
    \begin{subfigure}[t]{0.4\textwidth}
        \includegraphics{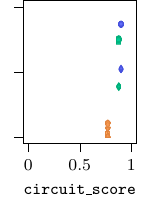}
        \;
        \includegraphics{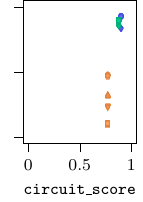}
        \caption{VQE, w/o (left) and w/ (right) dynamical decoupling.}
    \end{subfigure}
    \caption{        
        Number of $\ECR$ gates (top) and \circuitScore\ (bottom)
        versus fidelity $\fidelity$
        for the QAOA and VQE 
        circuits with $\repsQaoa = 1$ layer
        for all LamA use case examples executed on \ibmsherbrooke.
        The results stem from
        the same experiment as shown in
        Figure~\ref{fig:lama-fidelity-boxplots-ibm-sherbrooke}.
        The legend for the marker symbols can be found 
        in Figure~\ref{fig:legend-lama-examples}.
    }
    \label{fig:lama-fidelity-vs-metrics-ibm-sherbrooke}
\end{figure}

\begin{figure}
    \small
    \centering
    %
    % \begin{subfigure}[t]{0.9\textwidth}
    %     \includegraphics{figures/lama_trained_circuits_ibmq_ehningen/scatterplots_cnots/legend.pdf}
    %     \caption*{}
    % \end{subfigure}
    % %
    % \\
    %
    \begin{subfigure}[t]{0.55\textwidth}
        \includegraphics{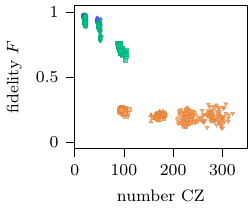}
        \;
        \includegraphics{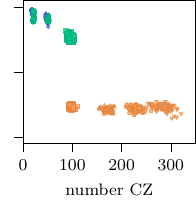}
        \caption{QAOA, w/o (left) and w/ (right) dynamical decoupling.}
    \end{subfigure}
    \;
    \begin{subfigure}[t]{0.4\textwidth}
        \includegraphics{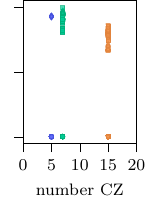}
        \;
        \includegraphics{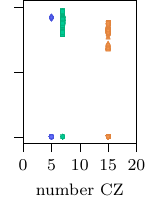}
        \caption{VQE, w/o (left) and w/ (right) dynamical decoupling.}
    \end{subfigure}
    \\[0.2cm]
    \begin{subfigure}[t]{0.55\textwidth}
        \includegraphics{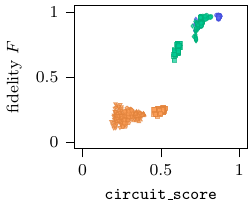}
        \;
        \includegraphics{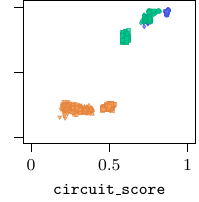}
        \caption{QAOA, w/o (left) and w/ (right) dynamical decoupling.}
    \end{subfigure}
    \; \;
    \begin{subfigure}[t]{0.4\textwidth}
        \includegraphics{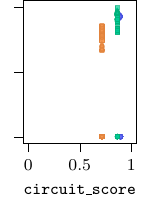}
        \;
        \includegraphics{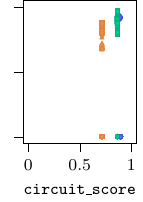}
        \caption{VQE, w/o (left) and w/ (right) dynamical decoupling.}
    \end{subfigure}
    \caption{
        Number of $\CZ$ gates (top) and \circuitScore\ (bottom)
        versus fidelity $\fidelity$
        for the QAOA and VQE 
        circuits with $\repsQaoa = 1$ layer
        for all LamA use case examples executed on \ibmtorino.
        The results stem from
        the same experiment as shown in
        Figure~\ref{fig:lama-fidelity-boxplots-ibm-torino}.
        The legend for the marker symbols can be found 
        in Figure~\ref{fig:legend-lama-examples}.
    }
    \label{fig:lama-fidelity-vs-metrics-ibm-torino}
\end{figure}

In Figures~\ref{fig:lama-fidelity-vs-metrics-ibm-cairo},
\ref{fig:lama-fidelity-vs-metrics-ibm-sherbrooke}, and 
\ref{fig:lama-fidelity-vs-metrics-ibm-torino} we show the two metrics versus
the fidelity $\fidelity$ for \ibmcairo, 
\\
\ibmsherbrooke, and \ibmtorino, respectively.
The results for \ibmqehningen\ are given in 
Figure~\ref{fig:lama-fidelity-vs-cnots-ehningen} for the number of $\CX$ gates 
and in Figure~\ref{fig:lama-fidelity-vs-circuit-score-ehningen} for the \circuitScore.
Note that for the other backends we are not showing simulations with noise models
of the backends.
Comparing the results from \ibmcairo\ with \ibmqehningen\ we see very similar
fidelities for QAOA. Only for \lamaEx{1}{3} and 
in the case with dynamical decoupling \ibmcairo\ achieves a bit higher fidelities.
Note that this improvement is not represented by the \circuitScore. For the VQE
circuits we observe better fidelities from \ibmqehningen\ and, again, note that
neither the number of $\CX$ gates nor the \circuitScore\ can reflect this difference.
Next, let us compare \ibmsherbrooke\ and \ibmqehningen. Most prominently, we see
that \ibmsherbrooke\ yields, for the same number of two qubit gates or the
same \circuitScore, fidelities with a much larger variance than \ibmqehningen.
Other than that the observations we did with \ibmcairo\ also hold for this backend.
For \ibmtorino\ the fidelities look quite different than for the other three backends.
In particular, we see high fidelities for example series \lamaSeries{0} and 
\lamaSeries{1} and that the number of $\CZ$ gates or the \circuitScore\ can predict
well the decrease of the fidelity for the different examples. However,
note that for QAOA the fidelities of \lamaEx{2}{1} are much lower than for
\lamaEx{1}{3} although the circuits have the same number of $\CZ$ gates and
nearly the same \circuitScore s. Here, we see that the different number of qubits
(and the associated error sources) are not well represented by neither of our two
metrics. Last, note that all examples of \lamaSeries{2}, and in particular
in the cases with dynamical decoupling, have fidelity values on a plateau of
around $0.22$ although they have strongly different numbers of $\CZ$ gates and
\circuitScore s.

% ibm_cairo, scatter plot number cx vs fidelity,
% circuit score vs fidelity
%

% ibm_sherbrooke, scatter plot number ecr vs fidelity
% circuit score vs fidelity
%

% ibm_torino, scatter plot number cz vs fidelity
% and circuit score vs fidelity

\subsection{Experiments on D-Wave \dwaveAdvantage}
\label{sec: Annealing for LamA}
\begin{figure}
	\centering
	\begin{subfigure}{0.2\textwidth}
		\centering
		\includegraphics{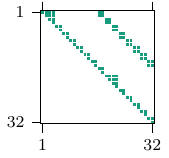}
		\caption{\lamaEx{3}{1}}
	\end{subfigure}
	\begin{subfigure}{0.2\textwidth}
		\includegraphics{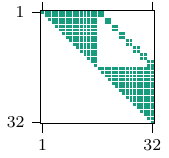}
		\caption{\lamaEx{3}{2}}
	\end{subfigure}
	\;
	\begin{subfigure}{0.2\textwidth}
		\includegraphics{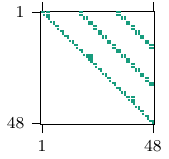}
		\caption{\lamaEx{4}{1}}
	\end{subfigure}
	\begin{subfigure}{0.2\textwidth}
		\includegraphics{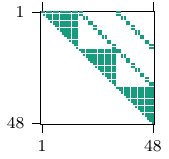}
		\caption{\lamaEx{4}{2}}
	\end{subfigure}
	\caption{
		Sparsity of QUBO matrix $\matQubo$ for example series
		\lamaSeries{3} and \lamaSeries{4}.}
	\label{matrix plots for 2 cars 8 slots and 3 cars 8 slots}
\end{figure}
This last section of the LamA use case is dedicated to experiments on the
quantum annealer D-Wave \dwaveAdvantage. Recall from 
Table~\ref{tab:specifications-dwave} that this quantum system has more
than 5000 qubits. Thus, we consider in this section, besides the
previously introduced example series,
% \lamaSeries{0}, \lamaSeries{1}, and \lamaSeries{2},
two additional
example series termed \lamaSeries{3} and \lamaSeries{4}. Both series model 
one charging station with four charging levels and $\numTimeslots = 8$ time slots.
In \lamaSeries{3} we have the case of $\numCars = 2$ cars, while in \lamaSeries{4}
we have $\numCars = 3$. This results in QUBOs with 32 and 48 variables, respectively.
In both series we consider two examples, where in the first example the time slots,
when the different cars charge energy, only overlap in one time slot, while in the
second one the overlap extends for all available slots. As a consequence the 
respective first example is only weakly coupled, while the second one
has a strong coupling between the 
variables. This can nicely be seen in the sparsity plots in 
Figure~\ref{matrix plots for 2 cars 8 slots and 3 cars 8 slots}.

\begin{table}
	\centering
	\begin{tabular}{ccccc}
		% \hline
		example & \# logical qubits & \# physical qubits 
		& \# chains & max chain length \\
		\hline \hline
		\lamaEx{0}{1}& 6&6 &0 & 0 \\

		\lamaEx{0}{2}& 6& 8&2 &2  \\
		\hline
		\lamaEx{1}{1}& 8& 8& 0&0  \\

		\lamaEx{1}{2}& 8& 10&2 &2  \\

		\lamaEx{1}{3}& 8&12 & 4&2 \\
		\hline
		\lamaEx{2}{1}& 16&16 & 0& 0 \\

		\lamaEx{2}{2}& 16&23 &7 &2  \\

		\lamaEx{2}{3}& 16&27 & 10&3  \\

		\lamaEx{2}{4}& 16&32  &16 &2 \\
		\hline
		\lamaEx{3}{1}& 32& 32&0 & 0 \\

		\lamaEx{3}{2}& 32& 98&32 &4  \\
		\hline
		\lamaEx{4}{1}& 48& 65& 17&2  \\
		\lamaEx{4}{2}& 48&192 &48 &5  \\
		% \hline
	\end{tabular}
	\caption{The number of logical qubits, required physical qubits, 
	number of chains, and maximum chain length for all LamA use case examples.}
	\label{Table for qubit and chain number}
\end{table}

\begin{figure}
	\centering
	\begin{subfigure}{0.3\textwidth}
		\centering
		\includegraphics[width=3.5cm]{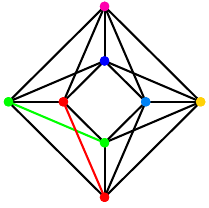}
		\caption{\lamaEx{0}{2}.}
	\end{subfigure}
	\,
	\begin{subfigure}{0.3\textwidth}
		\centering
		\includegraphics[width=4.5cm]{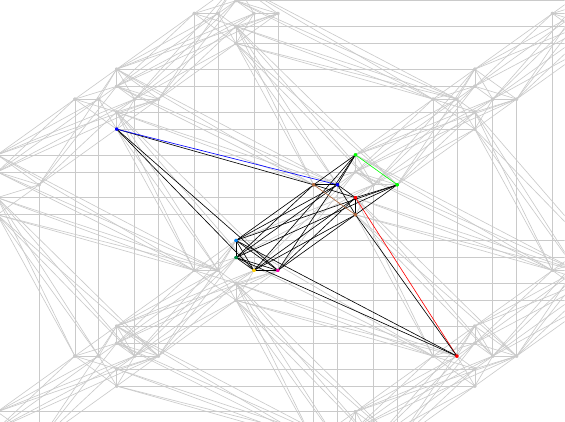}
		\caption{\lamaEx{1}{3}.}
	\end{subfigure}
	\quad
	\begin{subfigure}{0.3\textwidth}
		\centering
		\includegraphics[width=4.5cm]{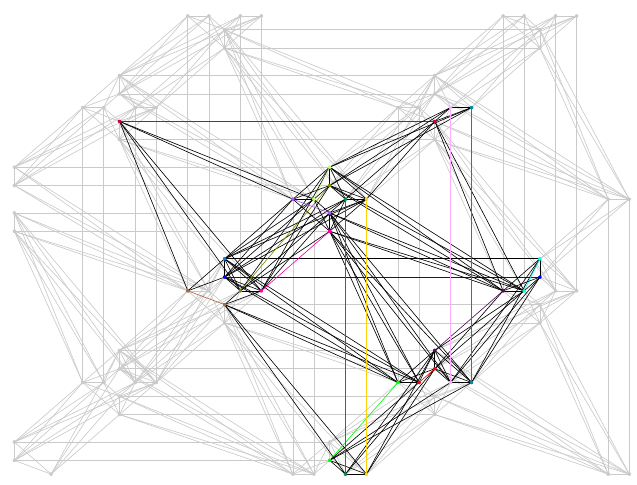}
		\caption{\lamaEx{2}{4}.}
	\end{subfigure}
	\caption{
		Embedding of the QUBO matrix $\matQubo$ of different LamA examples
		onto D-Wave \dwaveAdvantage. 
		The colored lines are chains,
		the black lines are hardware connections, 
		and the gray lines are couplers which are not used in the embedding.}
	\label{QUBO embedding on QPU for some examples}
\end{figure}

As we have discussed in Section~\ref{sec:quantum-systems-dwave} our logical
QUBOs have to be embedded onto the QPU. In 
Table~\ref{Table for qubit and chain number} we give the details on the
amount of utilized physical qubits, chains, and the maximum chain lengths for all
LamA use case examples. Note that in most cases more physical qubits are needed than
the problem has logical ones. This is due to the need for chains, which are
required to overcome the limited hardware connectivity in correctly representing the coupling of QUBO variables. We visualize this
in Figure~\ref{QUBO embedding on QPU for some examples}
exemplary for the embeddings of the examples \lamaEx{0}{2}, \lamaEx{1}{3},
and \lamaEx{2}{4}. It is instructive to compare these plots with the sparsity
pattern of the QUBO matrices given in Figure~\ref{fig:lama-sparsity-qubo}.
For example, we can see that \lamaEx{0}{2}
is mapped to the Pegasus unit cell, recall Figure~\ref{Dwave Pegasus 
architecture}, by directly embedding 4 logical qubits onto 4 physical qubits, while
representing the remaining 2 logical qubits with 2 chains containing 2 physical qubits each.
Thus, the embedding of \lamaEx{0}{2} requires 8 physical qubits
although the underlying QUBO only has 6 (logical) qubits.
By this reasoning, we see that \lamaEx{1}{3} and \lamaEx{2}{4}
require 12 and 32 physical qubits, respectively, to embed 
their 8 and 16 logical qubits.

Besides the embedding we pointed out in Section~\ref{sec:quantum-systems-dwave}
that the hardware parameters chain strength and anneal time can have a strong
impact on the quality of the annealing solution. In the next paragraphs
we report on experiments that illuminate this. In order to obtain 
statistically relevant results, we ran each experiment
10 times. For each such run we used either 400 or 1000 reads. We report on the quality of
the D-Wave results in terms of the percentage of feasible and optimal solutions
from the overall reads. Here, a bitstring is called feasible if it satisfies
the constraint \eqref{eq:qcio-2} of our underlying optimization problem.
This is understood in the sense that the integer vector stemming 
from the bitstring's transformation via the transformation matrix
\eqref{eq:qcio-2} satisfies the constraint.
%
% \todo{@Bharad or Leon: Define 'read'}

\begin{figure}
	\centering
	\includegraphics{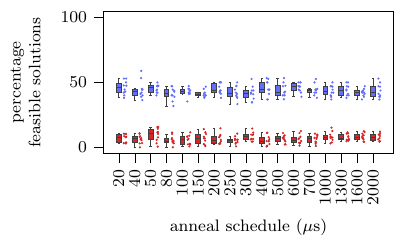}
	\qquad
	\includegraphics{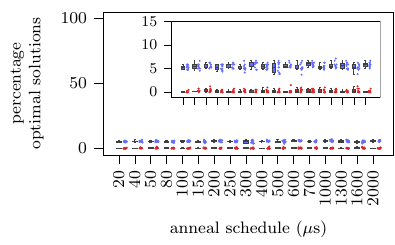}
	\caption{
		Percentage of feasible (left) and optimal (right) solutions
		for different anneal schedules for \lamaEx{0}{1} (blue markers)
		and \lamaEx{4}{2} (crimson markers).
		For the former the chain strength was chosen as 0.3 and
		for the latter as 0.8. For each run we used 400 reads.
	}
	\label{Anneal schedules for smallest and largest use case}
\end{figure}

\begin{figure}
	\centering
	\includegraphics{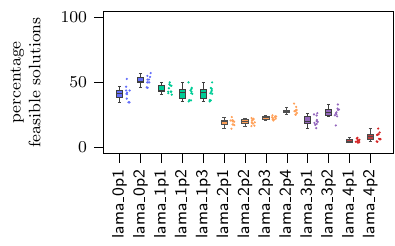}
	\qquad
	\includegraphics{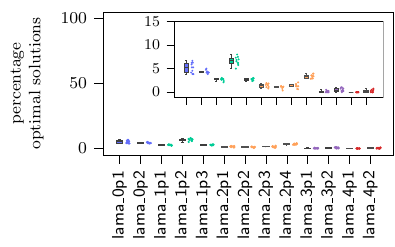}
	\caption{
		Percentage of feasible (left) and optimal (right) solutions for 
		all LamA examples. The anneal schedule was chosen as 300 $\mu$s.
		For each run we used 400 reads.}
	\label{Anneal schedule 300 for all systems}
\end{figure}

\paragraph{Anneal schedule}
We systematically studied the effect of the anneal schedule and of the
chain strength for all examples in the LamA use case.
Thereby, we observed an almost constant quality 
of the solutions with respect to the anneal time. We show this exemplarily
for \lamaEx{0}{1} and \lamaEx{4}{2} in
Figure~\ref{Anneal schedules for smallest and largest use case}. Since the precise
choice of the annealing schedule is only secondary for our LamA examples
we fixed it to 300 $\mu$s and report the solutions qualities in
Figure~\ref{Anneal schedule 300 for all systems}. 
There, we can nicely see how the percentages of feasible and optimal solutions
decrease when the problem size increases.

\begin{figure}
	\raggedright
	\begin{subfigure}{0.35\textwidth}
		\includegraphics{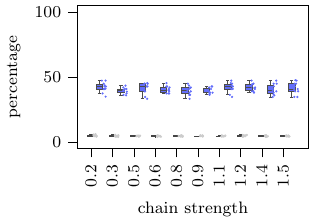}
		\caption{\lamaEx{0}{1}.}
		\label{fig:lama-dwave-chain-strength-ex0p1}
	\end{subfigure}
	\;
	\begin{subfigure}{0.3\textwidth}
		\includegraphics{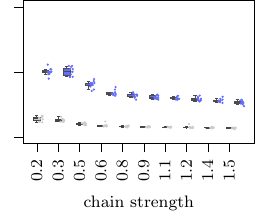}
		\caption{\lamaEx{0}{2}.}
		\label{fig:lama-dwave-chain-strength-ex0p2}
	\end{subfigure}
	\\[0.25cm]
	\begin{subfigure}{0.35\textwidth}
		\includegraphics{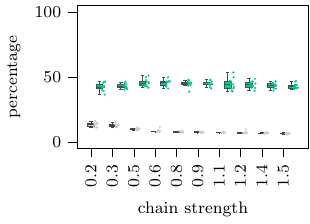}
		\caption{\lamaEx{1}{1}.}
		\label{fig:lama-dwave-chain-strength-ex1p1}
	\end{subfigure}
	\;
	\begin{subfigure}{0.3\textwidth}
		\includegraphics{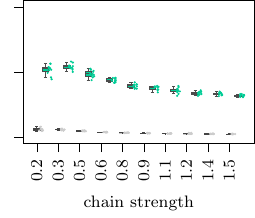}
		\caption{\lamaEx{1}{2}.}
		\label{fig:lama-dwave-chain-strength-ex1p2}
	\end{subfigure}
	\begin{subfigure}{0.3\textwidth}
		\includegraphics{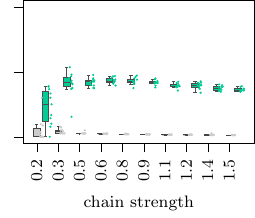}
		\caption{\lamaEx{1}{3}.}
		\label{fig:lama-dwave-chain-strength-ex1p3}
	\end{subfigure}
	\\[0.25cm]
	\begin{subfigure}{0.35\textwidth}
		\includegraphics{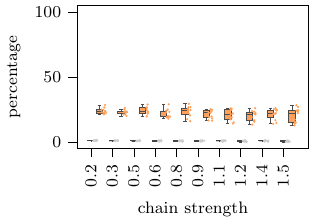}
		\caption{\lamaEx{2}{1}.}
		\label{fig:lama-dwave-chain-strength-ex2p1}
	\end{subfigure}
	\;
	\begin{subfigure}{0.3\textwidth}
		\includegraphics{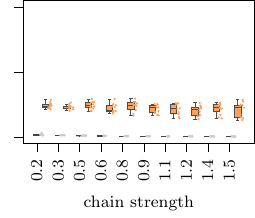}
		\caption{\lamaEx{2}{2}.}
		\label{fig:lama-dwave-chain-strength-ex2p2}
	\end{subfigure}
	\begin{subfigure}{0.3\textwidth}
		\includegraphics{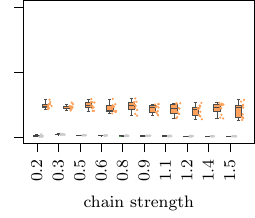}
		\caption{\lamaEx{2}{3}.}
		\label{fig:lama-dwave-chain-strength-ex2p3}
	\end{subfigure}
	\\[0.15cm]
	\begin{subfigure}{0.33\textwidth}
		\includegraphics{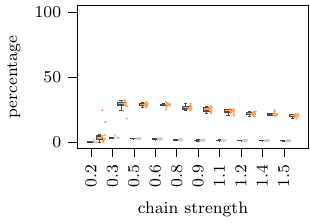}
		\caption{\lamaEx{2}{4}.}
		\label{fig:lama-dwave-chain-strength-ex2p4}
	\end{subfigure}
	\caption{
		Percentage of feasible (colored markers) and optimal (gray markers)
		solutions for different chain strength values for example series \lamaSeries{0} (first row), 
		\lamaSeries{1} (second row), and \lamaSeries{2} (third and fourth row).
		The anneal time was chosen as 300 $\mu$s and for each 
		run we used 1000 reads.
	}
	\label{Chain strength for 1 car with 3 and 4 slots}
\end{figure}

\begin{figure}
	\raggedright
	\begin{subfigure}{0.35\textwidth}
		\includegraphics{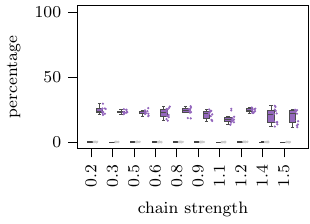}
		\caption{\lamaEx{3}{1}.}
		\label{fig:lama-dwave-chain-strength-ex3p1}
	\end{subfigure}
	\;
	\begin{subfigure}{0.3\textwidth}
		\includegraphics{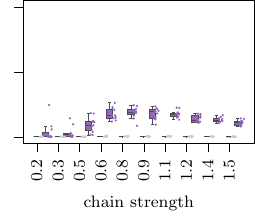}
		\caption{\lamaEx{3}{2}.}
		\label{fig:lama-dwave-chain-strength-ex3p2}
	\end{subfigure}
	\\[0.15cm]
	\begin{subfigure}{0.35\textwidth}
		\includegraphics{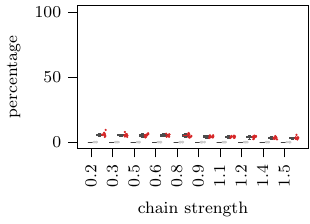}
		\caption{\lamaEx{4}{1}.}
		\label{fig:lama-dwave-chain-strength-ex4p1}
	\end{subfigure}
	\;
	\begin{subfigure}{0.3\textwidth}
		\includegraphics{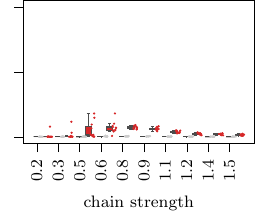}
		\caption{\lamaEx{4}{2}.}
		\label{fig:lama-dwave-chain-strength-ex4p2}
	\end{subfigure}
	\caption{
		Percentage of feasible (colored markers) and optimal (gray markers)
		solutions for different chain strength values for example
		series \lamaSeries{3} (first row) and \lamaSeries{4} (second row).
		The anneal time was chosen as 300 $\mu$s and for each 
		run we used 1000 reads.
	}
	\label{Chain strength for 2 cars with 4 and 8 slots}
\end{figure}

\paragraph{Chain strength}
In contrary to the anneal schedule, we observe an impact of the chain 
strength in the quality of our annealing solutions. For example, we see in 
Figure~\ref{fig:lama-dwave-chain-strength-ex0p2} that for \lamaEx{0}{2}
a higher chain strength decreases the fraction of both feasible and optimal
solutions. For \lamaEx{1}{2} and \lamaEx{1}{3} in
Figures~\ref{fig:lama-dwave-chain-strength-ex1p2} and 
\ref{fig:lama-dwave-chain-strength-ex1p3}, respectively, we first observe a
fast increase in quality with increasing anneal time until a certain threshold 
and afterwards a slow decline. The same behavior can be seen for \lamaEx{2}{4},
\lamaEx{3}{2}, and \lamaEx{4}{2} in 
Figures~\ref{fig:lama-dwave-chain-strength-ex2p4},
\ref{fig:lama-dwave-chain-strength-ex3p2}, and
\ref{fig:lama-dwave-chain-strength-ex4p2}, respectively. These are exactly the 
examples with the strongest coupling. Recalling how these problems have to 
be embedded onto the QPU it is clear that they are most sensitive to the chain strength.
Last, let us note the decrease in the fraction of feasible solutions from 
around 50 percent in \lamaSeries{0} to only a few percent in \lamaSeries{4}.
The same holds for the percentage of optimal solutions. Here, we can clearly observe
the limitations of current quantum hardware when dealing with large problems.

% Routing Part
\newcommand{\BCM}[1]{\textcolor{blue}{#1}}
\section{Use Case 2: Optimization of Truck Routes}
\label{sec: Annealing for Routing}
In the truck routing problem (TRP) we aim to find the route with the shortest travel distance between a given set of cities. This route has to start and to end in a particular city (which is usually the depot of the truck) and each city may be visited only once on the route. The TRP, notorious for its NP-hard complexity, represents a significant challenge in combinatorial optimization. Finding the shortest route to visit every city once becomes exponentially harder as the number of cities increases (combinatorial explosion), rendering traditional algorithms inefficient. Thus, it is interesting to explore other paradigms to solve this problem. The formulation of the TRP as a QUBO problem opens the door for its representation on D-Wave systems. We will explore how to represent the TRP using the binary quadratic model (BQM), which is suitable for D-Wave's  QPU.

\begin{figure}
	
	\centering
	\begin{subfigure}{0.32\textwidth}
		\includegraphics[width=\textwidth]{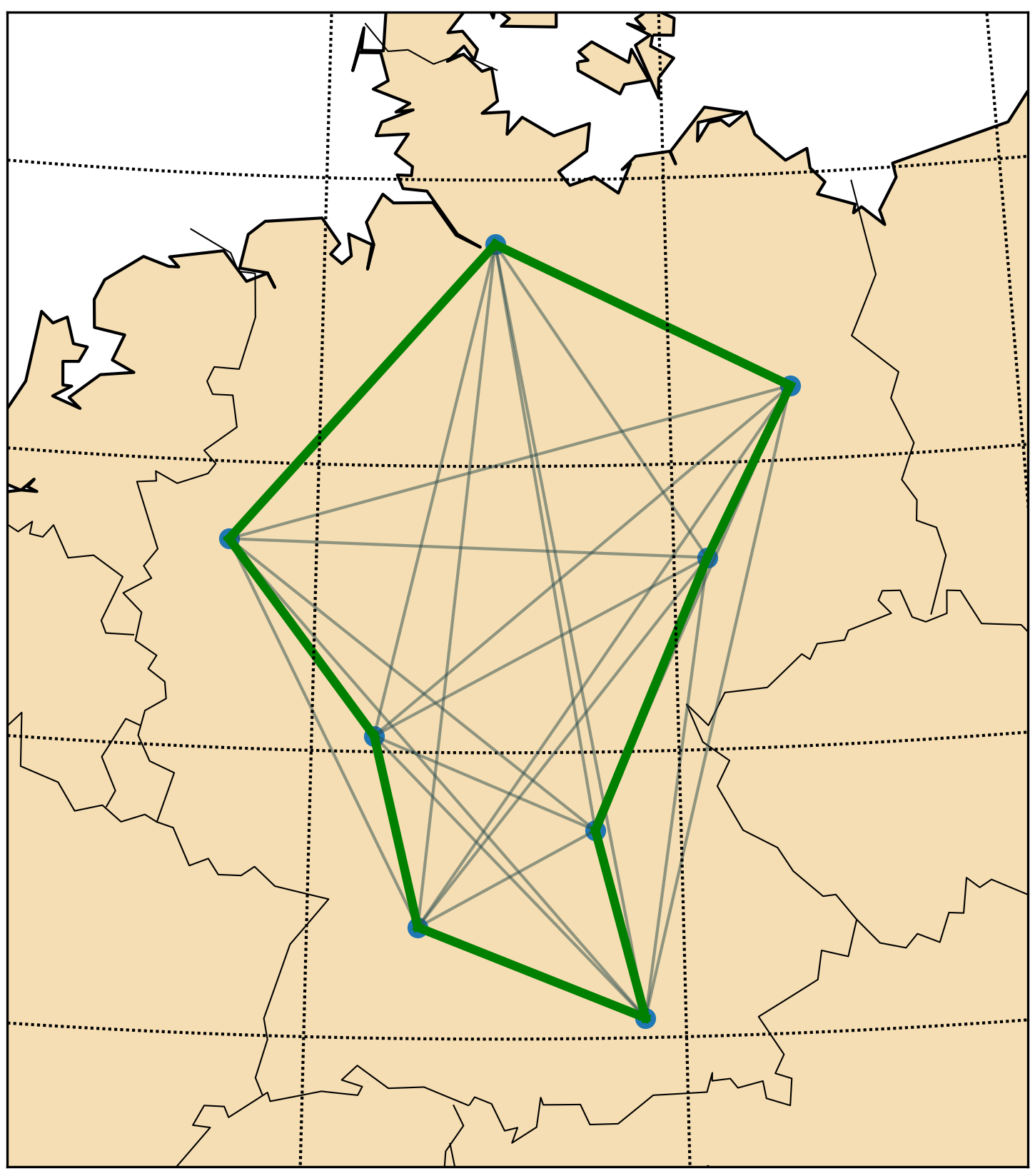}
		\caption{Asymmetrical (\trpasym{8}).}
	\end{subfigure}
	\hspace*{2cm}
	\begin{subfigure}{0.32\textwidth}
		\includegraphics[width=\textwidth]{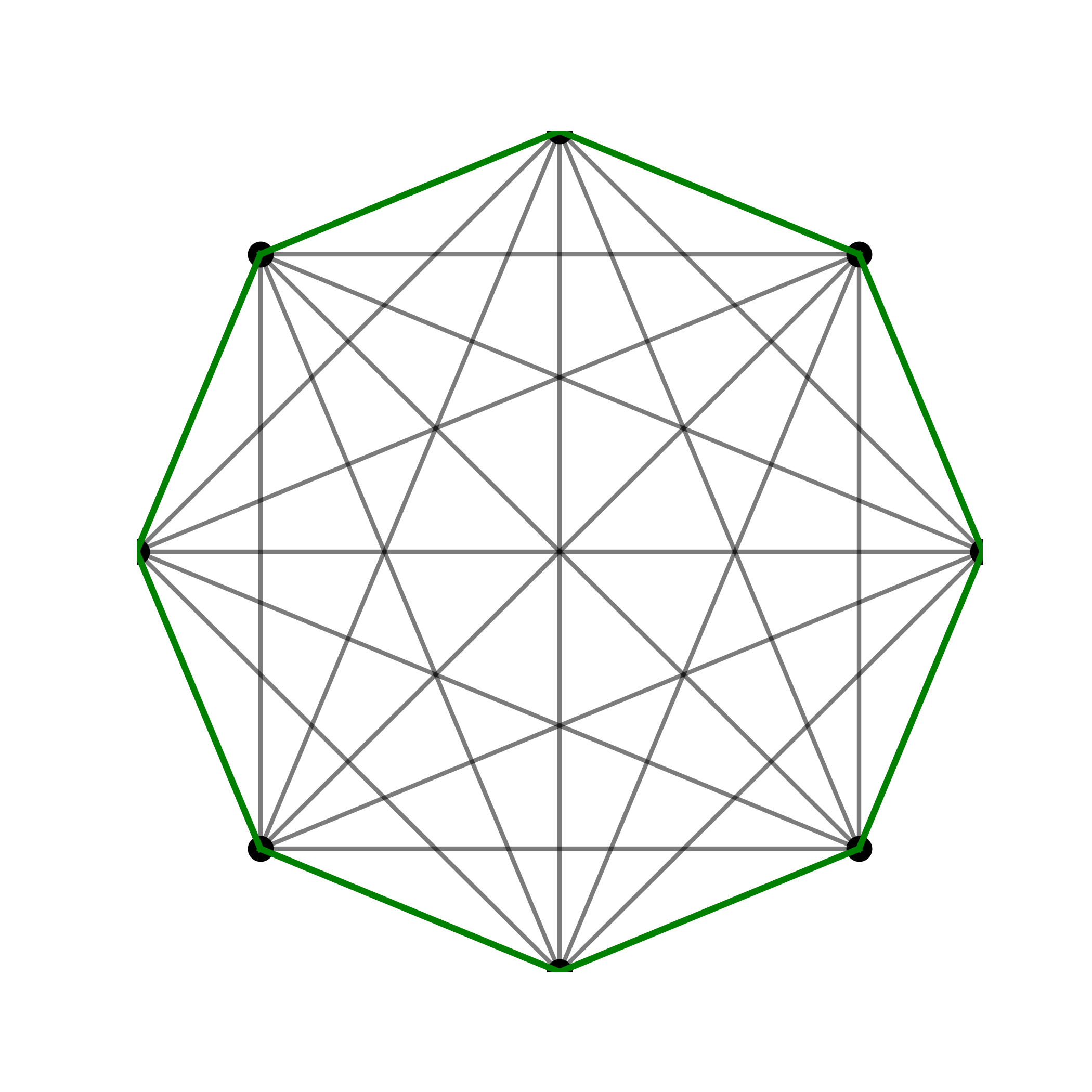}
		\caption{Symmetrical (\trpsym{8}).}
	\end{subfigure}
	\caption{Examples for the two types of distributions that we consider in this paper: asymmetrical \trpasym{8} (left) and symmetrical  (right).}
	\label{Symmetric and asymmetric distributions }
\end{figure}
%In recent years, quantum annealing has emerged as a promising avenue for tackling such problems, with D-Wave systems offering a unique hardware platform for exploration. D-Wave's quantum annealer leverages a physical process called quantum annealing to explore potential solutions probabilistically and efficiently. This approach aims to navigate the vast solution space and converge towards the shortest route, offering a potential advantage over classical algorithms for specific TRP instances.

\subsection{Problem formulation}
Consider a set of $n$ cities represented by indices from $1$ to $n$. Let $b_{i,t} \in \{ 0,1 \}$ be a binary variable indicating whether city $i \in \{1,...,n\}$ is visited ($b_{i,t} =1$) or not ($b_{i,t}=0$) at each time $t \in \{1,...,n\}$.\\

The QUBO formulation consists of two parts. Let $E$ be the set of tuples that
indicate if two cities are connected 
% $E= \{ (i,j) \in \{1, ...., n\}^n \ | \ \text{cities $i$ and $j$ are connected} \}$,
and let $d_{ij}$ denote their distances.
%Let $E$ be the set of all possible paths from city $i$ to city $j$ and $D_\mathrm{ij}$ their distance. 
The quadratic QUBO terms are then given by
\begin{align}
	Q_d=  \sum_{(i,j)\in E} d_{ij} \sum_{t=1}^n b_{i,t}b_{j,t+1} \ .
\end{align}
%with a sum over every path in $E$ and a sum over all time steps. 
This part of the QUBO encodes the distances. The constraints are included into the cost function via 
\begin{align*}
	Q_c = \varrho \cdot \left [ \sum_{i=1}^n \left ( 1 - \sum_{t=1}^n b_{i,t} \right )^2 + \sum_{t=1}^n \left ( 1 - \sum_{i=1}^n b_{i,t} \right )^2 \right ]. 
\end{align*}
% The penalty $\varrho$ parameter imposes the constraints more or less strictly on the total problem.
The first term imposes that in the overall route each city is only visited once while the second term favors solutions where each time step shows up exactly once.  
The combined QUBO problem with distance minimization and constraints can be expressed by
\begin{align}
	Q = Q_d+Q_c \ .
	\label{Hamiltonian of TRP }
\end{align}

Since there are $n$ possible locations for $n$ time steps, we need $n^2$ variables that can represent all possible combinations of time and city.  

\begin{figure}
	
	\centering
	\begin{subfigure}{0.32\textwidth}
		\centering
		\includegraphics[width=\textwidth]{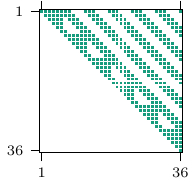}
		\caption{\trpasym{6} and \trpsym{6}.}
	\end{subfigure}
	\begin{subfigure}{0.32\textwidth}
		\centering
		\includegraphics[width=\textwidth]{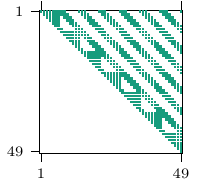}
		\caption{\trpasym{7}and \trpsym{7}.}
	\end{subfigure}
	\begin{subfigure}{0.32\textwidth}
		\includegraphics[width=\textwidth]{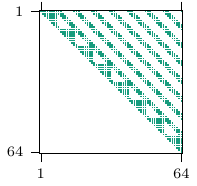}
		\caption{\trpasym{8} and \trpsym{8}.}
	\end{subfigure}
	\caption{Sparsity pattern of a QUBO matrix for 6, 7, and 8 cities.}
	\label{Sparsity plot for TRP}
\end{figure}

In this paper, we considered three different problem sizes. Each problem size has two different distributions of the cities, with one being asymmetric and the other radially symmetric. The asymmetrical distribution refers to a scenario where cities are placed randomly, resulting in a unique distance between each pair of cities. This randomness introduces complexity to the optimization problem, presenting a more challenging scenario for solution algorithms. On the other hand symmetrical distribution implies that cities are equidistantly placed on a ring. This uniformity simplifies the optimization landscape, potentially leading to more straightforward solutions. The asymmetric and symmetric distribution problem classes are referred as \trpasym{$m$} and \trpsym{$m$}, respectively, where $m$ corresponds to number of cities. The asymmetric and symmetric distributions of 8 cities are given in Figure~\ref{Symmetric and asymmetric distributions }. By using \eqref{Hamiltonian of TRP } we can build QUBO matrices for our problems. Their sparsity patterns are given in the Figure \ref{Sparsity plot for TRP}. We observe that the problems are strongly coupled. Note that the symmetric and asymmetric problems have the same sparsity pattern.
\subsection{Binary quadratic model}
In order to implement the QUBO formulation on the quantum annealer, it has to be reformulated as a binary quadratic model (BQM), that ideally has the form of an Ising Hamiltonian native to the hardware architecture. For that, a simple variable transformation from $b_{i,t} \in \{ 0,1 \}$ to $x_{i,t} \in \{ -1,1 \}$ is done, where the relationship between the variables is simply
\begin{equation*}
	b_{i,t} = \frac{1+x_{i,t}}{2} ~,
\end{equation*}
which results in
\begin{equation}
	H = \sum_{i=1}^{n} h_i x_i + \sum_{i=1}^{n} \sum_{j=1}^{n} J_{ij} x_i x_j + c ~.
\end{equation}
The coefficients $h_i$ (bias) and $J_{ij}$ (coupling strength) are related to the linear and quadratic QUBO terms, respectively, and are modified by the variable transformation. The constant term $c$ can be disregarded.

\subsection{Results} 
\label{sec:TRP results}
In this section, we present results of the TRP examples with respect to
different hardware parameters. As in Section~\ref{sec: Annealing for LamA} we
use the percentage of feasible and optimal solutions as quality metric.
Feasible solutions are those which satisfy the constraints imposed in the QUBO 
problem and optimal solutions are feasible solutions with minimum distance traveled. 
\paragraph{Setup}
We used clique embedding \cite{Choi2008, clique_embedding} to embed the the QUBO matrices onto the QPU  \dwaveAdvantage. This method is particularly useful for the Pegasus chip when the source graph (QUBO graph) is a clique, i.e. a fully connected graph. D-Wave's Ocean software provides a function, \textit{find\_clique\_embedding}, which uses a polynomial-time algorithm to find better embeddings for cliques than the default \textit{embeddingcomposite} method. We used \textit{CliqueSampler} to solve the embeded problems. The strength lies in its specificity, rendering it unsuitable for general optimization problems lacking pronounced clique structures. An advantage of this type of method is that it tries to keep the maximum chain length low, which is beneficial for highly interconnected problems like the ones at hand. 

\begin{figure}
	
	\centering
	\begin{subfigure}{0.33\textwidth}
		\centering
		\includegraphics[width=\textwidth]{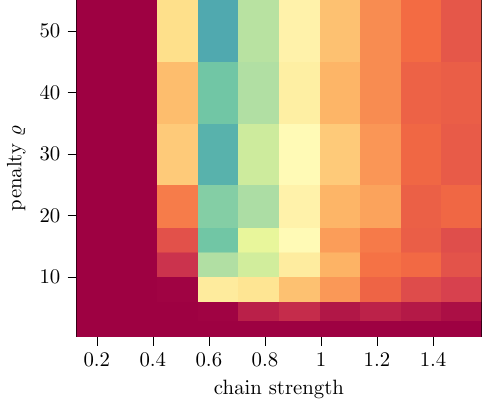}
		\caption{\trpasym{6}.}
	\end{subfigure}
	\begin{subfigure}{0.295\textwidth}
		\centering
		\includegraphics[width=\textwidth]{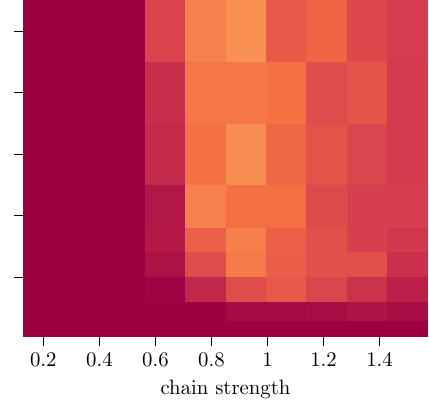}
		\caption{\trpasym{7}.}
	\end{subfigure}
	\begin{subfigure}{0.35\textwidth}
		\includegraphics[width=\textwidth]{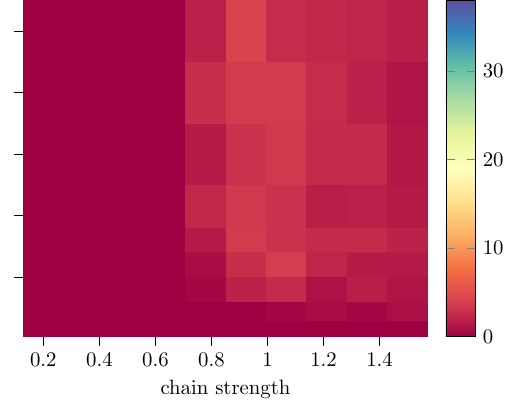}
		\caption{\trpasym{8}.}
	\end{subfigure}
	\vspace{1cm}
	\begin{subfigure}{0.33\textwidth}
		\centering
		\includegraphics[width=\textwidth]{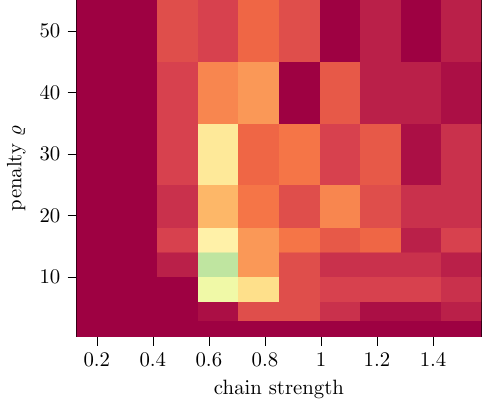}
		\caption{\trpasym{6}.}
	\end{subfigure}
	\begin{subfigure}{0.295\textwidth}
		\centering
		\includegraphics[width=\textwidth]{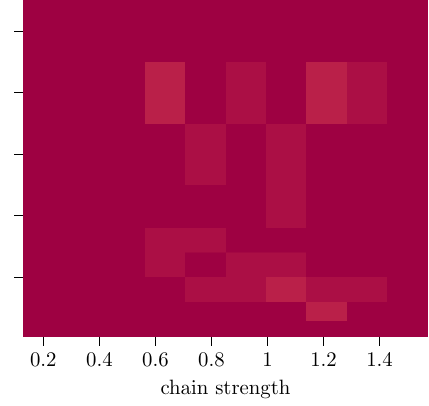}
		\caption{\trpasym{7}.}
	\end{subfigure}
	\begin{subfigure}{0.355\textwidth}
		\includegraphics[width=\textwidth]{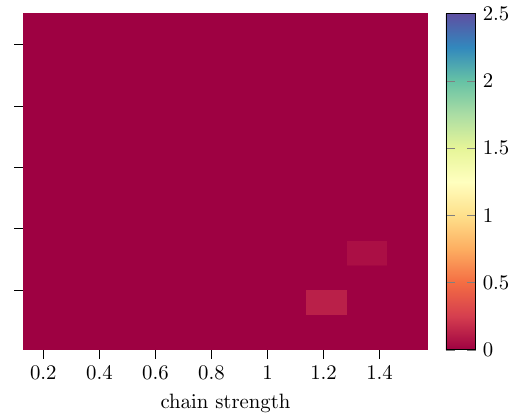}
		\caption{\trpasym{8}.}
	\end{subfigure}
	\caption{Percentage of feasible (upper row) and optimal (lower row) solutions for asymmetrical distribution of cities with variable chain strength and penalty values. The number of anneal reads are 1500 for each combination of penalty and chain strength. The annealing time was chosen to be 300 $\mu$s. Clique embedding was used to embed the QUBO. The calculations were executed on \dwaveAdvantage\ and results were obtained on 2023/11/06. 
	%(For the (a) -- (c) the color scale ranges from $0$ to $38$ and for (d) -- (f) from $0$ to $2.5$)
	}
	\label{Surface plot for asymmetry cities}
\end{figure}

\begin{figure}
	
	\centering
	\begin{subfigure}{0.33\textwidth}
		\includegraphics[width=\textwidth]{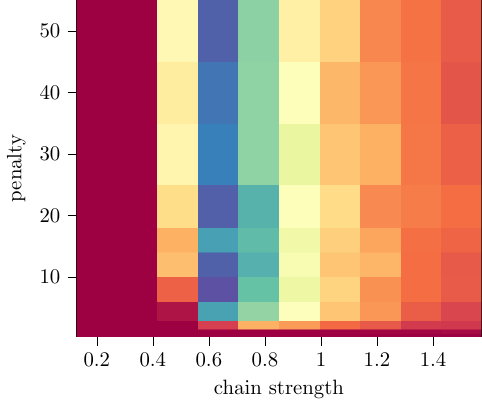}
		\caption{\trpsym{6}.}
	\end{subfigure}
	\begin{subfigure}{0.295\textwidth}
		\centering
		\includegraphics[width=\textwidth]{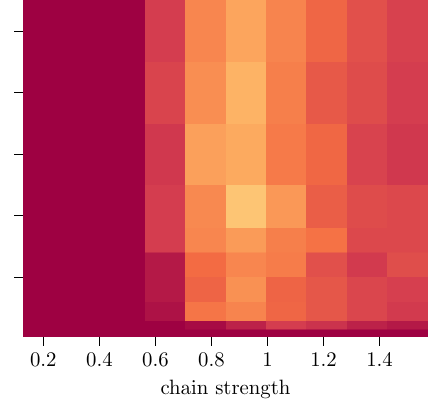}
		\caption{\trpsym{7}.}
	\end{subfigure}
	\begin{subfigure}{0.35\textwidth}
		\includegraphics[width=\textwidth]{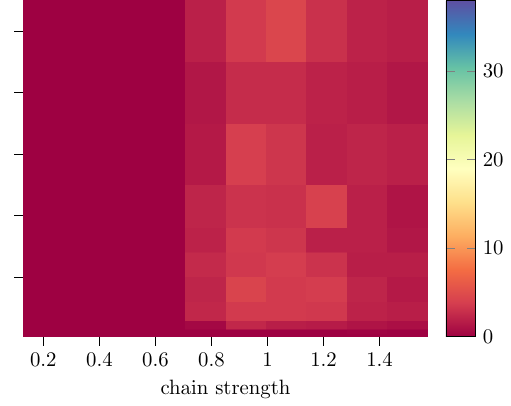}
		\caption{\trpsym{8}.}
	\end{subfigure}
	\vspace{1cm}
	\begin{subfigure}{0.33\textwidth}
		\includegraphics[width=\textwidth]{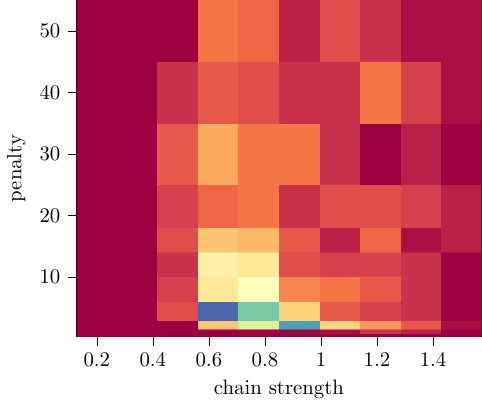}
		\caption{\trpsym{6}.}
	\end{subfigure}
	\begin{subfigure}{0.295\textwidth}
		\centering
		\includegraphics[width=\textwidth]{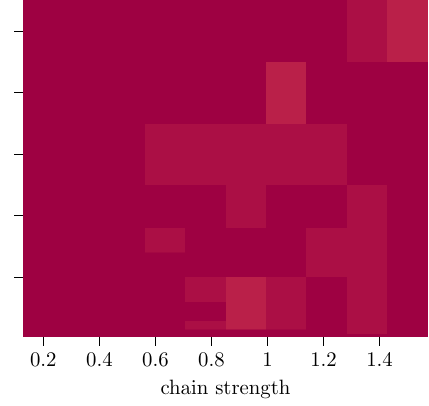}
		\caption{\trpsym{7}.}
	\end{subfigure}
	\begin{subfigure}{0.355\textwidth}
		\includegraphics[width=\textwidth]{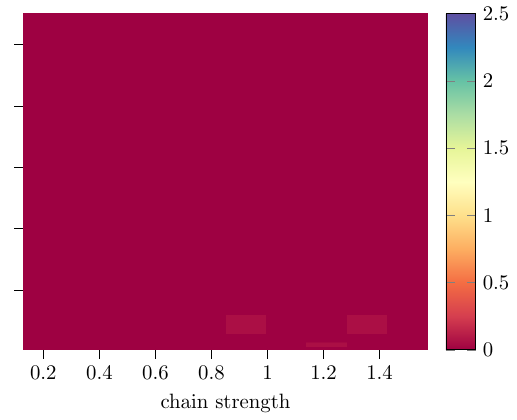}
		\caption{\trpsym{8}.}
	\end{subfigure}
	\caption{Percentage of feasible (upper row) and optimal (lower row) solutions for symmetrical distribution of cities with variable chain strength and penalty values. The number of anneal reads are 1500 for each value of penalty and chain strength. The annealing time was chosen to be 300 $\mu$s. Clique embedding was used to embed the QUBO. The calculations were executed on \dwaveAdvantage\ and results were obtained on 2023/11/11.
	%(For the (a) -- (c) the color scale ranges from $0$ to $38$ and for (d) -- (f) from $0$ to $2.5$) 
	}
	\label{Surface plot for symmetry cities}
\end{figure}

\paragraph{Penalty parameters}
We analyze two different penalty parameters effecting the quality of the solution. On the QUBO level we used $\varrho$ and on the hardware level the chian strength. The QUBO penalty parameter $\varrho$ controls the constraints in the problem formulation. The chain strength is responsible for keeping the chains consistent, penalizing chain breaks, and is implemented on a lower level concerning the actual Hamiltonian that describes the hardware operation.  The number of anneal reads were 1500 for each combination of penalty and chain strength value.
The heat map plot for the chain strength and the penalty parameter is instructive as these parameters ensure the integrity of the solution and the satisfaction of constraints. 
The percentage of feasible and optimal solutions for the asymmetrical distribution of cities are given in Figure \ref{Surface plot for asymmetry cities}. One common observation for all the different number of cities is that for small chain strength, quantum annealing almost cannot find any feasible solutions, irrespective of the strengths of penalty. In the case of \trpasym{6}, the chain strength of $0.6$ yields the highest percentage of solutions for nearly all penalties. On the contrary, the higher percentage of feasible solutions for \trpasym{7} and \trpasym{8} are obtained for chain strength of $0.9$ and only for larger penalty values. The problems \trpasym{7} and \trpasym{8} need to maintain the chain length of 6 and 7 to hold 284 and 436 physical qubits, respectively, see Table~\ref{Table for qubit and chain number TRP}. Since the stronger chain strength can keep the longer chains and thus the original problem intact, they lead to higher percentages of feasible and optimal solutions. As the problem size increases the samller chain strength values fail to yield any solutions. Higher percentage of optimal solutions for \trpasym{6} are obtained for the same values of chain strength values as for feasible solutions. On the other hand, for \trpasym{7} and \trpasym{8}, stronger chain strengths and smaller penalty values yield optimal solutions. The higher percentage of feasible and optimal solutions do not share the same chain strength values. Overall, the percentage of feasible and optimal solutions reduces as we increase the problem size. 
In the case of symmetrical distribution of the cities, the percentage of feasible solutions for all cities is higher than in the asymmetrical distribution of cities, see Figure \ref{Surface plot for symmetry cities}. For \trpsym{6}, with a chain strength of $0.6$  and for nearly all penalty values, we obtain the highest amount of feasible soultions. The same chain strength yields maximum optimal solutions. The feasible solutions for \trpsym{7} are at maximum for higher chain strength values which is similar to the asymmetrical distribution of cities. On the contrary to the asymmetrical distribution, the same chain strength is enough to obtain a higher percentage of both feasible and optimal solution for \trpsym{7} and \trpsym{8}.

% \FloatBarrier

% \FloatBarrier

\begin{figure}
	\centering
	\begin{subfigure}[b]{0.37\textwidth}
		\centering
		\includegraphics[width=\textwidth]{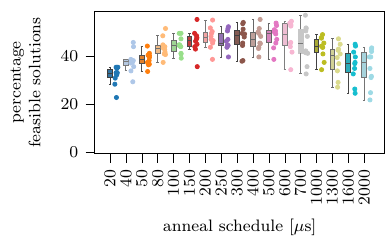}
		\caption{\trpasym{6}.}
	\end{subfigure}
	\begin{subfigure}{0.30\textwidth}
		\centering
		\includegraphics[width=\textwidth]{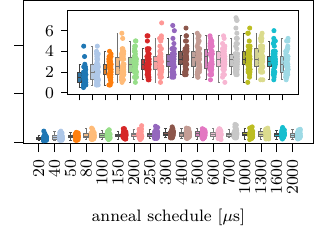}
		\caption{\trpasym{7}.}
	\end{subfigure}
	\begin{subfigure}{0.30\textwidth}
		\includegraphics[width=\textwidth]{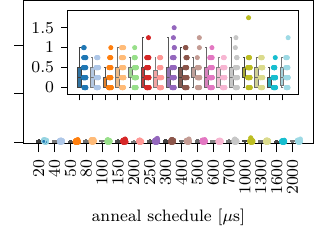}
		\caption{\trpasym{8}.}
	\end{subfigure}
	\begin{subfigure}{0.37\textwidth}
		\centering
		\includegraphics[width=\textwidth]{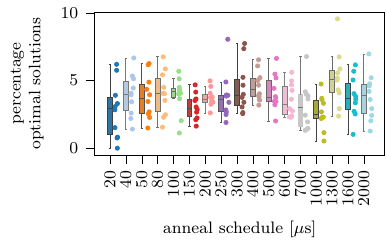}
		\caption{\trpasym{6}.}
	\end{subfigure}
	\begin{subfigure}{0.30\textwidth}
		\centering
		\includegraphics[width=\textwidth]{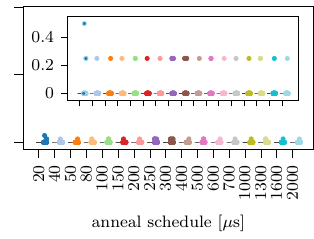}
		\caption{\trpasym{7}.}
	\end{subfigure}
	\begin{subfigure}{0.30\textwidth}
		\includegraphics[width=\textwidth]{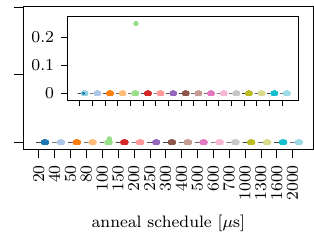}
		\caption{\trpasym{8}.}
	\end{subfigure}
	\caption{Percentage of feasible and optimal solutions for asymmetrical distribution of cities with different anneal schedules. The chain strength value is $0.6$ for \trpasym{6} and $0.9$ for \trpasym{7} and \trpasym{8}. The number of anneal reads are 20000 for each anneal schedule. Clique embedding was used to embed the QUBO. The calculations were executed on \dwaveAdvantage\ and results were obtained on 2023/11/15. }
	\label{Box plot for asymmetry cities of annealing schedule}
\end{figure}

\begin{figure}
	
	\centering
	\begin{subfigure}{0.37\textwidth}
		\includegraphics[width=\textwidth]{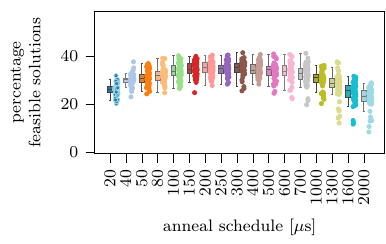}
		\caption{\trpsym{6}.}
	\end{subfigure}
	\begin{subfigure}{0.30\textwidth}
		\centering
		\includegraphics[width=\textwidth]{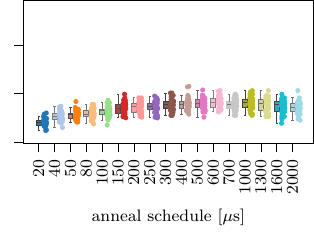}
		\caption{\trpsym{7}.}
	\end{subfigure}
	\begin{subfigure}{0.30\textwidth}
		\includegraphics[width=\textwidth]{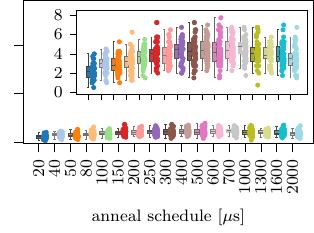}
		\caption{\trpsym{8}.}
	\end{subfigure}
	\begin{subfigure}{0.37\textwidth}
		\includegraphics[width=\textwidth]{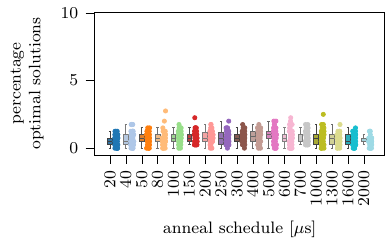}
		\caption{\trpsym{6}.}
	\end{subfigure}
	\begin{subfigure}{0.305\textwidth}
		\centering
		\includegraphics[width=\textwidth]{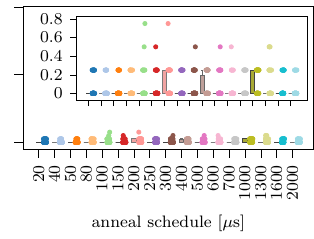}
		\caption{\trpsym{7}.}
	\end{subfigure}
	\begin{subfigure}{0.305\textwidth}
		\includegraphics[width=\textwidth]{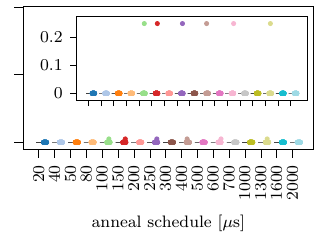}
		\caption{\trpsym{8}.}
	\end{subfigure}
	\caption{Percentage of feasible and optimal solutions for symmetrical distribution of cities with different anneal schedules. The chain strength values of $0.6$ is used for \trpsym{6} and $0.9$ for \trpsym{7}, and \trpsym{8}. The number of anneal reads are 20000 for each anneal schedule. Clique embedding was used to embed the QUBO. The calculations were executed on \dwaveAdvantage\ and results were obtained on 2023/11/24.}
	\label{Box plot for symmetry cities of annealing schedule}
\end{figure}

\paragraph{Anneal schedule}
As we have discussed in Section~\ref{sec:quantum-systems-dwave}, one of the most 
important parameters is the annealing time. Here, we discuss the impact of the annealing schedule on enhancing the likelihood of obtaining feasible and optimal solutions. We employed multiple anneal schedules for the annealing to observe the solution quality. Optimal values of chain strength and QUBO penalty were chosen from the heatmap analysis for all problem cases. Different anneal schedules were chosen from shorter times to longer times. We plotted the percentage of feasible and optimal solutions for asymmetric and symmetric distributions. Each marker in the Figures \ref{Box plot for asymmetry cities of annealing schedule} and \ref{Box plot for symmetry cities of annealing schedule} corresponds to the percentage of feasible solutions for 400 anneal reads. There are 50 such markers for each anneal schedule resulting in total of $20000$ anneal reads. The feasible solutions for \trpasym{6} has an inverted parabola type behavior, see Figure \ref{Box plot for asymmetry cities of annealing schedule}. Rapid annealing disrupts the system's ability to maintain the ground state, leading to transitions to higher energy states associated with suboptimal or infeasible solutions. Conversely, longer anneal schedules beyond $600 \, \mu s$ also reduce the percentage of feasible solutions, possibly due to factors like thermal relaxation or qubit noise. 
%Additionally, offer a higher likelihood of feasible solutions but do not guarantee maximal outcomes at all times.
We observe the same beahviour in \trpasym{7} but with a lesser intensity. For \trpasym{8} the impact of the anneal schedule is not as significant as in the other cases. The inverted parabola behaviour is less pronounced in optimal solutions.\\

\begin{table}
	\centering
	\begin{tabular}{c|ccccc}
		example & \# logical qubits & \# physical qubits & \# of chains & chain length \\
		\hline\hline
		\trpasym{6}, \trpsym{6} & 36&172 &36 & 5 \\
		\trpasym{7}, \trpsym{7}& 49& 284&49 &6  \\
		\trpasym{8}, \trpsym{8} & 64& 436& 64&7  \\	
	\end{tabular}
	\caption{The number of logical qubits, physical qubits, chains, and chain length for all the problem sizes in the TRP use case. The symmetric and asymmetric problems have the same numbers. }
	\label{Table for qubit and chain number TRP}
\end{table}

% \FloatBarrier

For the symmetric distribution we can see the inverted parabola behavior in symmetrical distribution of cities, see Figure~\ref{Box plot for symmetry cities of annealing schedule}. For all the problem sizes the feasible solutions peak at $300-400 \, \mu s$, and drop at shorter or longer anneal schedules. In contrast to the asymmetrical distribution, in the \trpsym{7} and \trpsym{8} cases the number of optimal solutions were found more often in anneal schedules which are longer in duration. This behavior is possibly caused by scaling effects of the hardware coupling strengths calculated from the quadratic terms in the QUBO. In the case of a radial symmetric placement of cities, the difference between shortest and longest route is smaller than for an asymmetric placement. After scaling the maximum QUBO coefficient (corresponding to the longest possible route) to the maximum hardware coupling term, the symmetric TRP will result in a larger coupling term for the shortest route than in the asymmetric case. This has the effect that the energy scale of all hardware couplings is larger and thus we have a larger spacing between ground and excited states. A wider gap is directly connected to the success of staying in the groundstate during annealing and finding a optimal solution. Therefore, symmetric distributions of cities have a higher percentages of optimal solutions \cite{Dwave_reformulation}.

% \FloatBarrier

% Conclusion
\section{Conclusion}
\label{sec:conclusion}
Quantum computers have undergone an impressive development in recent decades
as it can be observed, for example, by the continuously increasing qubit numbers,
the decreasing error rates, or from a plurality of component or system level 
benchmarks. This advancement has sparked interest in a large variety of domains
and the promise of solving the most challenging 
computational problems in each discipline is in the air. Both, the
current quantum computing systems as well as the quantum software stack,
are now mature enough to review this promise. 
It is critical to accomplish this not only on an academic level but more importantly
for industry relevant applications.

In this spirit, we considered in this paper two industry use cases from the field of 
discrete optimization, which is among the
most prominent candidates for the application of quantum computing. 
We carried out extensive and systematic experiments on the superconducting IBM
quantum computers as well as on a D-Wave quantum annealer to illustrate the 
current status of these technologies for optimization problems. Our main conclusion
is that 
there are still significant challenges on the way to a practical use. 
In particular, we observe that on the IBM quantum systems
only small proof-of-concept examples can be executed with an acceptable quality.
For larger problem instances the solution quality quickly deteriorates and throughout
all examples we find a lack of reproducibility and predictability. However,
comparing the different generations of IBM quantum processors we could clearly
observe the improvement of the result quality with the evolution of the
hardware. On the more mature quantum annealing system of D-Wave we could 
execute larger problem instances bringing us one step closer from proof-of-concept
to real-world examples. Nevertheless, again the solution quality decreases with
the problem size which poses a clear limit to the usability also of
quantum annealers for industry relevant problems.

In summary, this paper gave a comprehensive insight into the current state 
of quantum computing based on two industry relevant optimization problems. 
While we observed that today's quantum computers are still limited, the
examples and experiments in this paper provide a blueprint to track the progress
and the readiness for real world applications of the fast evolving quantum
computing ecosystem.

\section*{Code Availability}
The implementation of the LamA use case and of the
corresponding example series can be found 
\href{https://gitlab.cc-asp.fraunhofer.de/fraunhofer_iao_qc/sequoia_end-to-end/optimization-ev-charging-schedules}{here}.
The truck routing use case is available 
\href{https://gitlab.cc-asp.fraunhofer.de/fraunhofer_iao_qc/sequoia_end-to-end/truck-fleet-route-planning-in-supply-chain-management}{here}\footnote{Content will be available by end of April 2024.}.
The IBM and D-Wave experiment data is available on request.

\section*{Acknowledgement}
A.~S. thanks Niclas Schillo for support in the experiments on
the impact of cross talk in the QAOA circuits.
All authors thank Christian Tutschku for proofreading.
This work was funded by the Ministry of Economic Affairs, Labour, and Tourism 
Baden-W\"urttemberg in the frame of the Competence Center Quantum Computing
Baden-W\"urttemberg (project SEQUOIA End-to-End)

\bibliography{literature}
\bibliographystyle{quantum}

\appendix
\section{Additional Results for LamA Use Case}

In this part of the appendix, we provide the convergence plots of the
classical optimizer COBYLA for \lamaEx{0}{2} and \lamaEx{2}{3}. 
The setup is the same as described in detail in 
Section~\ref{sec:lama-classical-optimization} and in 
Figure~\ref{fig:convergence-cobyla-exp1p3-exact-simulation} for
\lamaEx{1}{3}.

\begin{figure}[h]
    \small
    \centering
    \begin{subfigure}{0.33\textwidth}
        \includegraphics{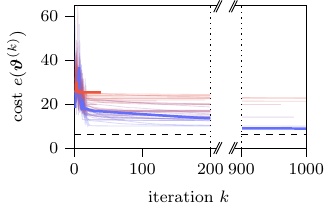}
        \caption{QAOA, $\repsQaoa = 1$, $\numQaoaParams = 2$.}
    \end{subfigure}
    \; \;
    \begin{subfigure}{0.3\textwidth}
        \includegraphics{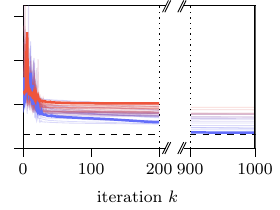}
        \caption{QAOA, $\repsQaoa = 2$, $\numQaoaParams = 4$.}
    \end{subfigure}
    \begin{subfigure}{0.3\textwidth}
        \includegraphics{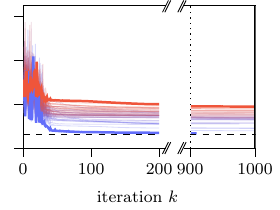}
        \caption{QAOA, $\repsQaoa = 3$, $\numQaoaParams = 6$.}
    \end{subfigure}
    \\[0.2cm]
    \begin{subfigure}{0.33\textwidth}
        \includegraphics{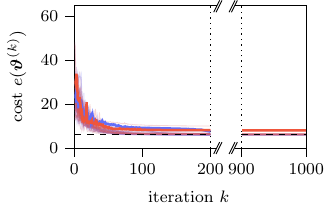}
        \caption{VQE, $\repsVqe = 1$, $\numVqeParams = 12$.}
    \end{subfigure}
    \; \;
    \begin{subfigure}{0.3\textwidth}
        \includegraphics{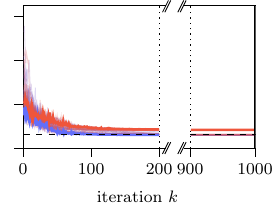}
        \caption{VQE, $\repsVqe = 2$, $\numVqeParams = 18$.}
    \end{subfigure}
    \begin{subfigure}{0.3\textwidth}
        \includegraphics{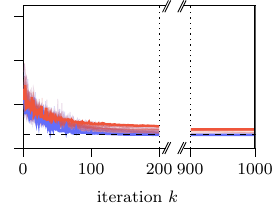}
        \caption{VQE, $\repsVqe = 3$, $\numVqeParams = 24$.}
    \end{subfigure}
    \\[0.4cm]
    \begin{subfigure}{0.33\textwidth}
        \includegraphics{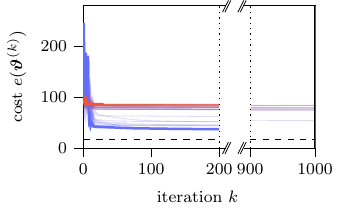}
        \caption{QAOA, $\repsQaoa = 1$, $\numQaoaParams = 2$.}
    \end{subfigure}
    \; \;
    \begin{subfigure}{0.3\textwidth}
        \includegraphics{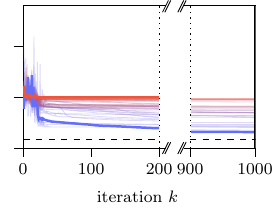}
        \caption{QAOA, $\repsQaoa = 2$, $\numQaoaParams = 4$.}
    \end{subfigure}
    \;
    \begin{subfigure}{0.3\textwidth}
        \includegraphics{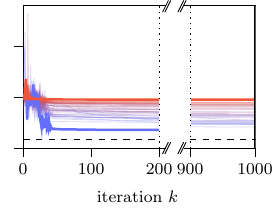}
        \caption{QAOA, $\repsQaoa = 3$, $\numQaoaParams = 6$.}
    \end{subfigure}
    \\[0.2cm]
    \begin{subfigure}{0.33\textwidth}
        \includegraphics{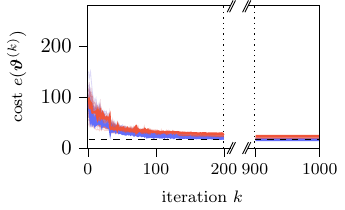}
        \caption{VQE, $\repsVqe = 1$, $\numVqeParams = 32$.}
    \end{subfigure}
    \; \;    %
    \begin{subfigure}{0.3\textwidth}
        \includegraphics{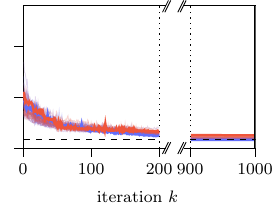}
        \caption{VQE, $\repsVqe = 2$, $\numVqeParams = 48$.}
    \end{subfigure}
    \;
    \begin{subfigure}{0.3\textwidth}
        \includegraphics{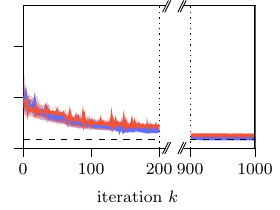}
        \caption{VQE, $\repsVqe = 3$, $\numVqeParams = 64$.}
    \end{subfigure}
    \caption{
        Convergence of the classical optimizer COBYLA for
        \lamaEx{0}{2} (two upper rows) and \lamaEx{2}{3} (two lower rows).
        The setup is the same as described in 
        Figure~\ref{fig:convergence-cobyla-exp1p3-exact-simulation}.
    }
    %
    % \label{fig:convergence-cobyla-exp0p2-exact-simulation}
    \label{fig:convergence-cobyla-ex0p2-ex2p3-exact-simulation}
\end{figure}

\section{Information on IBM Quantum Computing Resources}
In this part of the appendix, we provide background information on the IBM
quantum computing systems, on which we the experiments presented in
this paper were executed. The data stems from the dates
when these experiments where carried out.

\begin{figure}[h]
    \centering
    \begin{subfigure}{0.45\textwidth}
        \centering
        \includegraphics{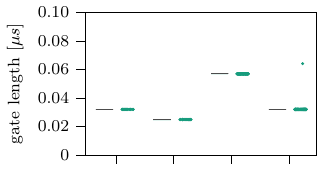}
        \caption{Gate length of $\pauliX$ gate.}
    \end{subfigure}
    \;
    \begin{subfigure}{0.45\textwidth}
        \centering
        \includegraphics{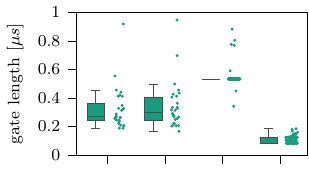}
        \caption{Gate length of two qubit gate.}
    \end{subfigure}
    \\[0.2cm]
    \begin{subfigure}{0.45\textwidth}
        \centering
        \includegraphics{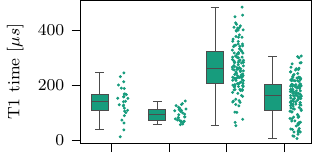}
        \caption{T1 time for each qubit.}
    \end{subfigure}
    \;
    \begin{subfigure}{0.45\textwidth}
        \centering
        \includegraphics{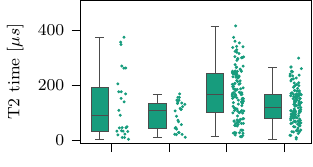}
        \caption{T2 time for each qubit.}
    \end{subfigure}
    \\[0.2cm]
    \begin{subfigure}{0.45\textwidth}
        \centering
        \includegraphics{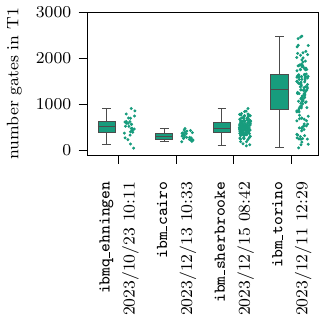}
        \caption{Number of two qubit gates in T1 time.}
    \end{subfigure}
    \;
    \begin{subfigure}{0.45\textwidth}
        \centering
        \includegraphics{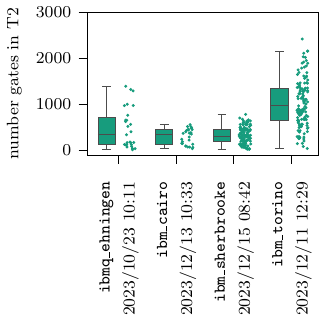}
        \caption{Number of two qubit gates in T2 time.}
    \end{subfigure}
    \caption{
        Gate lengths, coherence times, and
        number of two qubit gates, that can be executed in the coherence
        times, of the different backends used in this paper.
        For the number of gates in the coherence time we used the median of the
        gate times.
        The calibration data stems from the dates when the experiments in 
        Section~\ref{sec:lama-ibmq-ehningen} and \ref{sec:lama-other-ibmq}
        were executed.
        The respective two qubit gate of the backends can be found
        in Table~\ref{tab:ibm-quantum-systems}.
    }
    \label{fig:ibmq-coherence-times}
\end{figure}

\begin{figure}[t]
    \centering
    \begin{subfigure}{0.34\textwidth}
        \includegraphics{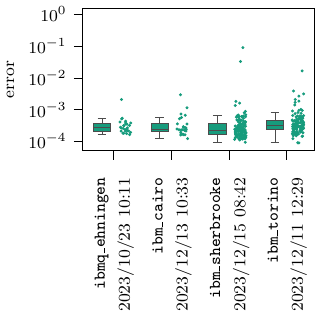}
        \caption{Error of the $\pauliX$ gate.}
    \end{subfigure}
    \; \;
    \begin{subfigure}{0.3\textwidth}
        \includegraphics{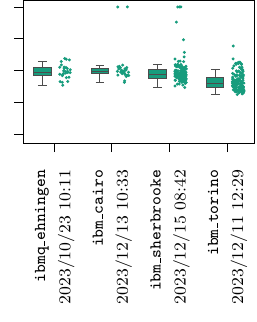}
        \caption{Error of the two qubit gate.}
    \end{subfigure}
    \begin{subfigure}{0.3\textwidth}
        \includegraphics{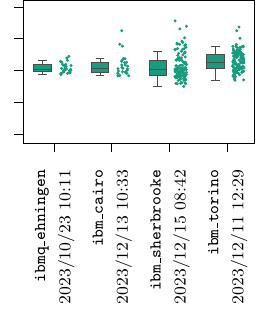}
        \caption{Readout error.}
    \end{subfigure}
    \caption{
        Errors of the hardware gates and of the readout 
        for the different backends used in this paper.
        We only give the error of the $\pauliX$ and the two qubit gate since
        the $\RZ$ gate is a virtual gate and thus does not have an 
        error and the error of the $\SX$ gate agrees with the one of the
        $\pauliX$ gate.
        The calibration data stems from the dates when the experiments in 
        Section~\ref{sec:lama-ibmq-ehningen} and \ref{sec:lama-other-ibmq}
        were executed.
        The respective two qubit gate of the backends can be found
        in Table~\ref{tab:ibm-quantum-systems}.
    }
    \label{fig:ibmq-gate-errors}
\end{figure}

\vfill

\end{document}